\documentclass[a4paper,fleqn]{cas-sc}
\usepackage[authoryear]{natbib}

\newif\ifhideprelim

\hideprelimtrue

\usepackage{amssymb}

\ExplSyntaxOn
\RenewDocumentCommand \eadauthor {} 
    { 
      \seq_map_inline:Nn \l_stm_au_seq 
        { 
            \regex_extract_once:nnNTF {(\w)\w*-(\w)} { ##1 } \l_stm_au_fn_seq
            { 
                \seq_pop_left:NN \l_stm_au_fn_seq \temp_var
                \seq_use:Nn \l_stm_au_fn_seq { .- }
                { . } 
            }
            { 
                \regex_match:nnTF { \. } { ##1 } 
                { ##1 }
                { \tl_head:n {##1}. }
            }
      }{ ~\l_stm_au_sn_seq }
    }
\ExplSyntaxOff
\ExplSyntaxOn
\makeatletter

\RenewDocumentEnvironment { PrelimsAbstract } { O{} }
  {\parindent=0pt 
   { \fontsize{14pt}{16pt}\selectfont #1 }\par
   \vskip12pt
   { \fontsize{12pt}{14pt}\bfseries\selectfont\casprelimstitle } \par
   \vskip6pt
   \ifnum\theblind>0\relax
     \vspace*{\the\baselineskip}
   \else  
     \seq_use:Nn \g_stm_prelimsau_seq { ,\ }
   \fi  
   \vskip12pt
   \par 
  } 
  {}
  
\makeatother
\ExplSyntaxOff
\newif\ifabbreviation
\pretocmd{\thebibliography}{\abbreviationfalse}{}{}
\AtBeginDocument{\abbreviationtrue}
\DeclareRobustCommand\acroauthor[2]{%
  \ifabbreviation
    \ifcsname acroused@#2\endcsname
      #2%
    \else
      #1%
      ~[\mbox{#2}]
      \expandafter\gdef\csname acroused@#2\endcsname{}%
    \fi
  \else
    \ifcsname bibacroused@#2\endcsname
        \mbox{#2}%
    \else
        \mbox{#1}~(\mbox{#2})%
        \expandafter\gdef\csname bibacroused@#2\endcsname{}%
    \fi
  \fi
}

\usepackage{lipsum}

\usepackage{tabularx}
\usepackage{booktabs}
\usepackage{pdflscape}
\usepackage{rotating}
\usepackage{longtable}
\usepackage{multirow}
\usepackage{caption}
\usepackage{float}
\usepackage{ragged2e}
\usepackage{caption}

\usepackage{cleveref}
\usepackage{subfigure}
\usepackage{siunitx}
\DeclareSIUnit \decibelA {dB(A)}
\DeclareSIUnit \decibelC {dB(C)}
\DeclareSIUnit \soneGF {soneGF}
\DeclareSIUnit \acum {acum}
\DeclareSIUnit \asper {asper}
\DeclareSIUnit \vacil {vacil}
\DeclareSIUnit \tuhms {tuHMS}

\DeclareUnicodeCharacter{202F}{FIX ME!!!!}

\usepackage{xcolor}
\definecolor{set1_1}{RGB}{228,26,28}
\definecolor{set1_2}{RGB}{55,126,184}
\definecolor{set1_3}{RGB}{77,175,74}
\definecolor{den_1}{RGB}{239,209,0}
\definecolor{den_2}{RGB}{78,184,123}
\definecolor{den_3}{RGB}{0,127,196}
\definecolor{loaclr}{RGB}{152, 78, 163}
\definecolor{mclr}{RGB}{255, 127, 0}

\usepackage{hyperref}
\hypersetup{
    colorlinks=true,
    linkcolor=blue,
    filecolor=magenta,      
    urlcolor=cyan,
    pdftitle={Overleaf Example},
    pdfpagemode=FullScreen,
    }

\usepackage{setspace}

\newif\ifshowchanges

\definecolor{lightgray}{RGB}{200,200,200}
\ifshowchanges
    \usepackage[commandnameprefix=always]{changes}
    \setcommentmarkup{\marginpar{\small\centering\color{purple}\textbf{[{#1}]}}}
    \setdeletedmarkup{ {\color{lightgray}\sout{#1}}}
\else
    \usepackage[final,commandnameprefix=always]{changes}
\fi

\begin{document}
\let\WriteBookmarks\relax
\def\floatpagepagefraction{1}
\def\textpagefraction{.001}

\newcommand*{\papertitle}{Anti-noise window: subjective perception of active noise reduction and effect of informational masking}
\shorttitle{\papertitle}    
\shortauthors{Lam et al.}  

\title[mode=title]{\papertitle}

\author[eee]{Bhan Lam}[
    auid=000,
    bioid=001,
    degree=PhD,
    orcid=0000-0001-5193-6560]
\ead{blam002@e.ntu.edu.sg}
\corref{c}\cortext[c]{Corresponding author}
\credit{Conceptualization, Methodology, Software, Validation, Formal analysis, Investigation, Project administration, Data Curation, Writing - Original Draft, Writing - Review \& Editing, Visualization, Supervision}

\author[eee]{Kelvin Chee Quan Lim}
\credit{Software, Validation, Investigation, Data Curation}

\author[eee]{Kenneth Ooi}[orcid=0000-0001-5629-6275]
\credit{Methodology, Formal analysis,Writing - Review \& Editing, Visualization}

\author[eee]{Zhen-Ting Ong}[orcid=0000-0002-1249-4760]
\credit{Project administration, Data Curation}

\author[eee]{Dongyuan Shi}[orcid=0000-0003-0768-6386,degree=PhD]
\credit{Methodology, Writing - Review \& Editing, Supervision}

\author[eee]{Woon-Seng Gan}[orcid=0000-0002-7143-1823, degree=PhD]
\credit{Conceptualization, Writing - Review \& Editing, Supervision}

\affiliation[eee]{organization={School of Electrical and Electronic Engineering, Nanyang Technological University},
            addressline={50 Nanyang Avenue}, 
            city={Singapore 639798},
            country={Singapore}}

\tnotemark[1]
\tnotetext[1]{The research protocols used in this research were approved by the institutional review board of Nanyang Technological University, Singapore [IRB-2021-386].}

\begin{abstract}
 Reviving natural ventilation (NV) for urban sustainability presents challenges for  indoor acoustic comfort. Active control and interference-based noise mitigation strategies, such as the use of loudspeakers, offer potential solutions to achieve acoustic comfort while maintaining NV. However, these approaches are not commonly integrated or evaluated from a perceptual standpoint. This study examines the perceptual and objective aspects of an active-noise-control (ANC)-based "anti-noise" window (ANW) and its integration with informational masking (IM) in a model bedroom. Forty participants assessed the ANW in a three-way interaction involving noise types (traffic, train, and aircraft), maskers (bird, water), and ANC (on, off). The evaluation focused on perceived annoyance (PAY; ISO/TS 15666), perceived affective quality (ISO/TS 12913-2), loudness (PLN), and included an open-ended qualitative assessment. Despite minimal objective reduction in decibel-based indicators and a slight increase in psychoacoustic sharpness, the ANW alone demonstrated significant reductions in PAY and PLN, as well as an improvement in ISO pleasantness across all noise types. The addition of maskers generally enhanced overall acoustic comfort, although water masking led to increased PLN. Furthermore, the combination of ANC with maskers showed interaction effects, with both maskers significantly reducing PAY compared to ANC alone.  

\end{abstract}


\begin{keywords}
active noise control \sep natural sounds \sep indoor soundscape \sep auditory masking \sep subjective listening \sep soundscape augmentation 
\end{keywords}

\maketitle


\doublespacing

\section{Introduction}
\label{sec:Introduction}

\subsection{Background and Motivation} \label{sec:intro-bg}

The emergence of increased public health risks and burden of disease from environmental noise exposure have led to the recent reduction in recommended noise limits by the World Health Organisation \citep{WorldHealthOrganizationRegionalOfficeforEurope2018}. This pervasiveness of noise, especially in urban areas, and its myriad of associated severe health risks is akin to being exposed to a "new secondhand smoke" \citep{Fink2019}. Noise exposure is the most severe in urbanised areas \citep{Basu2021,Mir2023} , especially in high-density, high-rise topographies \citep{Yuan2019ExaminingChina}. The continued adoption of remote work post-pandemic has brought about renewed impetus and urgency for noise abatement in now dual-function dwellings (i.e. for work and rest). 

Environmental noise mitigation is trifold by nature and is most effective at the source, followed by along the propagation path, and finally at the receivers' end \citep{Lam2021,EuropeanEnvironmentAgency2020}. Land-scarce, high-density and high-rise urban areas, however, reduce the feasibility and efficacy of traditional noise management measures along the propagation path (e.g. low high-rise efficacy of noise barriers, unavailable land for building setback guidelines) and at the source (e.g. limited airspace) \citep{Lam2021,Bin2019}. Hence, despite the reduced efficacy, mitigation at the receivers' end (e.g., windows) plays an increasingly important role in the overall noise management strategy. 

\chadded[comment=R0.1]{However, the path towards sustainable urban development, as outlined in the United Nations General Assembly’s
2015 United National Sustainable Development Goals \citep{UnitedNationsGeneralAssembly2015}, has brought forth additional challenges in effectively mitigating noise in residential dwellings. One prominent solution is natural ventilation (NV), which has the potential to decrease building energy consumption by up to \SI{50}{\percent} \citep{Tong2016,Cao2016BuildingDecade,Spandagos2017EquivalentCities,Zhong2023}. The COVID-19 pandemic has further highlighted the importance of NV for infection control and safeguarding public health \citep{Nathalie2011HealthSector,Morawska2020HowMinimised,Amirzadeh2023}. Nonetheless, the increased emphasis on NV necessitates a trade-off between NV and noise insulation, particularly concerning the openings in the building facade, such as windows.  Additionally, research suggests that access to outside sounds as a result of natural ventilation increases the overall sense of contentment \textit{content} \citep{Torresin2020IndoorBuildings}. This perception of contentment aligns with the concepts of arousal and eventfulness in the circumplex models of affect \citep{Russell1980AAffect.}, and perceived affective quality \citep{iso12913-3}, respectively. The combination of increased contentment and comfort in the presence of outdoor natural sounds creates an engaging environment suitable for both work-from-home scenarios and relaxation \citep{Torresin2022}. Therefore, it is crucial to implement noise mitigation strategies that preserve NV, thus addressing the inherent trade-off of increased noise when opening windows for NV.} 

\chadded[comment=R2.1]{\subsection{Active and passive control strategies for naturally-ventilated façade openings}} \label{sec:intro-ap}

Of the NV-preserving noise control measures for façade openings \citep{Tang2017,Torresin2019a}, active noise control (ANC) strategies have demonstrated significant noise reduction performance with minimal impact on airflow \citep{Lam2021}. In contrast to passive control strategies, wherein physical elements (e.g. metastructures \citep{Fusaro2019a,Fusaro2020,Pan2020}, plenum windows \citep{Du2020,Li2020a}) are used to disrupt the sound waves, ANC utilises active elements (e.g. microphones, processing units) that drive actuator(s) (e.g. loudspeaker) to generate an ``anti-noise'' wave that destructively interferes with the impinging noise to attenuate it. Recently, proof-of-principle ``anti-noise'' window systems demonstrated substantial noise mitigation potential on open façade apertures for NV \citep{Lam2020c,Lam2020}. The anti-noise window (ANW) employs sensors (e.g. microphones) outside the window to obtain time-advanced information of the impinging noise, and subsequently computes an equal and opposing anti-noise wave that is reproduced from an array of spatially-optimised loudspeakers, resulting in destructive interference (i.e. acoustic attenuation). 

\chadded[comment={R2.1,\\R2.2}]{While the distinction between active and passive noise control strategies is elucidated in \citet{Lam2021}, it is important to recognize their complementary nature, particularly in terms of their efficacy within specific frequency ranges. This relationship is exemplified in ANC headphones, where the sealed earcup provides passive insulation to high frequencies ($>\SI{1000}{\hertz}$) and ANC effectively attenuates residual low-frequencies ($<\SI{1000}{\hertz}$). However, in a recent evaluation of passive metawindows for noise control, it is plausible that the attenuation of high frequencies could potentially heighten perceptual annoyance due to the innate human affinity and sensitivity to sounds within this range  \citep{Fusaro2022AssessmentPerception,Haapakangas2020}. Therefore, the ANW is aptly suited to mitigate low-frequency urban transportation noise while allowing crucial high-frequency sounds to pass through, such as the melodies of birds and the resonance of human voices. Moreover, active components offer the added benefit of flexibility in control, allowing for selective mitigation of specific noises or programming to enable the passage of important sounds, such as alarms \citep{Shi2020h,Shi2022a,Shi2023a}.}

\subsection{Informational Masking} \label{sec:intro-IM}

Noise control measures for façade openings that preserve NV, with active or passive techniques, are still technically challenging and are effective for a limited range of acoustic frequencies \citep{Lam2021}. Hence, active interference techniques -- an active sound source that projects audio material to ``mask'' the noise source -- that have been widely employed in office spaces to provide speech privacy could potentially be extended to dwellings \citep{Lenne2020}. Similar to ANC, active interference employs an active sound source to project the masking sounds (``maskers'') and would also allow for NV. 

When these maskers overlap spectrotemporally such that the target sound (``noise'') can no longer be audible, ``energetic masking'' occurs \citep{Brungart2001,Culling2017a}. This is akin to the principles in the psychoacoustic masking models employed in audio compression standards \citep{InternationalOrganizationforStandardization1993,Kidd2008}, where only audible sounds are encoded. In contrast, ``informational masking'' (IM) is a broad categorization of phenomena that result in the reduced audibility of noise even in the absence of spectrotemporal overlap with the maskers \citep{Kidd2017}. With the adoption of the soundscape ISO standard \citep{InternationalOrganizationforStandardization2014ISOFramework,InternationalOrganizationforStandardization2018,iso12913-3}, there is increasing evidence showing that IM or augmentation of outdoor urban noise with biophilic maskers positively impacts affect and health outcomes \citep{Buxton2021,Hong2020h,Leung2017,Coensel2011b,VanRenterghem2020,Jeon2010,Li2022}.

Owing to the dependency of the soundscape perception on context \citep{InternationalOrganizationforStandardization2014ISOFramework}, however, biophilic maskers that were effective in outdoor urban environments may not be perceived equally in indoor spaces \citep{Torresin2019,Torresin2020IndoorBuildings}. With limited research on biophilic maskers for indoor soundscapes, the suitability and efficacy of IM as an NV-preserving noise mitigation method in indoor spaces still remain an open question  \citep{Hasegawa2021AudiovisualReview}.

\subsection{Human perception in noise control} \label{intro-human}

Since both ANC and active interference techniques require similar active elements, they have been integrated to form a class of perceptually-driven ANC techniques. However, perceptual ANC systems are usually based on energetic masking principles to shape the residual audio after ANC to achieve desired perceptual objectives \citep{Kajikawa2012c,Jiang2018,Lam2021}. For instance, most of the perceptually-driven ANC systems were designed to meet industrial demand for automobile cabin sound quality \citep{Rees2006AdaptiveSound-profiling,Patel2017ModifiedProfiling,Jiang2018}, and music playback quality in ANC headphones \citep{Belyi2019}. With only a single reported instance of integration of ANC with stream sounds to mitigate snoring noise \citep{Liu2019}, the combination of biophilic maskers with ANC remains underexplored, at least for NV indoor spaces.

Due to the widespread adoption of A-weighted SPL [i.e. \si{\decibelA}] in noise regulations and WHO guidelines, the effectiveness of noise abatement measures is almost always evaluated with \si{\decibelA}-based measures such as insertion loss (IL) and reduction in A-weighted equivalent sound pressure levels (SPL). However, there is growing evidence pointing to the shortcomings of A-weighted SPL in its lack of association with annoyance \citep{Kang2017,Lercher2017CommunityAnalysis}, and its underestimation of low-frequency noise and its potential health effects \citep{AraujoAlves2020a,Baliatsas2016}. It is worth highlighting that suppression of low-frequencies in \si{\decibelA}-based metrics also potentially undermines the perceptual benefit of ANC, which is predominantly effective at mitigating low-frequency noise \citep{Cheer2016ActiveYacht}. Moreover, the overall annoyance perceived in dwellings is modulated by different noise types (e.g. aircraft, traffic, train) and their interactions that are difficult to be captured in \si{\decibelA}-based metrics \citep{Munzel2014CardiovascularExposure,Miedema2004a, Lercher2017CommunityAnalysis}. With the introduction of maskers, an increase in SPL is to be expected, and thus alternative evaluation methodologies that evaluate the overall perception of the acoustic environment, such as the triangulation method in ISO 12913-3, should be adopted \citep{iso12913-3,Fusaro2022AssessmentPerception}.

\subsection{Research questions} \label{sec:intro-RQ}

To address the dearth of research regarding the perception of ANC and its interaction with biophilic maskers in indoor soundscapes, this study evaluates the perceptual effects of an ANC-based ANW in mitigating typical urban noise and its interaction with a proposed introduction of biophilic maskers. In particular, the following research questions would be addressed:

\begin{enumerate}
  \item[RQ1.] Does the active reduction of low-frequencies in typical urban noise types by an ANW translate to increased comfort (i.e. perceived annoyance, perceived loudness, perceived affective quality, sentiment) in the context of dwellings?
  \item[RQ2.] In terms of objective performance and comfort perception, what is the effect of IM with representative biophilic maskers on the ANW in mitigating typical urban noise?
  \item[RQ3.] In the context of evaluating novel acoustic solutions to improve indoor soundscape, can perceived acoustic comfort be reliably estimated with only objective parameters?
\end{enumerate}

\section{Method}
\label{sec:Method}

\subsection{Study site and administration}

The experiment was conducted in a model bedroom with a fully-opened two-pane sliding window that is housed in a recording studio. The full-sized sliding window was mounted with a security grille, to which the active noise control system was secured. A detailed description of the construction of the model bedroom and the active noise control system can be found in the referenced PhD thesis and prior work \citep{Lam2020c,Lam2019}. 

The recording studio was temperature controlled  at approximately $\SI{25}{\degreeCelsius}$ and significantly isolated from outside noise, with a noise floor of $32.16\pm0.05~\si{\decibelA}$. \chadded[comment=R2.11]{To facilitate interaction with the ANW, a custom web-based graphical user interface (GUI; G Web Development Software 2021, National Instruments Corporation, TX, USA) was implemented for the questionnaire. The GUI was presented on an electronic tablet, and an accompanying stylus was provided for more accurate interaction with the input elements, such as moving sliders and handwriting recognition for open-ended input.}

Formal ethical approval was obtained from the Institutional Review Board of Nanyang Technological University (IRB 2021-386) before participant recruitment and the conduct of experiments.

\subsection{Stimuli}

Taking reference from the WHO noise guidelines \citep{EuropeanEnvironmentAgency2020}, three of the most prevalent transportation noise sources (i.e. aircraft, train and traffic) were examined in the local context. For comparability, three stimuli used to evaluate the proof-of-principle anti-noise window prototype was employed in this study \citep{Lam2020c}. The aircraft noise (\texttt{AIR}) was jet-powered and recorded from an apartment window under the landing flight path. Traffic noise (\texttt{TRA}) was recorded from an apartment with direct line of sight (LoS) to a 6-lane expressway in central Singapore. The train noise was from the Mass Rapid Transit (\texttt{MRT}) system and was also recorded from an apartment with LoS to the overground tracks. 

\chadded[comment=R2.4]{Due to recent upgrades to \texttt{MRT} trains and railways, it is worth noting that at the time of recording, the \texttt{MRT} train was traversing an overground stretch of the \textit{East--West} railway supported by wooden sleepers. The \texttt{MRT} train consisted of electrical multiple unit rolling stock capable of reaching top speeds of \SI{80}{\kilo\meter/\hour}, with an average speed of \SI{41}{\kilo\meter/\hour} \citep{Lee2009}. The acoustic signature of the \texttt{MRT} train is characterised by a strong tonal component at 700 Hz (see Figure 3 in \citet{Lam2020c}) and typically registers sound pressure levels of around \SI{72.5}{\decibelA} at nearby residential units \citep{Lee2009}.}

The ANC was based on the same fixed-filter implementation as in \citet{Lam2020c}, where the control filters were pre-trained with the respectively bandlimited (0.1 to \SI{1}{\kilo\hertz}) noise tracks. 
Unlike the referenced prior study and most ANC studies, unfiltered full frequency bandwidth noise stimuli were evaluated in this study to provide a better approximation of real-world perception. Therefore, the pre-trained filters were subsequently used to control the full bandwidth noise with the anti-noise window prototype (i.e. active control only occurs from 0.1 to \SI{1}{\kilo\hertz} in the full-band noise).

To investigate the informational masking effects of natural sounds, two representative biophilic masker types were employed: birdsong and water sound. Both maskers were among the most pleasant birdsong and water sounds as recommended by a masker recommendation system to improve the pleasantness [i.e. formula A.1 in \citet{iso12913-3}] of traffic noise \citep{Ooi2023a}. The artificial-intelligence-based autonomous masker selection system was trained on \num[]{12600} subjective responses to \num[]{12000} unique soundscapes, which included 56 unique maskers (i.e. birdsongs, traffic, construction, water, wind). The birdsong used in this study was originally retrieved from xeno-canto: ``Orchard Oriole (Icterus spurius)'' by \textit{Matt Wistrand} (\url{https://xeno-canto.org/482053}) licensed under \texttt{CC BY-NC-SA 4.0}, while the water sound was from freesound: "small fountain" by \textit{roman\_cgr} (\url{https://freesound.org/s/415027/}) licensed under \texttt{CC0 1.0}.

A total of 2 ANC conditions (i.e. on, off), 3 noise types (i.e. \texttt{TRA}, \texttt{MRT} and \texttt{AIR}), and 3 masking conditions (i.e. none, water, bird) resulted in $2\times3\times3=18$ stimulus combinations at a noise SPL level of \SI{65}{\decibelA}. Additionally, the noise-only conditions across the 3 noise types were evaluated with both ANC conditions at \SI{70}{\decibelA}, resulting in an additional $3\times2=6$ combinations. Lastly, a \SI{60}{\decibelA} \texttt{TRA} stimulus was introduced as a control and test-retest reliability check. In total, there were 25 unique stimuli generated for this study. 

\begin{figure}[t!]
    \centering
    \subfigure[]{\includegraphics[trim=0 500 0 500,clip=true,width=0.45\linewidth]{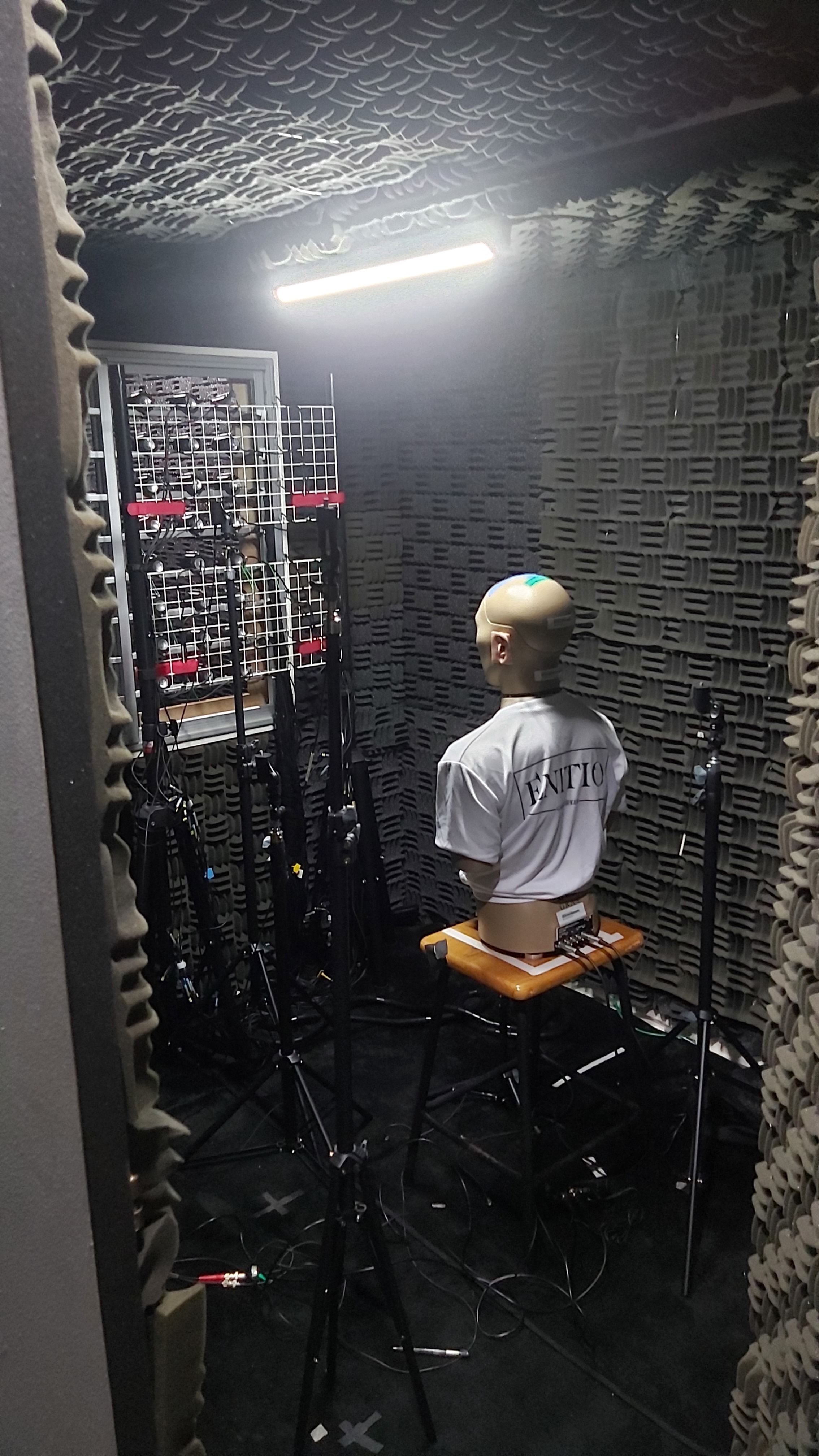}
    \label{fig:HATSroom}}%
    ~~
    \subfigure[]{\includegraphics[width=0.45\linewidth]{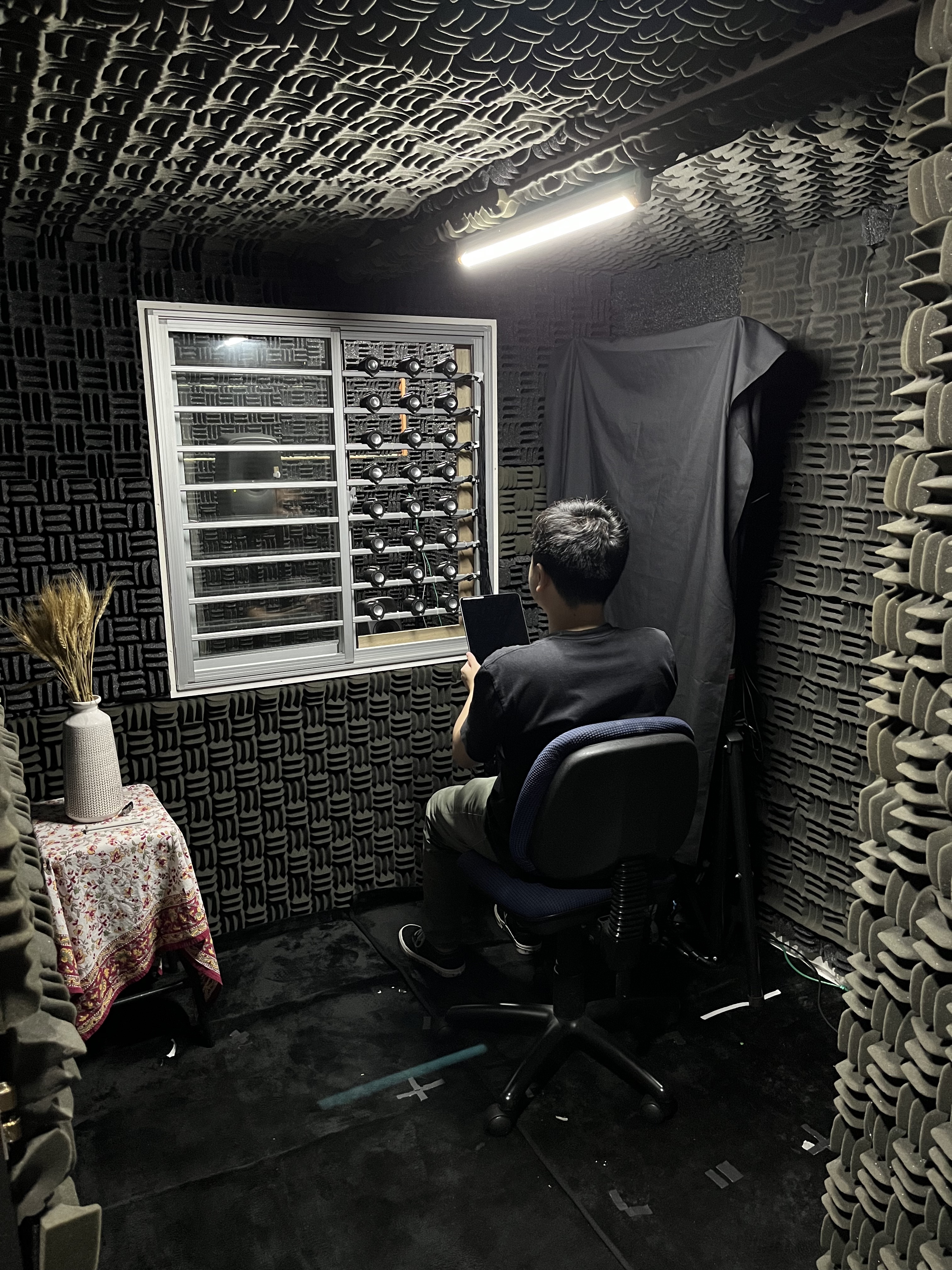}
    \label{fig:Subjroom}}%
    \\
    \subfigure[]{\includegraphics[width=0.95\linewidth]{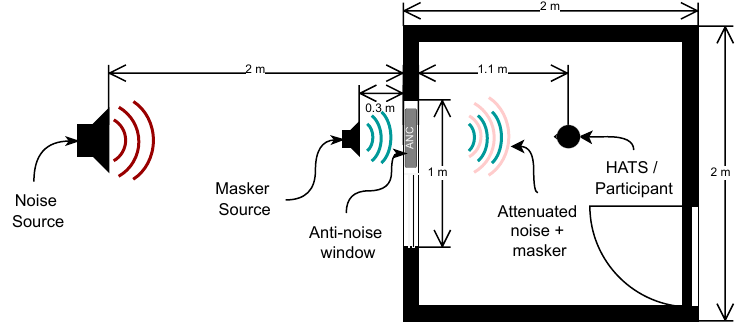}
    \label{fig:roomExpsetup}}%
    \caption{\small Interior view of the model bedroom (a) with the EA microphones and HATS setup during calibration and measurement, and (b) during subjective experiment; (c) top-view block diagram of the experiment setup with noise and masker sources placed just outside the open window of the model bedroom.}
    \label{fig:roominterior}
\end{figure}

Maskers were set to \SI{3}{\decibelA} below the noise levels in the both ANC conditions, as masker-to-noise ratio of \SI{3}{\decibelA} was previously found to invoke the greatest positive effect \citep{Galbrun2013,RadstenEkman2015a,Hong2019d,Hong2020h}. All stimuli were excerpted or concatenated to \SI{30}{\second}, as \SI{30}{\second} was chosen for perceptual consistency and to minimize participant fatigue \citep{Zacharov2018}. All 30-s stimuli were calibrated to the target continuous equivalent A-weighted sound pressure levels, $L_\text{Aeq,30s}$, measured with a calibrated head and torso simulator (HATS) based on energetic average between both ears. The HATS (45BB-5 KEMAR Head and Torso, G.R.A.S. Sound \& Vibration A/S, Holte, Denmark) was positioned \SI{1.1}{\meter} from the window at the ear height of \SI{1.2}{\meter} from the ground, where the listening test participants would be seated, as shown in \Cref{fig:HATSroom}. The calibration was accurate to within \SI{0.5}{\decibelA} via an automated calibration system \citep{Ooi2021b}. 

For the ease of referencing and for brevity, all decibel, \si{\decibel}, values shall herein refer to the $L_\text{eq,30s}$ values unless otherwise specified, wherein frequency weightings are indicated in parentheses, i.e. \si{\decibelA}. Stimuli combinations are also henceforth coded as $\mathcal{N}^{\{*\}}_{l}+\mathcal{M}$, where $\mathcal{N}$ is the noise type (i.e. \texttt{AIR, MRT, TRA}), $\mathcal{M}$ is the masker type (i.e. \texttt{B, W, None}), $l$ is the $L_\text{Aeq,30s}$ level of the noise in \si{\decibelA}, and the presence and absence of $^{\{*\}}$ reflects whether ANC is present or not, respectively. Hence, the traffic noise stimuli at \SI{65}{\decibelA} with ANC and with the bird masker would read: \texttt{$\text{TRA}^{*}_{\text{65dB(A)}}+\text{B}$}.

\chadded[comment={R2.5\\R2.6}]{The noise samples, including traffic, train, and aircraft sounds, were faithfully reproduced using a three-way monitor loudspeaker (8341A, Genelec Oy, Iisalmi, Finland), as shown in \Cref{fig:roomExpsetup}. This particular loudspeaker boasts exceptional features, including a large wavefront, higher dynamic range, and wider flat frequency response range, effectively emulating the planar waves emitted by environmental noise sources in the far-field, such as those generated by traffic, trains, and aircraft. These planar waves cannot be faithfully reproduced by the commonly employed omnidirectional loudspeakers used in room acoustic measurements. The suitability of the emitted planar acoustic wavefront was previously validated by \citet{Lam2019} using an acoustic camera (Type 9712-W-FEN, Brüel Kjær, Nærum, Denmark).} 

\chadded[comment={R2.5}]{ To mitigate the potential impact of reduced audio fidelity from the small loudspeakers on the ANW, the water and bird maskers were reproduced using two-way monitors (8320A, Genelec Oy, Iisalmi, Finland). The loudspeakers were placed \SI{0.3}{\meter} outside the window to emulate sound emitting from the vicinity of the window opening, as illustrated in \Cref{fig:roomExpsetup}.}

\subsection{Objective acoustic and psychoacoustic analysis}

Objective observation measurements of all stimuli, including those with active noise control activated, were made using the HATS in the same physical orientation as the calibration.  Additionally, six microphones (40PH, G.R.A.S. Sound \& Vibration A/S, Holte, Denmark) arranged according to ISO 16283-3 for energy-average SPL measurements \citep{InternationalOrganizationforStandardization2016}. hese six microphones correspond to the physical locations of microphones used in a previous study (i.e. microphones 5, 13, 14, 15, 17, and 18 in \citet{Lam2020c}). A total of 8 channels of acoustic data from the HATS and microphones were recorded with two 4-channel data recorders (SQobold, HEAD acoustics GmbH, Herzogenrath, Germany).

To account for sound pressure, temporal and spectral characteristics of the auditory sensation, both acoustic and psychoacoustic parameters were computed from the binaural HATS channels \citep{InternationalOrganizationforStandardization2018}. All parameters were calculated as the mean of both binaural channels for consistency with the calibration. The continuous equivalent A- and C-weighted acoustic indicators (i.e. $L_\text{Aeq}$, $L_\text{Ceq}$) represent the time-averaged sound pressure levels based on the 40-phon Fletcher-Munson ``equal loudness'' contour, and the approximated 100-phon equal loudness contour, respectively \citep{InternationalElectrotechnicalCommission2013a}.  Psychacoustic parameters such as loudness (\textit{N};  \citet{InternationalOrganizationforStandardization2017a}), sharpness (\textit{S}; \citet{GermanInstituteforStandardisationDeutschesInstitutfurNormung2009b}), roughness (\textit{R}; \citet{International2020}), tonality based on Sottek's hearing model \citep{EcmaInternational2009a}, fluctuation strength (\textit{FS}; \citet{Zwicker2013}), and perceived annoyance (\textit{PA}; \citet{Zwicker2013}) as well as acoustic derivatives ($L_{\text{C,eq}}-L_{\text{A,eq}}$, $L_{\text{A10}}-L_{\text{A90}}$) were included to account for temporal and spectral influences on perception. Common summary statistics such as percentage exceedance levels at \{5, 10, 50, 90, and 95\}\si{\percent}, as well as maximum levels were also included \citep{InternationalOrganizationforStandardization2016d,InternationalOrganizationforStandardization2018}, as summarised in \Cref{tab:acoustIndsummary}.
\begin{table}[t]
        \scriptsize
	\centering
	\caption{\scriptsize Acoustic and psychoacoustic indicators computed and mean across HATS channels and 5 repeated measurements for each stimuli. The set of summary statistics indicated as ``common'' were the mean, and exceedance levels for the $\{5,10,50,90,95\}$\si{\percent} percentage exceedance levels.}
	\label{tab:acoustIndsummary}
	\setlength{\tabcolsep}{3pt}
	\begin{tabularx}{\linewidth}{XllX}
		\toprule
		Indicator
		& Symbol
		& Unit
		& Summary statistics \\
		\midrule
		Sharpness
		    & $S$
		    & acum            
		    & common \\
		Loudness          
		    & $N$
		    & sone       
		    & common - mean\\
		Fluctuation strength  
		    & $\textit{FS}$
		    & vacil       
		    & common - mean\\
		Roughness             
		    & $R$
		    & asper     
		    & common \\
		Tonality               
		    & $T$
		    & tuHMS  
		    & common - mean\\ 
		$L_{\text{Aeq}}$
		    & $L_\text{Aeq}$
		    & dB 
		    & common \\
		$L_{\text{A10}}-L_{\text{A90}}$
		    & $L_{\text{A10}}-L_{\text{A90}}$
		    & dB 
		    & -- \\
		$L_{\text{C,eq}}$
		    & $L_\text{C,eq}$
            & dB 
            & common \\
        $L_{\text{Ceq}}-L_{\text{Aeq}}$
            & $L_{\text{Ceq}}-L_{\text{Aeq}}$
            & dB 
            & --\\
        Psychoacoustic Annoyance
            & $\textit{PA}$
            & --
            & --\\
		\bottomrule
	\end{tabularx}
\end{table}

\subsection{Subjective experimental design}

The participants were first required to provide basic demographic information (gender, race, nationality, age range) along with a pre-test assessment of (1) individual noise sensitivity (INS) \chadded[comment=R2.7]{via the Weinstein noise sensitivity scale (WNSS; \citet{Weinstein1978IndividualDormitory})}, (2) a baseline noise annoyance (BNA) \chadded[]{guided by ISO/TS 15666 \citep{InternationalOrganizationforStandardization2021}}, (3) Perceived Stress Scale (PSS; \citet{Cohen1983}), and (4) the WHO-Five Well-being Index (WHO-5; \citet{WorldHealthOrganization1998}). The pretest questionnaires were selected to assess potential confounding factors such as stress and mood on soundscape perception \citep{Mitchell2020,Ooi2023a}. The demographic and pre-test assessments took about \SI{5}{\minute} to complete and their exact wordings are detailed in \Cref{tab:question} in \labelcref{sec:questions}.

Since the short-form 5-item WNSS (WNSS-5) \citep{Aletta2018a,Benfield2014,Zhong2018} was deemed unreliable for the local Asian context \citep{Lam2022c}, the original 21-item WNSS was adopted \chadded[comment=R2.7]{to evaluate the INS} in this study \citep{Weinstein1978IndividualDormitory}.  

The BNA was assessed based on guidelines in ISO/TS 15666. Both the five-point categorical scales (from ``not at all'' to ``extremely'') and the 101-point numeric scale were employed to assess the baseline annoyance of 6 noise typical noise types, namely road traffic, aircraft, mass rapid transit (MRT), construction work (work site), construction work (renovations), any other noises \citep{InternationalOrganizationforStandardization2021}. 

A shortened 10-item version of the PSS scale (PSS-10) was employed to assess the baseline stress levels of the participants before the listening experiment \citep{Cohen1983}. In addition, the WHO-5 questionnaire assessed the baseline mental wellbeing of the participants \citep{WorldHealthOrganization1998}.

Upon completion of the demographic information and pre-test assessment, a short training session was provided to familiarise the participants with the environment within the chamber [\Cref{fig:Subjroom}], and with the graphical user interface (GUI) and the stylus operation. The GUI is a web application (G Web Development Software, NI, Texas, United States) that interfaces with the LabVIEW-based anti-noise window system in \citet{Lam2020c}, along with hosting the listening test questionnaire.

For test-retest reliability, the \SI{60}{\decibelA} traffic noise track was evaluated as the first and last stimuli for every participant, whereas the rest of the 24 stimuli combinations were presented in random order to all the participants. The stimuli playback is controlled via the GUI, wherein questionnaire items were hidden from the participants throughout the 30-s stimuli duration. Thereafter, participants were able to repeat the stimuli as many times as required to complete the questionnaire. 

Each stimulus was evaluated on its (1) perceived annoyance (PAY), (2) perceived affective quality (PAQ), (3) perceived loudness (PLN), and (4) an open-ended description, as detailed in \Cref{tab:stimuliquestion}. In contrast to the pre-test assessment of annoyance in the absence of acoustic stimuli, the assessment of PAY is directed to the \textit{noise} in the presented stimulus. Both the 5-point verbal categorical scale and 101-point numerical scale were used for the stimuli evaluation, whereby the question now reads:
\begin{quote}
    \textit{\small Thinking about the noise you just heard, how much does the noise bother, disturb or annoy you?}
\end{quote}
\noindent \chadded[comment=R.10]{The usage of both scales follows the recommendations of ISO/TS 15666 for increased reliability \citep{Brink2016EffectsExperiment,InternationalOrganizationforStandardization2021}. The adoption of the 101-point scale, as opposed to the 11-point scale, was chosen to capture subtle differences. Additionally, a visual analog scale with marked regular intervals was employed for the 101-point scale to increase response precision \citep{Liu2016}.}

The PAQ attributes proposed in ISO/TS 12913-2 were adopted for a structured evaluation of the affective responses to the indoor soundscape \textit{in its entirety}, as modified by the stimuli. The PAQ consists of 8 attributes (i.e. \textit{eventful}, \textit{vibrant}, \textit{pleasant}, \textit{calm}, \textit{uneventful}, \textit{monotonous}, \textit{annoying}, \textit{chaotic}),  which form an octant circumplex model \citep{iso12913-3}. On a 0-to-100 opinion scale, where 0 indicates ``strongly disagree'' and 100 represents ``strongly agree'', the 8 attributes were evaluated through:
\begin{quote}
    \textit{\small For each of the 8 scales below, to what extent do you agree or disagree that the present surrounding sound environment is [...]?}
\end{quote}
where \textit{[...]} is one of the 8 PAQ attributes. \chadded[comment=R.10]{The utilization of the 101-point scale in assessing the PAQ attributes aimed to ensure consistency during analysis.}

To judge the PLN, a relative magnitude estimation method was adopted \citep{InternationalOrganizationforStandardization2017a,Zwicker2013,Park2017b,Torija2015a,Hong2020c}. Participants were instructed to compare PLN of the stimulus under test (SUT) to a reference track. A numerical score is assigned to the stimulus in linear relativity to the reference, which has a modulus score of 100. For instance, if the PLN of the SUT is exactly half as loud as the reference, then the stimulus should be assigned a score of 50. No limits were set on the magnitude estimation scores, and the participants were allowed to repeat the tracks as many times as required. For comparability, the reference tracks were fixed at \SI{65}{\decibelA} but were set to the respective noise type of the SUT. The question reads:
\begin{quote}
    \small \textit{This task compares the test track you just heard to a reference track. If the perceived loudness of the reference track is assigned a value of 100, rate the perceived loudness of the test track in comparison with the reference track. For example, if the test track is twice as loud, it should be rated as 200. And if it is half as loud, it should be rated as 50. You are free to listen to the tracks as many times as required}
\end{quote}

Lastly, an open-ended question was included to capture free-form qualitative assessments of the SUT relative to the reference track, i.e.
\begin{quote}
    \textit{\small Describe the test track in comparison to the reference track in a few descriptive words.}
\end{quote}

\subsection{Participants}

 A total of 45 participants were recruited through advertisements on online messaging platforms. Before the commencement of the experiment, all participants underwent an audiometric test (Interacoustics AD629). With the exception of one participant, all others had normal hearing for all the frequencies tested (mean threshold of hearing < 15 dB at 0.125, 0.5, 1, 2, 3, 4, 6, and 8 kHz)). 

Of the 44 participants who participated in the listening experiment, four participants did not meet the recommended within-subject test-retest reliability score of 0.7. \citep{Wood2017}. Therefore, the analysis included test results from 40 participants. The majority of the participants were between 21 to 30 years old ($n=\si{32}$), while the remaining participants were distributed among age ranges: less than 21 years old ($n=\si{2}$), between 31 and 40 years old ($n=\si{2}$), between 41 and 50 years old ($n=\si{1}$), and more than 60 years old ($n=\si{2}$).  The gender distribution was even, with 20 male participants and 20 female participants. More than half of the participants were either Singapore citizens ($n=\si{23}$) or permanent residents ($n=\si{2}$), and the rest were foreigners ($n=\si{15}$).

\subsection{Data analysis}

The acoustic and psychoacoustic indicators were computed with a commercial software package (ArtemiS \textsc{suite}, HEAD acoustics GmbH, Herzogenrath, Germany).

The reliability of the pre-test WNSS, PSS, and WHO-5 questionnaires were evaluated with both the Cronbach's Alpha ($\alpha$) and Mcdonald's Omega ($\omega$) \citep{Peters2014,Dunn2014}, based on polychoric correlations due to the ordinal nature of the Likert scales \citep{Fox2019,OConnor2021}.

Since both the baseline ISO/TS 15666 annoyance and the stimuli-based PAY was assessed with both a 5-point verbal, as well as a 101-point numerical scale \citep{InternationalOrganizationforStandardization2021}, Bland-Altman statistics and plots were employed to determine the agreement, and thus reliability of the annoyance ratings \citep{MartinBland1986,Brink2016EffectsExperiment, Lam2022ICA}. The three-way ART ANOVA (3W-ART ANOVA) was then conducted with the numerical-scale annoyance as the independent variable, and with three dependent variables (i.e noise type, masker type, ANC condition). 

The circumplexity of the PAQ attributes was examined to assess the generality of the PAQ model, particularly for indoor soundscapes \citep{Torresin2020IndoorBuildings}. This examination involved the randomized test of hypothesized order relations (RTHOR)  (RTHOR) \citep{Tracey2000}, correlation inequality criteria \citep{Locke2019}, and sinusoidality of the first two PCA loadings \citep{Lam2022c}. Pearson correlation was used for the circumplexity analysis due to the interval nature of the PAQ scales \citep{freedman2007statistics}. Since the derived \textit{Pleasantness} (\textit{ISOPL}) and \textit{Eventfulness} (\textit{ISOEV}) from the 8 PAQ attributes were found to be somewhat normally distributed via Mardia's multivariate normality tests \citep{Mardia1970,Mardia1974}, a three-way multivariate analysis of variance (3W-MANOVA) was conducted with \textit{ISOPL}, and \textit{ISOEV} as the two dependent variables, and masker type (\texttt{B}, \texttt{W}, \texttt{None}), ANC (yes, no), noise type (\texttt{AIR}, \texttt{MRT}, \texttt{TRA}) as the three independent variables. If significant differences were found at \SI{5}{\percent} significance level, post-hoc univariate three-way ANOVA (3W-ANOVA) tests were conducted for each dependent variable, followed by a pairwise post-hoc Tukey's honest significant differences (HSD) test with Bonferonni correction. 

Loudness estimation was investigated independently for each noise type since the reference track for loudness varied depending on the noise type.  Non-normality was detected using Shapiro-Wilk's test \citep{Shapiro1965AnSamples}, and a two-way Aligned Rank Transform Analysis of Variance (2W-ART ANOVA) was perfomed with \textit{PLN} as the independent variable \citep{Wobbrock2011}. The two dependent variables were masker type (\texttt{B}, \texttt{W}, \texttt{None}) and ANC conditions (i.e. yes, no) as the two dependent variables. Posthoc pairwise ART contrast tests with Bonferroni correction for were conducted at \SI{5}{\percent} significance level \citep{Elkin2021}.

Descriptive comparisons with noise-type dependent reference tracks were examined using sentence-level sentiment analysis \citep{Rinker2021}. Non-normality was addressed using the 2W-ART ANOVA to examine differences in sentiment scores between masker types and ANC conditions. Posthoc ART contrast tests with Bonferroni correction were performed, similar to the loudness analysis.

All data analyses were conducted with the R programming language \citep{RCoreTeam2021} on a 64-bit ARM environment. Specifically, the analyses were performed with the following packages: 3WANOVA, Tukey's HSD, and SWT with \texttt{stats} \citep{RCoreTeam2021}; Mardia's multivariate normality tests with \texttt{MVN} \citep{Korkmaz2014MVN:Normality}; ART ANOVA and ART contrast with \texttt{ARTool} \citep{Kay2021}; and sentiment analysis with \texttt{sentimentr} \citep{Rinker2021}.

\section{Results}
\label{sec:results}

\subsection{Pre-test assessment}

Surprisingly, the WNSS, WNSS-5, and PSS-10 items showed low inter-relatedness and reliability based on $\alpha$ and $\omega$ scores. The full and shortened INS questionnaires had marginal reliability ($\omega_\text{WNSS}=0.65$; $\omega_\text{WNSS-5}=0.68$), whereas suppressed $\alpha$ scores ($\alpha_\text{WNSS}=0.57$, $\alpha_\text{WNSS-5}=0.53$) signal a violation of $\tau$-equivalence in both WNSS questionnaires. The WHO-5 scores were reliable without redundancy ($\alpha_\text{WHO-5}=0.81$, $\omega_\text{WHO-5}=0.8$) \citep{Nunnally1994,Cho2015CronbachsAlpha}.

Participants were relatively neutral in terms of INS, perceived stress, and general well-being ($\mu_\text{WNSS-5}=0.63$, $\sigma_\text{WNSS-5}=0.17$, $\mu_\text{WNSS}=0.58$, $\sigma_\text{WNSS}=0.09$;  $\mu_\text{PSS-10}=0.54$, $\sigma_\text{PSS-10}=0.10$; $\mu_\text{WHO-5}=0.53$, $\sigma_\text{WHO-5}=0.14$), as summarised in \Cref{tab:pretestSummary}. From the BNA based on ISO/TS 15666, they were most annoyed by renovation noises ($\mu_\text{CON-R}=0.54$, $\sigma_\text{CON-R}=0.33$), followed by worksite construction noises ($\mu_\text{CON-W}=0.45$, $\sigma_\text{CON-W}=0.34$), \texttt{TRA} noise ($\mu_\text{TRA}=0.40$, $\sigma_\text{TRA}=0.30$), and \texttt{AIR} noise ($\mu_\text{AIR}=0.32$, $\sigma_\text{AIR}=0.32$). \texttt{MRT} noise caused the least annoyance ($\mu_\text{MRT}=0.22$, $\sigma_\text{MRT}=0.31$), as shown in \Cref{tab:pretestSummary}. 


\begin{table}[b]

\small


\centering
\caption{Summary statistics of pre-test indices.}
\label{tab:pretestSummary}
\begin{tabularx}{\linewidth}{@{\extracolsep{\fill}}
>{\raggedright\arraybackslash}X
*{2}{>{\raggedleft\arraybackslash}X}
@{}}
\toprule
Index 
& Mean, $\mu$ 
& SD, $\sigma$ \\
\midrule
15666 (MRT) & 0.22 & 0.31\\
15666 (AIR) & 0.32 & 0.32\\
15666 (TRA) & 0.40 & 0.30\\
15666 (CON-R) & 0.45 & 0.34\\
15666 (CON-W) & 0.54 & 0.33\\
WNSS-5 & 0.63 & 0.17\\
WNSS & 0.58 & 0.09\\
PSS-10 & 0.54 & 0.10\\
WHO-5 & 0.53 & 0.14\\
\bottomrule
\end{tabularx}
\end{table}

\subsection{Energy-based assessment of active control and masking}

\begin{figure}
    \centering
    \includegraphics[width=1\linewidth]{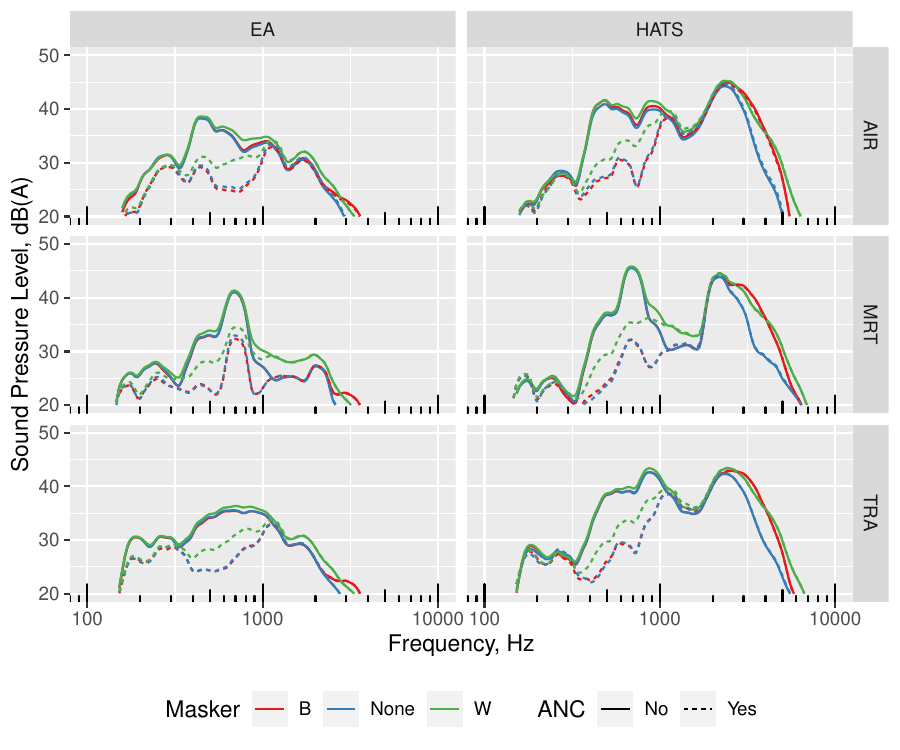}
    \caption{\small A-weighted energy-average (left) and binaural energetic average (right) spectra, across \texttt{AIR} (top), \texttt{MRT} (middle) and \texttt{TRA} (bottom) noise types. The solid and dashed lines respectively indicate cases before and after ANC, across cases with bird (\textcolor{set1_1}{---}), water (\textcolor{set1_3}{---}), and no maskers (\textcolor{set1_2}{---}). }
    \label{fig:65dBAHATS}
\end{figure}

The reduction in sound pressure due to the ANW is evidenced in both the spectral plots of the A-weighted energy-average across 6 microphones, and the binaural energetic average of the HATS, as shown in \Cref{fig:65dBAHATS}. About \SI{10}{\decibelA} reduction was achieved between \SI{300}{\hertz} and \SI{1}{\kilo\hertz} across all noise types in both measurement methods. This corroborates with the results that were reported previously \citep{Lam2020c}, and further illustrates that the ANW did not amplify frequencies beyond its control range.

Without ANC, the maskers only slightly increased the sound pressure beyond \SI{2.5}{\kilo\hertz}. With ANC, the broadband water masker slightly increased the SPL between \SI{300}{\hertz} and \SI{1}{\kilo\hertz}, as well as beyond \SI{2.5}{\kilo\hertz}, whereas the bird maskers only increased the SPL beyond \SI{2.5}{\kilo\hertz} due to dominant high frequencies. Since the ANW had limited performance below \SI{300}{\hertz}, the C-weighted spectra exhibited similar trends.

\subsubsection{Decibel-based assessment of ANC without maskers}

Despite about \SI{10}{\decibelA} reduction in the frequency spectra between \SI{300}{\hertz} to \SI{1}{\kilo\hertz}, decibel-based indicators showed modest attenuation across the full audio bandwidth (i.e. \SI{20}{\hertz} to \SI{20}{\kilo\hertz}) across measurement methods, noise types, and SPL levels, as shown in \Cref{tab:attHATSEA_SV}. For instance, the EA reduction across all A-weighted indicators (i.e. $L_\text{Aeq,EA}$, $L_\text{Amax,EA}$, $L_\text{A95,EA}$, $L_\text{A90,EA}$, $L_\text{A50,EA}$, $L_\text{A10,EA}$, $L_\text{A5,EA}$) was between $[1.95,3.07]$ \si{\decibelA}, $[3.68,4.97]$ \si{\decibelA} and $[2.68,2.95]$ \si{\decibelA}, across \texttt{AIR}, \texttt{MRT}, and \texttt{TRA} noise, respectively, at \SI{65}{\decibelA} and without maskers. The lack of active attenuation below \SI{300}{\hertz} is evidenced in the lower EA reduction across all C-weighted indicators in \texttt{AIR}, \texttt{MRT}, and \texttt{TRA} noise between $[0.77,1.26]$ \si{\decibelC}, $[1.20,2.74]$ \si{\decibelC} and $[-0.68,1.66]$ \si{\decibelC}, respectively. Similar trends were observed in the EA A- and C-weighted indicators across the \SI{70}{\decibelA} stimuli in the absence of maskers. In contrast, minimal attenuation was observed in the HATS \si{\decibelA} \{i.e. $[0.15,1.17]$, $[0.61,1.24]$, $[0.78,0.85]$ \si{\decibelA}\} and \si{\decibelC} \{i.e. $[0.39,1.06]$, $[0.91,2.47]$, $[0.38,1.35]$ \si{\decibelC}\} indices, across \texttt{AIR}, \texttt{MRT}, and \texttt{TRA} noise types, respectively. 

\subsubsection{Decibel-based assessment of overall sound levels with a bird masker}
The bird masker had minimal impact on EA measurements across all noise types, with maximum increments below \SI{1}{\decibel} in both $L_\text{Aeq}$ and $L_\text{Ceq}$, and below \SI{2}{\decibel} across other A- and C-weighted indicators. 

Although increments observed in both $L_\text{Aeq}$ and $L_\text{Ceq}$ HATS measurements were below \SI{2}{\decibel}, modest increments of up to \SI{7}{\decibelA} and \SI{4}{\decibelC} was observed in the other decibel-based indicators regardless of ANC condition. This discrepancy in the decibel-based indicators associated with temporality could be attributed to the greater emphasis of high-frequencies in the HATS measurements, which models after the increased high-frequency sensitivity of the human hearing system, and also due to the intermittent nature of bird sounds. For instance, the bird masker was 0.93 and \SI{4.04}{\decibelA} louder at $L_\text{Amax}$ and $L_\text{A50}$, respectively, with \texttt{AIR}, indicating temporal dominance about half of the time and had the loudest peaks, but not as loud for short time periods; 5.09, 2.38 and \SI{3.27}{\decibelA} louder at $L_\text{Amax}$, $L_\text{A5}$ and $L_\text{A90}$, respectively with \texttt{MRT}, indicating louder peaks and temporal dominance; and 5 and \SI{10}{\percent} of the time with \texttt{TRA}, indicating prominent intermittent peaks but a lack of temporal dominance, as shown in \Cref{tab:attHATSEA_SV}.

It is worth noting that the addition of bird maskers had minimal impact on ANC performance as shown in \Cref{fig:65dBAHATS}, but resulted in a slight overall energetic increase as shown in the A- and C-weighted indices in \Cref{tab:attHATSEA_SV}.


\subsubsection{Decibel-based assessment of overall sound levels with a water masker}

Without ANC, the water masker resulted at most \SI{2}{\decibelA} and \SI{0.5}{\decibelC} increments in the EA SPL levels, across all noise types. Nonetheless, the broadband nature of the water masker resulted in up to \SI{5.33}{\decibelA} increment in EA SPL levels at $L_\text{A50}$, $L_\text{A90}$, $L_\text{A95}$ with \texttt{AIR}; up to \SI{4.48}{\decibelA} increment in $L_\text{A90}$ and $L_\text{A95}$ with \texttt{MRT}; but less than \SI{0.5}{\decibelA} increment with \texttt{TRA} due to spectrotemporal similarities, as shown in \Cref{tab:attHATSEA_SV}. Increments in C-weighted EA SPL levels in the other indices were at most \SI{1.13}{\decibelC} in $L^{\texttt{AIR}_\texttt{65dB(A)}+\texttt{W}}_\text{A95,EA}$, which reflect lesser low-frequency content in water than in all noise types.


The increment in the same set of A-weighted indices from the HATS measurements were substantially greater in \texttt{AIR} [$L^{\texttt{AIR}_\texttt{65dB(A)}+\texttt{W}}_\text{A50,HATS}=-4.53$, $L^{\texttt{AIR}_\texttt{65dB(A)}+\texttt{W}}_\text{A90,HATS}=-10.31$ and $L^{\texttt{AIR}_\texttt{65dB(A)}+\texttt{W}}_\text{A95,HATS}=-11.32$ \si{\decibelA}]; and \texttt{MRT} [$L^{\texttt{MRT}_\texttt{65dB(A)}+\texttt{W}}_\text{A90,HATS}=-6.30$ and $L^{\texttt{MRT}_\texttt{65dB(A)}+\texttt{W}}_\text{A95,HATS}=-7.38$ \si{\decibelA}]. Increment in HATS C-weighted indices were also observed with the water masker in \texttt{AIR} [$L^{\texttt{AIR}_\texttt{65dB(A)}+\texttt{W}}_\text{C50,HATS}=-3.30$, $L^{\texttt{AIR}_\texttt{65dB(A)}+\texttt{W}}_\text{C90,HATS}=-4.62$ and $L^{\texttt{AIR}_\texttt{65dB(A)}+\texttt{W}}_\text{C95,HATS}=-4.94$ \si{\decibelC}]; and \texttt{MRT} [$L^{\texttt{MRT}_\texttt{65dB(A)}+\texttt{W}}_\text{C90,HATS}=-3.92$ and $L^{\texttt{MRT}_\texttt{65dB(A)}+\texttt{W}}_\text{C95,HATS}=-4.30$ \si{\decibelC}]. Similar to the EA measurements, the water masker had minimal impact on A- and C-weighted indices \{$[-1.62,-1.98]$ \si{\decibelA}, $[-0.10,-1.29]$ \si{\decibelC}\} with \texttt{TRA}.

\begin{landscape}
\begin{table}[ht]
\tiny
\setlength{\tabcolsep}{1pt}
\caption{\label{tab:attHATSEA_SV}Difference between mean acoustic and psychoacoustic indices before and after intervention, across both KEMAR HATS and energetic average measurements.}

\centering
\begin{tabularx}{\linewidth}
{@{\extracolsep{\fill}}>{\raggedright\arraybackslash}p{0.6cm}
>{\raggedleft\arraybackslash}p{0.5cm}
>{\raggedleft\arraybackslash}p{0.6cm}
*{2}{>{\raggedleft\arraybackslash}p{0.5cm}}
>{\raggedleft\arraybackslash}p{0.6cm}
*{32}{>{\raggedleft\arraybackslash}p{0.5cm}}@{}}

\toprule
\multicolumn{1}{c}{\textbf{ }} 
& \multicolumn{18}{c}{\textbf{HATS}} 
& \multicolumn{18}{c}{\textbf{Energetic Average}} \\

\cmidrule(l{1pt}r{1pt}){2-19} 
\cmidrule(l{1pt}r{1pt}){20-37}

\multicolumn{1}{c}{\textbf{ }} 
& \multicolumn{6}{c}{\textbf{Aircraft}} 
& \multicolumn{6}{c}{\textbf{MRT}} 
& \multicolumn{6}{c}{\textbf{Traffic}} 
& \multicolumn{6}{c}{\textbf{Aircraft}} 
& \multicolumn{6}{c}{\textbf{MRT}} 
& \multicolumn{6}{c}{\textbf{Traffic}} \\

\cmidrule(l{1pt}r{1pt}){2-7} 
\cmidrule(l{1pt}r{1pt}){8-13} 
\cmidrule(l{1pt}r{1pt}){14-19} 
\cmidrule(l{1pt}r{1pt}){20-25} 
\cmidrule(l{1pt}r{1pt}){26-31} 
\cmidrule(l{1pt}r{1pt}){32-37}

\multicolumn{1}{c}{\textbf{ }} 
& \multicolumn{2}{c}{\textbf{ANC:NO}} 
& \multicolumn{4}{c}{\textbf{ANC:YES}} 
& \multicolumn{2}{c}{\textbf{ANC:NO}} 
& \multicolumn{4}{c}{\textbf{ANC:YES}} 
& \multicolumn{2}{c}{\textbf{ANC:NO}} 
& \multicolumn{4}{c}{\textbf{ANC:YES}} 
& \multicolumn{2}{c}{\textbf{ANC:NO}} 
& \multicolumn{4}{c}{\textbf{ANC:YES}} 
& \multicolumn{2}{c}{\textbf{ANC:NO}} 
& \multicolumn{4}{c}{\textbf{ANC:YES}} 
& \multicolumn{2}{c}{\textbf{ANC:NO}} 
& \multicolumn{4}{c}{\textbf{ANC:YES}} \\

\cmidrule(l{1pt}r{1pt}){2-3} \cmidrule(l{1pt}r{1pt}){4-7} \cmidrule(l{1pt}r{1pt}){8-9} \cmidrule(l{1pt}r{1pt}){10-13} \cmidrule(l{1pt}r{1pt}){14-15} \cmidrule(l{1pt}r{1pt}){16-19} \cmidrule(l{1pt}r{1pt}){20-21} \cmidrule(l{1pt}r{1pt}){22-25} \cmidrule(l{1pt}r{1pt}){26-27} \cmidrule(l{1pt}r{1pt}){28-31} \cmidrule(l{1pt}r{1pt}){32-33} \cmidrule(l{1pt}r{1pt}){34-37}

\multicolumn{1}{c}{\textbf{ }} 
& \multicolumn{5}{c}{\textbf{\texttt{65dB(A)}}} 
& \multicolumn{1}{c}{\textbf{\texttt{70dB(A)}}} 
& \multicolumn{5}{c}{\textbf{\texttt{65dB(A)}}} 
& \multicolumn{1}{c}{\textbf{\texttt{70dB(A)}}} 
& \multicolumn{5}{c}{\textbf{\texttt{65dB(A)}}} 
& \multicolumn{1}{c}{\textbf{\texttt{70dB(A)}}} 
& \multicolumn{5}{c}{\textbf{\texttt{65dB(A)}}} 
& \multicolumn{1}{c}{\textbf{\texttt{70dB(A)}}} 
& \multicolumn{5}{c}{\textbf{\texttt{65dB(A)}}} 
& \multicolumn{1}{c}{\textbf{\texttt{70dB(A)}}} 
& \multicolumn{5}{c}{\textbf{\texttt{65dB(A)}}} 
& \multicolumn{1}{c}{\textbf{\texttt{70dB(A)}}} \\

\cmidrule(l{1pt}r{1pt}){2-6} 
\cmidrule(l{1pt}r{1pt}){7-7} 
\cmidrule(l{1pt}r{1pt}){8-12} 
\cmidrule(l{1pt}r{1pt}){13-13} 
\cmidrule(l{1pt}r{1pt}){14-18} 
\cmidrule(l{1pt}r{1pt}){19-19} 
\cmidrule(l{1pt}r{1pt}){20-24} 
\cmidrule(l{1pt}r{1pt}){25-25} 
\cmidrule(l{1pt}r{1pt}){26-30} 
\cmidrule(l{1pt}r{1pt}){31-31} 
\cmidrule(l{1pt}r{1pt}){32-36} 
\cmidrule(l{1pt}r{1pt}){37-37}

& B & W & $\emptyset$ & B & W & $\emptyset$
& B & W & $\emptyset$ & B & W & $\emptyset$
& B & W & $\emptyset$ & B & W & $\emptyset$
& B & W & $\emptyset$ & B & W & $\emptyset$
& B & W & $\emptyset$ & B & W & $\emptyset$
& B & W & $\emptyset$ & B & W & $\emptyset$\\

\midrule

$L_{\text{Aeq}}$ 
& -1.62 & -1.73 & 0.16 & -0.86 & -1.36 & 0.65 & -1.69 & -1.87 & 1.02 & -0.92 & -1.16 & 1.09 & -1.46 & -1.68 & 0.82 & -0.89 & -1.13 & 0.80 & -0.45 & -1.29 & 2.29 & 2.41 & 0.60 & 2.68 & -0.26 & -1.62 & 4.13 & 3.74 & 1.20 & 4.32 & -0.18 & -1.11 & 2.75 & 2.44 & 0.92 & 2.74\\
$L_{\text{Amax}}$ 
& -0.93 & -0.49 & 0.15 & -0.77 & -0.34 & 0.19 & -5.08 & -1.51 & 0.84 & -4.81 & -0.72 & 0.84 & -6.74 & -1.98 & 0.83 & -6.58 & -1.42 & 0.90 & -0.07 & -0.14 & 2.95 & 2.88 & 2.66 & 2.98 & -0.31 & -0.97 & 4.97 & 3.38 & 2.32 & 5.16 & -0.59 & -1.46 & 2.95 & 1.25 & 0.44 & 2.86\\
$L_{\text{A5}}$ 
& -1.47 & -0.52 & 0.40 & -0.98 & -0.13 & 0.47 & -2.38 & -1.20 & 1.00 & -1.96 & -0.47 & 0.97 & -3.50 & -1.74 & 0.86 & -3.21 & -1.22 & 0.83 & -0.17 & -0.35 & 2.31 & 2.29 & 1.79 & 2.39 & -0.20 & -1.20 & 4.22 & 3.82 & 1.96 & 4.34 & -0.20 & -1.03 & 2.89 & 2.34 & 0.98 & 2.85\\
$L_{\text{A10}}$ & -0.77 & -0.54 & 0.41 & -0.23 & -0.13 & 0.47 & -1.93 & -1.21 & 1.04 & -1.42 & -0.43 & 1.03 & -2.95 & -1.70 & 0.85 & -2.59 & -1.18 & 0.82 & -0.17 & -0.44 & 2.41 & 2.43 & 1.75 & 2.49 & -0.16 & -1.26 & 4.12 & 3.82 & 1.83 & 4.24 & -0.16 & -1.00 & 2.86 & 2.45 & 1.01 & 2.82\\
$L_{\text{A50}}$ 
& -4.04 & -4.53 & 0.48 & -2.37 & -4.27 & 0.93 & -1.07 & -1.62 & 1.24 & -0.13 & -0.80 & 1.27 & -0.49 & -1.65 & 0.78 & 0.30 & -1.09 & 0.76 & -1.52 & -3.80 & 1.95 & 1.47 & -2.78 & 3.06 & -0.23 & -1.51 & 4.15 & 3.85 & 1.43 & 4.29 & -0.18 & -1.13 & 2.68 & 2.43 & 0.88 & 2.68\\
$L_{\text{A90}}$ 
& -1.03 & -10.31 & 0.89 & 0.50 & -10.14 & 2.04 & -3.27 & -6.30 & 0.61 & -1.90 & -5.99 & 1.11 & -0.34 & -1.67 & 0.84 & 0.53 & -1.08 & 0.81 & -0.31 & -5.08 & 2.93 & 3.06 & -4.24 & 3.59 & -0.62 & -4.13 & 3.73 & 2.86 & -2.71 & 4.18 & -0.15 & -1.19 & 2.75 & 2.58 & 0.92 & 2.74\\
$L_{\text{A95}}$ 
& -0.12 & -11.32 & 1.17 & 1.41 & -11.16 & 1.98 & -1.45 & -7.38 & 0.80 & -0.49 & -7.06 & 1.23 & -0.28 & -1.62 & 0.85 & 0.59 & -0.97 & 0.82 & -0.05 & -5.33 & 3.07 & 3.27 & -4.54 & 3.58 & -0.17 & -4.48 & 3.68 & 3.47 & -3.25 & 3.93 & -0.10 & -1.14 & 2.78 & 2.64 & 1.01 & 2.77\\
\midrule
$L_{\text{Ceq}}$
& -1.21 & -1.40 & 0.72 & -0.05 & -0.59 & 1.23 & -1.16 & -1.39 & 1.66 & 0.14 & -0.20 & 1.81 & -1.01 & -1.24 & 1.21 & -0.08 & -0.35 & 1.26 & 0.01 & -0.45 & 1.66 & 1.88 & 1.19 & 2.34 & 0.01 & -0.46 & 1.95 & 2.02 & 1.20 & 2.48 & -0.06 & -0.36 & 1.37 & 1.28 & 0.87 & 1.65\\
$L_{\text{Cmax}}$ 
& -0.75 & -0.44 & 0.39 & -0.39 & -0.06 & 0.44 & -2.81 & -0.11 & 2.47 & -2.26 & 1.30 & 1.87 & -3.77 & -0.10 & 0.38 & -3.62 & -0.11 & 1.50 & 0.06 & -0.61 & 2.19 & 3.22 & 3.52 & 3.79 & 0.30 & 1.15 & 2.74 & 2.66 & 2.08 & 2.42 & -0.61 & 0.84 & -0.68 & -0.50 & 0.42 & 1.54\\
$L_{\text{C5}}$ 
& -1.01 & -0.44 & 1.06 & -0.16 & 0.56 & 1.16 & -1.62 & -0.93 & 1.69 & -0.82 & 0.38 & 1.76 & -2.34 & -1.17 & 1.35 & -1.83 & -0.30 & 1.44 & 0.02 & -0.17 & 2.26 & 2.41 & 2.12 & 2.78 & -0.01 & -0.27 & 2.02 & 2.07 & 1.53 & 2.56 & -0.12 & -0.30 & 1.21 & 1.11 & 0.84 & 1.53\\
$L_{\text{C10}}$ 
& -0.75 & -0.49 & 0.98 & 0.43 & 0.45 & 1.07 & -1.30 & -0.97 & 1.70 & -0.37 & 0.38 & 1.73 & -1.97 & -1.20 & 1.32 & -1.36 & -0.32 & 1.39 & -0.04 & -0.19 & 2.13 & 2.29 & 1.90 & 2.69 & 0.00 & -0.35 & 2.02 & 2.07 & 1.48 & 2.55 & -0.10 & -0.31 & 1.27 & 1.16 & 0.84 & 1.56\\
$L_{\text{C50}}$ 
& -2.83 & -3.30 & 0.62 & -1.20 & -2.67 & 1.66 & -0.80 & -1.26 & 1.80 & 0.78 & 0.08 & 1.95 & -0.51 & -1.27 & 1.19 & 0.68 & -0.39 & 1.22 & 0.08 & -0.60 & 1.26 & 1.45 & 0.55 & 1.84 & 0.01 & -0.49 & 2.04 & 2.10 & 1.27 & 2.52 & -0.03 & -0.39 & 1.42 & 1.34 & 0.87 & 1.65\\
$L_{\text{C90}}$ 
& -0.25 & -4.62 & 0.50 & 0.36 & -4.19 & 1.61 & -1.98 & -3.92 & 0.97 & -0.40 & -3.50 & 1.56 & -0.33 & -1.29 & 1.12 & 0.88 & -0.34 & 1.20 & 0.22 & -1.02 & 0.78 & 1.08 & -0.17 & 1.36 & -0.08 & -0.92 & 1.23 & 1.33 & 0.07 & 1.85 & -0.06 & -0.47 & 1.57 & 1.48 & 0.93 & 1.80\\
$L_{\text{C95}}$ 
& -0.11 & -4.94 & 0.67 & 0.73 & -4.51 & 1.66 & -0.61 & -4.30 & 0.91 & 0.69 & -3.84 & 1.27 & -0.32 & -1.28 & 1.16 & 0.92 & -0.25 & 1.27 & 0.19 & -1.14 & 0.77 & 1.11 & -0.29 & 1.68 & -0.07 & -0.95 & 1.20 & 1.32 & -0.07 & 1.71 & -0.05 & -0.45 & 1.66 & 1.54 & 0.97 & 1.87\\
\midrule
$N_{\text{max}}$ 
& -1.60 & -1.12 & 2.38 & 0.65 & 1.16 & 3.15 & -5.00 & -3.29 & 2.35 & -2.48 & -0.76 & 3.00 & -5.81 & -3.43 & 2.71 & -3.85 & -1.87 & 3.65 & -0.50 & -0.84 & 1.93 & 1.47 & 1.28 & 2.62 & -0.93 & -1.32 & 1.79 & 0.55 & -0.11 & 2.47 & -1.06 & -1.48 & 1.54 & 0.22 & -0.36 & 2.20\\
$N_{\text{5}}$ 
& -1.79 & -1.49 & 2.45 & 1.12 & 0.84 & 3.37 & -1.73 & -2.12 & 2.68 & 0.91 & -0.06 & 3.57 & -2.73 & -2.31 & 2.09 & -0.72 & -0.72 & 2.72 & -0.47 & -0.83 & 1.62 & 1.36 & 0.67 & 2.18 & -0.27 & -1.30 & 1.55 & 1.35 & 0.01 & 2.07 & -0.47 & -1.25 & 1.36 & 0.87 & -0.18 & 1.84\\
$N_{\text{10}}$ 
& -1.21 & -1.52 & 2.29 & 1.55 & 0.62 & 3.21 & -1.11 & -2.08 & 2.63 & 1.56 & -0.03 & 3.54 & -1.79 & -2.27 & 2.03 & 0.18 & -0.68 & 2.64 & -0.36 & -0.89 & 1.54 & 1.40 & 0.55 & 2.11 & -0.24 & -1.30 & 1.49 & 1.33 & -0.04 & 2.01 & -0.33 & -1.20 & 1.35 & 1.01 & -0.13 & 1.81\\
$N_{\text{50}}$ 
& -1.33 & -4.38 & 1.02 & 0.60 & -3.56 & 1.91 & -0.75 & -2.47 & 2.43 & 1.71 & -0.66 & 3.29 & -0.43 & -2.09 & 1.93 & 1.49 & -0.55 & 2.47 & -0.57 & -2.36 & 0.79 & 0.64 & -1.71 & 1.36 & -0.21 & -1.40 & 1.34 & 1.21 & -0.32 & 1.83 & -0.15 & -1.15 & 1.28 & 1.13 & -0.09 & 1.73\\
$N_{\text{90}}$ 
& -0.08 & -5.79 & 0.76 & 0.83 & -5.19 & 1.38 & -0.72 & -4.64 & 1.27 & 0.68 & -3.80 & 2.02 & -0.25 & -1.99 & 1.80 & 1.61 & -0.47 & 2.36 & -0.03 & -2.48 & 0.67 & 0.70 & -2.00 & 1.05 & -0.18 & -2.30 & 0.89 & 0.78 & -1.67 & 1.35 & -0.07 & -1.09 & 1.22 & 1.15 & -0.04 & 1.66\\
$N_{\text{95}}$ 
& 0.00 & -5.93 & 0.72 & 0.81 & -5.33 & 1.22 & -0.26 & -4.80 & 1.22 & 1.00 & -4.07 & 1.87 & -0.21 & -1.98 & 1.78 & 1.59 & -0.41 & 2.30 & 0.01 & -2.50 & 0.64 & 0.69 & -2.02 & 0.97 & -0.07 & -2.35 & 0.83 & 0.80 & -1.79 & 1.24 & -0.06 & -1.06 & 1.20 & 1.15 & -0.02 & 1.64\\
\midrule
$S_{\text{max}}$  
& -0.11 & -0.08 & -0.12 & -0.22 & -0.14 & -0.12 & -0.09 & -0.03 & -0.14 & -0.25 & -0.07 & -0.14 & -0.14 & -0.03 & -0.09 & -0.27 & -0.12 & -0.06 & -0.11 & -0.31 & -0.22 & -0.24 & -0.41 & -0.14 & -0.09 & -0.22 & -0.15 & -0.26 & -0.28 & -0.13 & -0.17 & -0.18 & -0.09 & -0.29 & -0.28 & -0.10\\
$S$ 
& -0.03 & -0.16 & -0.14 & -0.18 & -0.23 & -0.13 & -0.03 & -0.12 & -0.18 & -0.20 & -0.23 & -0.16 & -0.03 & -0.13 & -0.12 & -0.15 & -0.22 & -0.11 & -0.03 & -0.25 & -0.14 & -0.16 & -0.33 & -0.11 & -0.03 & -0.21 & -0.14 & -0.17 & -0.30 & -0.11 & -0.03 & -0.19 & -0.10 & -0.13 & -0.28 & -0.09\\
$S_{\text{5}}$ 
& -0.07 & -0.18 & -0.19 & -0.23 & -0.25 & -0.13 & -0.08 & -0.15 & -0.17 & -0.24 & -0.22 & -0.16 & -0.14 & -0.15 & -0.13 & -0.25 & -0.25 & -0.12 & -0.05 & -0.29 & -0.17 & -0.20 & -0.36 & -0.12 & -0.07 & -0.23 & -0.16 & -0.22 & -0.31 & -0.13 & -0.09 & -0.21 & -0.11 & -0.20 & -0.31 & -0.10\\
$S_{\text{10}}$  
& -0.04 & -0.18 & -0.17 & -0.20 & -0.24 & -0.13 & -0.06 & -0.15 & -0.17 & -0.21 & -0.22 & -0.15 & -0.07 & -0.14 & -0.13 & -0.18 & -0.24 & -0.12 & -0.03 & -0.28 & -0.16 & -0.18 & -0.35 & -0.12 & -0.05 & -0.22 & -0.15 & -0.21 & -0.31 & -0.13 & -0.06 & -0.21 & -0.11 & -0.16 & -0.30 & -0.10\\
$S_{\text{50}}$  
& -0.04 & -0.17 & -0.13 & -0.19 & -0.24 & -0.12 & -0.03 & -0.13 & -0.18 & -0.19 & -0.24 & -0.16 & -0.01 & -0.13 & -0.13 & -0.14 & -0.23 & -0.11 & -0.04 & -0.24 & -0.13 & -0.16 & -0.32 & -0.10 & -0.02 & -0.21 & -0.13 & -0.16 & -0.31 & -0.11 & -0.02 & -0.19 & -0.10 & -0.12 & -0.28 & -0.09\\
$S_{\text{90}}$  
& -0.01 & -0.12 & -0.13 & -0.15 & -0.23 & -0.13 & -0.01 & -0.11 & -0.18 & -0.18 & -0.24 & -0.16 & -0.01 & -0.12 & -0.12 & -0.14 & -0.22 & -0.11 & -0.03 & -0.23 & -0.12 & -0.15 & -0.32 & -0.10 & -0.01 & -0.18 & -0.12 & -0.13 & -0.28 & -0.10 & -0.01 & -0.17 & -0.10 & -0.11 & -0.26 & -0.09\\
$S_{\text{95}}$  
& -0.02 & -0.13 & -0.14 & -0.16 & -0.24 & -0.14 & -0.01 & -0.11 & -0.18 & -0.18 & -0.24 & -0.16 & -0.01 & -0.11 & -0.12 & -0.13 & -0.21 & -0.11 & -0.03 & -0.22 & -0.12 & -0.15 & -0.32 & -0.10 & -0.01 & -0.17 & -0.12 & -0.13 & -0.27 & -0.10 & -0.01 & -0.16 & -0.09 & -0.10 & -0.25 & -0.08\\
\midrule
$R_{\text{max}}$  
& 0.00 & 0.00 & 0.02 & 0.00 & 0.02 & 0.02 & -0.01 & -0.01 & 0.01 & 0.00 & 0.00 & 0.01 & 0.00 & 0.00 & 0.02 & 0.01 & 0.02 & 0.01 & 0.00 & 0.00 & 0.00 & 0.01 & 0.01 & 0.01 & 0.00 & 0.00 & 0.02 & 0.02 & 0.02 & 0.02 & 0.00 & 0.00 & 0.02 & 0.01 & 0.01 & 0.02\\
$R$  
& 0.00 & 0.00 & 0.00 & 0.00 & 0.00 & 0.01 & 0.00 & 0.00 & 0.01 & 0.01 & 0.00 & 0.01 & 0.00 & 0.00 & 0.01 & 0.01 & 0.01 & 0.01 & 0.00 & 0.00 & 0.00 & 0.00 & 0.00 & 0.00 & 0.00 & 0.00 & 0.01 & 0.01 & 0.00 & 0.01 & 0.00 & 0.00 & 0.01 & 0.01 & 0.00 & 0.01\\
$R_{\text{5}}$ 
& 0.00 & 0.00 & 0.01 & 0.01 & 0.01 & 0.01 & 0.00 & 0.00 & 0.01 & 0.01 & 0.00 & 0.01 & 0.00 & 0.00 & 0.01 & 0.01 & 0.01 & 0.01 & 0.00 & 0.00 & 0.01 & 0.01 & 0.01 & 0.01 & 0.00 & 0.00 & 0.01 & 0.02 & 0.01 & 0.02 & 0.00 & 0.00 & 0.01 & 0.01 & 0.01 & 0.01\\
$R_{\text{10}}$  
& 0.00 & 0.00 & 0.01 & 0.01 & 0.00 & 0.01 & 0.00 & 0.00 & 0.01 & 0.01 & 0.00 & 0.01 & 0.00 & 0.00 & 0.01 & 0.01 & 0.01 & 0.01 & 0.00 & 0.00 & 0.01 & 0.01 & 0.00 & 0.01 & 0.00 & 0.00 & 0.01 & 0.01 & 0.01 & 0.01 & 0.00 & 0.00 & 0.01 & 0.01 & 0.01 & 0.01\\
$R_{\text{50}}$  
& 0.00 & 0.00 & 0.00 & 0.00 & 0.00 & 0.01 & 0.00 & 0.00 & 0.01 & 0.01 & 0.00 & 0.01 & 0.00 & 0.00 & 0.01 & 0.01 & 0.01 & 0.01 & 0.00 & 0.00 & 0.00 & 0.00 & 0.00 & 0.00 & 0.00 & 0.00 & 0.01 & 0.01 & 0.00 & 0.01 & 0.00 & 0.00 & 0.01 & 0.01 & 0.00 & 0.01\\
$R_{\text{90}}$  
& 0.00 & 0.00 & 0.00 & 0.00 & 0.00 & 0.00 & 0.00 & 0.00 & 0.00 & 0.00 & 0.00 & 0.01 & 0.00 & 0.00 & 0.01 & 0.01 & 0.00 & 0.01 & 0.00 & 0.00 & 0.00 & 0.00 & 0.00 & 0.00 & 0.00 & 0.00 & 0.00 & 0.00 & 0.00 & 0.01 & 0.00 & 0.00 & 0.00 & 0.00 & 0.00 & 0.00\\
$R_{\text{95}}$  
& 0.00 & 0.00 & 0.00 & 0.00 & 0.00 & 0.00 & 0.00 & 0.00 & 0.00 & 0.00 & 0.00 & 0.01 & 0.00 & 0.00 & 0.01 & 0.01 & 0.00 & 0.01 & 0.00 & 0.00 & 0.00 & 0.00 & 0.00 & 0.00 & 0.00 & 0.00 & 0.00 & 0.00 & 0.00 & 0.01 & 0.00 & 0.00 & 0.00 & 0.00 & 0.00 & 0.00\\
\midrule
$T_{\text{max}}$  
& -1.09 & 0.14 & 0.25 & -1.17 & 0.31 & 0.36 & -0.98 & 0.15 & -0.07 & -1.05 & 0.06 & -0.07 & -1.38 & -0.02 & 0.01 & -1.47 & 0.03 & 0.00 & -0.14 & 0.05 & 0.16 & -0.11 & 0.27 & 0.20 & -0.14 & 0.14 & 0.08 & -0.10 & 0.29 & 0.14 & -0.28 & 0.00 & 0.02 & -0.34 & 0.05 & 0.01\\
$T_{\text{5}}$ 
& -0.53 & 0.04 & -0.04 & -0.53 & 0.03 & -0.02 & -0.17 & 0.15 & -0.08 & -0.23 & 0.09 & -0.09 & -0.50 & 0.02 & 0.01 & -0.55 & 0.06 & 0.01 & -0.10 & 0.04 & 0.00 & -0.07 & 0.07 & 0.00 & -0.03 & 0.12 & 0.02 & 0.01 & 0.19 & 0.06 & -0.08 & 0.03 & 0.03 & -0.07 & 0.07 & 0.02\\
$T_{\text{10}}$ 
& -0.32 & 0.03 & -0.03 & -0.32 & 0.04 & 0.00 & -0.09 & 0.14 & -0.09 & -0.16 & 0.09 & -0.09 & -0.26 & 0.01 & -0.01 & -0.29 & 0.04 & -0.01 & -0.08 & 0.04 & 0.00 & -0.06 & 0.06 & 0.00 & -0.03 & 0.11 & 0.00 & 0.00 & 0.17 & 0.04 & -0.05 & 0.03 & 0.04 & -0.03 & 0.08 & 0.03\\
$T_{\text{50}}$ 
& -0.05 & 0.02 & -0.02 & -0.05 & 0.02 & -0.02 & -0.02 & 0.11 & -0.05 & -0.07 & 0.08 & -0.05 & -0.03 & 0.01 & -0.01 & -0.04 & 0.02 & -0.02 & -0.02 & 0.03 & 0.00 & -0.01 & 0.03 & 0.00 & -0.01 & 0.07 & -0.02 & -0.01 & 0.09 & -0.01 & -0.02 & 0.02 & 0.00 & -0.01 & 0.03 & 0.00\\
$T_{\text{90}}$ 
& -0.01 & 0.01 & -0.01 & -0.01 & 0.01 & -0.01 & -0.02 & 0.10 & -0.03 & -0.04 & 0.09 & -0.03 & -0.01 & 0.01 & -0.01 & -0.02 & 0.01 & -0.01 & -0.01 & 0.01 & 0.00 & 0.00 & 0.01 & 0.00 & -0.02 & 0.03 & -0.01 & -0.02 & 0.03 & -0.01 & -0.01 & 0.01 & 0.00 & -0.01 & 0.01 & -0.01\\
$T_{\text{95}}$ 
& 0.00 & 0.01 & -0.01 & -0.01 & 0.01 & -0.01 & -0.01 & 0.10 & -0.03 & -0.04 & 0.09 & -0.03 & -0.01 & 0.01 & -0.01 & -0.02 & 0.01 & -0.01 & -0.01 & 0.01 & 0.00 & 0.00 & 0.01 & 0.00 & -0.01 & 0.02 & -0.01 & -0.01 & 0.02 & 0.00 & 0.00 & 0.01 & 0.00 & -0.01 & 0.01 & 0.00\\
\midrule
$FS_{\text{max}}$ 
& -0.02 & 0.01 & 0.01 & -0.02 & 0.01 & 0.00 & -0.01 & 0.00 & 0.01 & -0.01 & 0.00 & 0.01 & 0.00 & -0.01 & 0.00 & 0.00 & 0.00 & 0.01 & 0.01 & 0.00 & -0.01 & 0.00 & 0.00 & 0.00 & 0.00 & 0.00 & 0.00 & 0.01 & 0.00 & 0.00 & 0.00 & -0.01 & -0.01 & -0.01 & -0.01 & 0.01\\
$FS_{\text{5}}$ 
& -0.03 & 0.00 & 0.00 & -0.04 & 0.01 & 0.01 & -0.03 & 0.00 & 0.01 & -0.03 & 0.00 & 0.01 & -0.02 & 0.00 & 0.00 & -0.02 & 0.00 & 0.01 & 0.00 & 0.00 & -0.01 & 0.00 & 0.00 & 0.01 & 0.00 & 0.00 & 0.00 & 0.00 & 0.00 & 0.00 & 0.00 & 0.00 & 0.00 & 0.00 & 0.00 & 0.00\\
$FS_{\text{10}}$ 
& -0.03 & 0.01 & 0.01 & -0.04 & 0.02 & 0.02 & -0.04 & 0.01 & 0.00 & -0.04 & 0.00 & 0.00 & -0.04 & 0.00 & 0.00 & -0.04 & 0.00 & 0.00 & 0.00 & 0.00 & 0.00 & 0.00 & 0.01 & 0.00 & -0.01 & 0.00 & -0.01 & -0.01 & 0.00 & 0.00 & -0.01 & 0.00 & -0.01 & -0.01 & 0.00 & 0.00\\
$FS_{\text{50}}$ 
& -0.02 & 0.00 & 0.00 & -0.02 & 0.00 & 0.00 & -0.02 & 0.00 & 0.00 & -0.02 & 0.00 & 0.00 & -0.01 & 0.00 & 0.00 & -0.01 & 0.00 & 0.00 & 0.00 & 0.00 & 0.00 & -0.01 & 0.00 & 0.00 & -0.01 & 0.00 & 0.00 & 0.00 & 0.00 & 0.00 & 0.00 & 0.00 & 0.00 & -0.01 & 0.00 & 0.00\\
$FS_{\text{90}}$ 
& 0.00 & 0.00 & 0.00 & 0.00 & 0.00 & 0.00 & 0.00 & 0.00 & 0.00 & 0.00 & 0.00 & 0.00 & 0.00 & 0.00 & 0.00 & 0.00 & 0.00 & 0.00 & 0.00 & 0.00 & 0.00 & 0.00 & 0.00 & 0.00 & 0.00 & 0.00 & 0.00 & 0.00 & 0.00 & 0.00 & 0.00 & 0.00 & 0.00 & 0.00 & 0.00 & 0.00\\
$FS_{\text{95}}$ 
& 0.00 & 0.00 & 0.00 & 0.00 & 0.00 & 0.00 & 0.00 & 0.00 & 0.00 & 0.00 & 0.00 & 0.00 & 0.00 & 0.00 & 0.00 & 0.00 & 0.00 & 0.00 & 0.00 & 0.00 & 0.00 & 0.00 & 0.00 & 0.00 & 0.00 & 0.00 & 0.00 & 0.00 & 0.00 & 0.00 & 0.00 & 0.00 & 0.00 & 0.00 & 0.00 & 0.00\\
\midrule
\textit{PA} & -1.83 & -1.72 & 2.43 & 0.83 & 0.09 & 3.45 & -1.75 & -2.31 & 2.35 & 0.38 & -0.88 & 3.61 & -2.75 & -2.32 & 2.15 & -0.67 & -1.00 & 2.81 & -0.48 & -0.86 & 1.64 & 1.38 & 0.66 & 2.22 & -0.28 & -1.31 & 1.60 & 1.40 & 0.03 & 2.15 & -0.48 & -1.26 & 1.40 & 0.90 & -0.16 & 1.89\\
$L_{\text{A10}}-L_{\text{A90}}$  
& 0.25 & 9.78 & -0.47 & -0.73 & 10.01 & -1.57 & 1.35 & 5.09 & 0.43 & 0.48 & 5.56 & -0.08 & -2.60 & -0.03 & 0.01 & -3.13 & -0.11 & 0.01 & 0.14 & 4.63 & -0.52 & -0.63 & 5.99 & -1.10 & 0.46 & 2.88 & 0.39 & 0.96 & 4.54 & 0.06 & -0.02 & 0.19 & 0.11 & -0.13 & 0.10 & 0.08\\
$L_{\text{Ceq}}-L_{\text{Aeq}}$
& 0.86 & 1.24 & 0.82 & 1.29 & 1.81 & 0.42 & 0.39 & 0.90 & 0.68 & 0.55 & 1.55 & 0.64 & -0.50 & 0.47 & 0.49 & -0.47 & 0.80 & 0.59 & 0.41 & 1.11 & -0.17 & -0.12 & 1.30 & 0.01 & 0.26 & 1.27 & -2.11 & -1.67 & 0.28 & -1.77 & 0.08 & 0.81 & -1.47 & -1.28 & -0.08 & -1.18\\

\bottomrule
\end{tabularx}
\end{table}
\end{landscape}

\subsection{Single-value psychoacoustic indicators}

Although there is an increasing availability and accessibility of psychoacoustic indices for analysis, there is a lack of consensus on which indicator or its combinations could best represent perceptual effects such as perceived loudness and annoyance. A comprehensive list of traditional as well as suggested acoustic and psychoacoustic indices as described in \Cref{tab:acoustIndsummary} were computed, as shown in \Cref{tab:meanHATS_SV}. Since psychoacoustic parameters closely models the human perception, only the indices computed using the HATS measurements were shown in \Cref{tab:meanHATS_SV}, while psychoacoustic indices for EA measurements were shown in \Cref{tab:attHATSEA_SV} only for completeness.

\begin{table*}[ht]
\tiny
\setlength{\tabcolsep}{1pt}
\caption{\label{tab:meanHATS_SV}Mean acoustic and psychoacoustic parameters across both left and right channels of the KEMAR HATS for each stimuli}

\centering
\begin{tabularx}{\textwidth}{@{\extracolsep{\fill}}>{\raggedright\arraybackslash}p{1cm}*{25}{>{\raggedleft\arraybackslash}X}@{}}
\toprule
\multicolumn{1}{c}{\textbf{ }} 
& \multicolumn{13}{c}{\textbf{Masker:None}} 
& \multicolumn{6}{c}{\textbf{Masker:B}} 
& \multicolumn{6}{c}{\textbf{Masker:W}} \\
\cmidrule(l{2pt}r{2pt}){2-14} \cmidrule(l{2pt}r{2pt}){15-20} \cmidrule(l{2pt}r{2pt}){21-26}
\multicolumn{1}{c}{\textbf{ }} 
& \multicolumn{7}{c}{\textbf{ANC:NO}} 
& \multicolumn{6}{c}{\textbf{ANC:YES}} 
& \multicolumn{3}{c}{\textbf{ANC:NO}} 
& \multicolumn{3}{c}{\textbf{ANC:YES}} 
& \multicolumn{3}{c}{\textbf{ANC:NO}} 
& \multicolumn{3}{c}{\textbf{ANC:YES}} \\
\cmidrule(l{2pt}r{2pt}){2-8} \cmidrule(l{2pt}r{2pt}){9-14} \cmidrule(l{2pt}r{2pt}){15-17} \cmidrule(l{2pt}r{2pt}){18-20} \cmidrule(l{2pt}r{2pt}){21-23} \cmidrule(l{2pt}r{2pt}){24-26}
\multicolumn{1}{c}{\textbf{ }} 
& \multicolumn{1}{c}{\textbf{60dB(A)}} 
& \multicolumn{3}{c}{\textbf{65dB(A)}} 
& \multicolumn{3}{c}{\textbf{70dB(A)}} 
& \multicolumn{3}{c}{\textbf{65dB(A)}} 
& \multicolumn{3}{c}{\textbf{70dB(A)}} 
& \multicolumn{12}{c}{\textbf{65dB(A)}} \\
\cmidrule(l{2pt}r{2pt}){2-2} \cmidrule(l{2pt}r{2pt}){3-5} \cmidrule(l{2pt}r{2pt}){6-8} \cmidrule(l{2pt}r{2pt}){9-11} \cmidrule(l{2pt}r{2pt}){12-14} \cmidrule(l{2pt}r{2pt}){15-26}

& \texttt{TRA} & \texttt{AIR} & \texttt{MRT} & \texttt{TRA}
& \texttt{AIR} & \texttt{MRT} & \texttt{TRA}
& \texttt{AIR} & \texttt{MRT} & \texttt{TRA}
& \texttt{AIR} & \texttt{MRT} & \texttt{TRA}
& \texttt{AIR} & \texttt{MRT} & \texttt{TRA}
& \texttt{AIR} & \texttt{MRT} & \texttt{TRA}
& \texttt{AIR} & \texttt{MRT} & \texttt{TRA}
& \texttt{AIR} & \texttt{MRT} & \texttt{TRA} \\

\midrule
$L_{\text{Aeq}}$ 
& 60.08 & 66.02 & 65.24 & 65.25 & 70.90 & 70.25 & 70.00 
& 65.85 & 64.22 & 64.43 & 70.26 & 69.16 & 69.21 & 67.63 
& 66.94 & 66.72 & 66.88 & 66.17 & 66.14 & 67.75 & 67.11 
& 66.93 & 67.38 & 66.41 & 66.38\\
$L_{\text{Amax}}$ 
& 61.08 & 74.17 & 68.25 & 66.26 & 78.75 & 73.24 
& 71.01 & 74.02 & 67.40 & 65.43 & 78.56 & 72.41 & 70.11 
& 75.10 & 73.33 & 73.00 & 74.94 & 73.06 & 72.84 & 74.66 
& 69.76 & 68.25 & 74.51 & 68.97 & 67.68\\
$L_{\text{A5}}$
& 60.70 & 71.17 & 67.64 & 65.88 & 75.75 & 72.62 & 70.63 
& 70.77 & 66.65 & 65.01 & 75.28 & 71.65 & 69.80 & 72.64 
& 70.03 & 69.37 & 72.15 & 69.61 & 69.09 & 71.68 & 68.84 
& 67.61 & 71.30 & 68.11 & 67.09\\
$L_{\text{A10}}$ 
& 60.60 & 70.81 & 67.40 & 65.77 & 75.40 & 72.39 & 70.53 
& 70.40 & 66.36 & 64.92 & 74.93 & 71.36 & 69.71 & 71.58 
& 69.33 & 68.72 & 71.04 & 68.83 & 68.37 & 71.35 & 68.61 
& 67.47 & 70.94 & 67.83 & 66.96\\
$L_{\text{A50}}$ 
& 60.14 & 60.08 & 65.61 & 65.31 & 65.07 & 70.56 & 70.06 
& 59.61 & 64.37 & 64.53 & 64.14 & 69.29 & 69.31 & 64.12 
& 66.67 & 65.80 & 62.46 & 65.73 & 65.01 & 64.62 & 67.22 
& 66.96 & 64.35 & 66.41 & 66.40\\
$L_{\text{A90}}$ 
& 59.41 & 52.27 & 58.03 & 64.57 & 56.85 & 63.19 & 69.32 
& 51.38 & 57.42 & 63.74 & 54.81 & 62.09 & 68.51 & 53.30 
& 61.31 & 64.92 & 51.77 & 59.94 & 64.04 & 62.58 & 64.33 
& 66.24 & 62.41 & 64.02 & 65.65\\
$L_{\text{A95}}$ 
& 59.24 & 50.99 & 56.38 & 64.40 & 55.12 & 61.39 & 69.14 
& 49.82 & 55.58 & 63.55 & 53.14 & 60.16 & 68.32 & 51.11 
& 57.83 & 64.68 & 49.58 & 56.87 & 63.81 & 62.30 & 63.76 
& 66.02 & 62.15 & 63.44 & 65.37\\
\midrule
$L_{\text{Ceq}}$ 
& 61.02 & 65.88 & 65.43 & 65.47 & 70.52 & 70.16 & 69.97 
& 65.16 & 63.76 & 64.26 & 69.28 & 68.35 & 68.71 & 67.09 
& 66.59 & 66.48 & 65.93 & 65.28 & 65.55 & 67.28 & 66.82 
& 66.71 & 66.47 & 65.63 & 65.82\\
$L_{\text{Cmax}}$ 
& 63.41 & 73.13 & 69.23 & 67.89 & 77.67 & 73.09 & 71.12 
& 72.74 & 66.76 & 67.51 & 77.23 & 71.22 & 69.62 & 73.88 
& 72.04 & 71.66 & 73.52 & 71.49 & 71.51 & 73.57 & 69.34 
& 67.99 & 73.19 & 67.93 & 68.00\\
$L_{\text{C5}}$
& 61.97 & 70.69 & 67.52 & 66.17 & 75.22 & 72.34 & 70.67 
& 69.63 & 65.83 & 64.83 & 74.06 & 70.58 & 69.23 & 71.70 
& 69.14 & 68.52 & 70.85 & 68.34 & 68.00 & 71.14 & 68.45 
& 67.35 & 70.13 & 67.14 & 66.47\\
$L_{\text{C10}}$ 
& 61.71 & 70.23 & 67.26 & 66.02 & 74.76 & 72.07 & 70.53 
& 69.25 & 65.56 & 64.70 & 73.69 & 70.34 & 69.14 & 70.98 
& 68.57 & 67.98 & 69.80 & 67.64 & 67.38 & 70.72 & 68.23 
& 67.22 & 69.78 & 66.88 & 66.34\\
$L_{\text{C50}}$  
& 60.94 & 61.25 & 65.73 & 65.44 & 65.65 & 70.47 & 69.96 
& 60.63 & 63.93 & 64.24 & 63.98 & 68.52 & 68.74 & 64.08 
& 66.53 & 65.94 & 62.45 & 64.95 & 64.76 & 64.55 & 66.98 
& 66.71 & 63.92 & 65.65 & 65.83\\
$L_{\text{C90}}$ 
& 60.30 & 57.97 & 60.11 & 64.84 & 61.30 & 64.18 & 69.40 
& 57.47 & 59.14 & 63.72 & 59.69 & 62.62 & 68.19 & 58.22 
& 62.08 & 65.17 & 57.61 & 60.51 & 63.96 & 62.58 & 64.03 
& 66.13 & 62.16 & 63.61 & 65.18\\
$L_{\text{C95}}$ 
& 60.15 & 57.37 & 59.25 & 64.68 & 60.65 & 63.14 & 69.18 
& 56.70 & 58.34 & 63.53 & 58.99 & 61.87 & 67.92 & 57.48 
& 59.86 & 65.00 & 56.64 & 58.56 & 63.77 & 62.31 & 63.55 
& 65.97 & 61.87 & 63.08 & 64.94\\
\midrule
$N_{\text{max}}$ 
& 13.89 & 28.36 & 20.26 & 19.24 & 37.60 & 27.67 & 25.94 
& 25.98 & 17.91 & 16.53 & 34.45 & 24.67 & 22.29 & 29.96 
& 25.26 & 25.05 & 27.71 & 22.74 & 23.09 & 29.48 & 23.55 
& 22.67 & 27.20 & 21.02 & 21.11\\
$N_{\text{5}}$  
& 12.86 & 22.46 & 18.53 & 17.83 & 29.86 & 25.40 & 24.02 
& 20.01 & 15.85 & 15.74 & 26.49 & 21.83 & 21.30 & 24.25 
& 20.26 & 20.56 & 21.34 & 17.62 & 18.55 & 23.95 & 20.65 
& 20.14 & 21.62 & 18.59 & 18.55\\
$N_{\text{10}}$   
& 12.70 & 21.42 & 18.12 & 17.57 & 28.51 & 24.84 & 23.67 
& 19.13 & 15.49 & 15.54 & 25.30 & 21.30 & 21.03 & 22.63 
& 19.23 & 19.36 & 19.87 & 16.56 & 17.39 & 22.94 & 20.20 
& 19.84 & 20.80 & 18.15 & 18.25\\
$N_{\text{50}}$   
& 12.07 & 11.70 & 15.80 & 16.71 & 15.85 & 21.62 & 22.49 
& 10.68 & 13.37 & 14.78 & 13.94 & 18.33 & 20.02 & 13.03 
& 16.55 & 17.14 & 11.10 & 14.09 & 15.22 & 16.08 & 18.27 
& 18.80 & 15.26 & 16.46 & 17.26\\
$N_{\text{90}}$  
& 11.50 & 7.96 & 10.63 & 15.89 & 10.39 & 14.68 & 21.39 
& 7.20 & 9.36 & 14.09 & 9.01 & 12.66 & 19.03 & 8.04 
& 11.35 & 16.14 & 7.14 & 9.95 & 14.28 & 13.75 & 15.27 
& 17.88 & 13.15 & 14.43 & 16.36\\
$N_{\text{95}}$ 
& 11.33 & 7.50 & 9.82 & 15.68 & 9.69 & 13.46 & 21.11 
& 6.78 & 8.59 & 13.90 & 8.47 & 11.59 & 18.81 & 7.50 
& 10.07 & 15.89 & 6.69 & 8.82 & 14.09 & 13.43 & 14.62 
& 17.66 & 12.83 & 13.89 & 16.09\\
\midrule
$S_{\text{max}}$  
& 2.00 & 1.98 & 2.00 & 1.82 & 1.80 & 1.83 & 1.70 & 2.10 
& 2.14 & 1.92 & 1.93 & 1.97 & 1.76 & 2.09 & 2.09 & 1.96 
& 2.21 & 2.25 & 2.10 & 2.06 & 2.03 & 1.85 & 2.12 & 2.07 & 1.94\\
$S$ 
& 1.69 & 1.63 & 1.67 & 1.60 & 1.54 & 1.61 & 1.55 & 1.77 
& 1.85 & 1.72 & 1.66 & 1.77 & 1.66 & 1.67 & 1.70 & 1.62 
& 1.81 & 1.87 & 1.75 & 1.79 & 1.79 & 1.72 & 1.87 & 1.90 & 1.82\\
$S_{\text{5}}$ 
& 1.73 & 1.77 & 1.79 & 1.63 & 1.64 & 1.72 & 1.58 & 1.95 & 1.96 
& 1.76 & 1.77 & 1.87 & 1.71 & 1.84 & 1.87 & 1.77 & 2.00 & 2.03 
& 1.88 & 1.95 & 1.93 & 1.78 & 2.01 & 2.01 & 1.88\\
$S_{\text{10}}$ 
& 1.71 & 1.74 & 1.76 & 1.62 & 1.63 & 1.69 & 1.58 & 1.91 & 1.93 
& 1.75 & 1.75 & 1.84 & 1.70 & 1.78 & 1.82 & 1.69 & 1.95 & 1.97 
& 1.81 & 1.93 & 1.90 & 1.77 & 1.98 & 1.98 & 1.86\\
$S_{\text{50}}$ 
& 1.69 & 1.63 & 1.66 & 1.60 & 1.54 & 1.61 & 1.55 & 1.76 & 1.84 
& 1.72 & 1.66 & 1.77 & 1.66 & 1.67 & 1.69 & 1.61 & 1.82 & 1.85 
& 1.73 & 1.80 & 1.78 & 1.73 & 1.87 & 1.90 & 1.82\\
$S_{\text{90}}$ 
& 1.65 & 1.52 & 1.60 & 1.56 & 1.45 & 1.55 & 1.51 & 1.65 & 1.77 
& 1.69 & 1.59 & 1.71 & 1.62 & 1.54 & 1.61 & 1.58 & 1.68 & 1.78 
& 1.70 & 1.65 & 1.71 & 1.68 & 1.76 & 1.84 & 1.78\\
$S_{\text{95}}$ 
& 1.65 & 1.50 & 1.58 & 1.56 & 1.42 & 1.54 & 1.50 & 1.64 & 1.76 
& 1.68 & 1.57 & 1.70 & 1.61 & 1.52 & 1.59 & 1.56 & 1.66 & 1.76 
& 1.69 & 1.63 & 1.69 & 1.67 & 1.74 & 1.82 & 1.77\\
\midrule
$R_{\text{max}}$ 
& 0.04 & 0.06 & 0.05 & 0.06 & 0.07 & 0.06 & 0.08 & 0.04 & 0.04 
& 0.04 & 0.05 & 0.05 & 0.06 & 0.06 & 0.05 & 0.06 & 0.07 & 0.05 
& 0.05 & 0.06 & 0.05 & 0.06 & 0.04 & 0.05 & 0.04\\
$R$ 
& 0.02 & 0.02 & 0.03 & 0.03 & 0.03 & 0.03 & 0.03 & 0.02 & 0.02 
& 0.02 & 0.02 & 0.02 & 0.03 & 0.02 & 0.03 & 0.03 & 0.02 & 0.02 
& 0.02 & 0.03 & 0.03 & 0.03 & 0.02 & 0.02 & 0.02\\
$R_{\text{5}}$ 
& 0.03 & 0.04 & 0.04 & 0.04 & 0.04 & 0.04 & 0.05 & 0.03 & 0.03 
& 0.03 & 0.03 & 0.03 & 0.04 & 0.04 & 0.04 & 0.04 & 0.03 & 0.03 
& 0.03 & 0.04 & 0.04 & 0.04 & 0.03 & 0.03 & 0.03\\
$R_{\text{10}}$ 
& 0.03 & 0.03 & 0.03 & 0.04 & 0.04 & 0.04 & 0.04 & 0.03 & 0.02 
& 0.03 & 0.03 & 0.03 & 0.03 & 0.03 & 0.03 & 0.04 & 0.03 & 0.03 
& 0.03 & 0.03 & 0.03 & 0.04 & 0.03 & 0.03 & 0.03\\
$R_{\text{50}}$ 
& 0.02 & 0.02 & 0.02 & 0.03 & 0.03 & 0.03 & 0.03 & 0.02 & 0.02 
& 0.02 & 0.02 & 0.02 & 0.02 & 0.02 & 0.02 & 0.03 & 0.02 & 0.02
& 0.02 & 0.02 & 0.03 & 0.03 & 0.02 & 0.02 & 0.02\\
$R_{\text{90}}$ 
& 0.02 & 0.01 & 0.02 & 0.02 & 0.02 & 0.02 & 0.02 & 0.01 & 0.01 
& 0.02 & 0.01 & 0.01 & 0.02 & 0.01 & 0.02 & 0.02 & 0.01 & 0.01 
& 0.01 & 0.02 & 0.02 & 0.02 & 0.02 & 0.02 & 0.02\\
$R_{\text{95}}$ 
& 0.02 & 0.01 & 0.02 & 0.02 & 0.02 & 0.02 & 0.02 & 0.01 & 0.01 
& 0.01 & 0.01 & 0.01 & 0.02 & 0.01 & 0.02 & 0.02 & 0.01 & 0.01 
& 0.01 & 0.02 & 0.02 & 0.02 & 0.02 & 0.02 & 0.02\\
\midrule
$FS_{\text{max}}$ 
& 0.08 & 0.09 & 0.07 & 0.08 & 0.09 & 0.08 & 0.09 & 0.09 & 0.07 
& 0.08 & 0.09 & 0.08 & 0.08 & 0.11 & 0.08 & 0.08 & 0.11 & 0.08 
& 0.08 & 0.09 & 0.08 & 0.09 & 0.08 & 0.08 & 0.08\\
$FS_{\text{5}}$ 
& 0.05 & 0.05 & 0.04 & 0.05 & 0.06 & 0.05 & 0.05 & 0.05 & 0.04 
& 0.05 & 0.05 & 0.04 & 0.05 & 0.08 & 0.07 & 0.07 & 0.09 & 0.07 
& 0.07 & 0.05 & 0.05 & 0.05 & 0.05 & 0.04 & 0.05\\
$FS_{\text{10}}$ 
& 0.03 & 0.05 & 0.02 & 0.01 & 0.05 & 0.02 & 0.01 & 0.03 & 0.02 
& 0.02 & 0.03 & 0.02 & 0.01 & 0.08 & 0.06 & 0.05 & 0.08 & 0.07 
& 0.05 & 0.04 & 0.02 & 0.01 & 0.02 & 0.02 & 0.01\\
$FS_{\text{50}}$ 
& 0.01 & 0.02 & 0.01 & 0.01 & 0.02 & 0.01 & 0.01 & 0.02 & 0.01 
& 0.01 & 0.02 & 0.01 & 0.01 & 0.04 & 0.03 & 0.02 & 0.03 & 0.03 
& 0.02 & 0.01 & 0.01 & 0.01 & 0.01 & 0.01 & 0.01\\
$FS_{\text{90}}$ 
& 0.00 & 0.01 & 0.01 & 0.00 & 0.01 & 0.01 & 0.00 & 0.01 & 0.01 
& 0.00 & 0.01 & 0.01 & 0.00 & 0.01 & 0.01 & 0.01 & 0.01 & 0.01 
& 0.01 & 0.01 & 0.01 & 0.00 & 0.01 & 0.01 & 0.00\\
$FS_{\text{95}}$ 
& 0.00 & 0.01 & 0.01 & 0.00 & 0.00 & 0.01 & 0.00 & 0.01 & 0.00 
& 0.00 & 0.01 & 0.01 & 0.00 & 0.01 & 0.01 & 0.00 & 0.01 & 0.01 
& 0.00 & 0.01 & 0.01 & 0.00 & 0.01 & 0.00 & 0.00\\
\midrule
$T_{\text{max}}$ 
& 0.36 & 0.97 & 1.09 & 0.43 & 1.17 & 1.31 & 0.52 & 0.72 & 1.16 
& 0.42 & 0.81 & 1.39 & 0.52 & 2.06 & 2.06 & 1.81 & 2.14 & 2.14 
& 1.90 & 0.83 & 0.94 & 0.45 & 0.66 & 1.03 & 0.40\\
$T_{\text{5}}$ 
& 0.24 & 0.37 & 0.80 & 0.28 & 0.46 & 0.96 & 0.33 & 0.41 & 0.88 
& 0.27 & 0.48 & 1.05 & 0.32 & 0.89 & 0.97 & 0.78 & 0.90 & 1.03 
& 0.83 & 0.33 & 0.64 & 0.26 & 0.34 & 0.70 & 0.22\\
$T_{\text{10}}$ 
& 0.19 & 0.29 & 0.70 & 0.23 & 0.36 & 0.85 & 0.27 & 0.32 & 0.79 
& 0.24 & 0.36 & 0.94 & 0.28 & 0.61 & 0.80 & 0.49 & 0.61 & 0.87 
& 0.52 & 0.26 & 0.57 & 0.22 & 0.25 & 0.62 & 0.19\\
$T_{\text{50}}$ 
& 0.10 & 0.12 & 0.43 & 0.12 & 0.15 & 0.52 & 0.14 & 0.14 & 0.49 
& 0.13 & 0.17 & 0.57 & 0.16 & 0.16 & 0.45 & 0.15 & 0.17 & 0.50 
& 0.16 & 0.10 & 0.33 & 0.11 & 0.10 & 0.35 & 0.10\\
$T_{\text{90}}$ 
& 0.05 & 0.05 & 0.22 & 0.06 & 0.06 & 0.26 & 0.07 & 0.06 & 0.25 
& 0.07 & 0.07 & 0.29 & 0.08 & 0.06 & 0.23 & 0.07 & 0.06 & 0.26 
& 0.08 & 0.04 & 0.12 & 0.05 & 0.04 & 0.13 & 0.05\\
$T_{\text{95}}$ 
& 0.04 & 0.04 & 0.17 & 0.05 & 0.05 & 0.21 & 0.06 & 0.05 & 0.20 
& 0.06 & 0.06 & 0.24 & 0.07 & 0.04 & 0.18 & 0.06 & 0.05 & 0.21 
& 0.06 & 0.03 & 0.07 & 0.04 & 0.03 & 0.08 & 0.04\\
\midrule
\textit{PA} 
& 13.01 & 22.68 & 18.76 & 18.06 & 30.17 & 25.72 & 24.33 & 20.25 
& 16.41 & 15.90 & 26.72 & 22.11 & 21.53 & 24.52 & 20.51 & 20.80 
& 21.86 & 18.38 & 18.73 & 24.40 & 21.07 & 20.38 & 22.60 & 19.64 
& 19.05\\
\bottomrule
\end{tabularx}
\end{table*}

\subsubsection{Psychoacoustic loudness} 
\label{sec:psyloud}

Since psychoacoustic loudness is a relative linear scale (i.e. doubling of loudness in Sones is equivalent to the doubling of the loudness sensation), and $N_5$ closely represents the human perception of loudness \citep{InternationalOrganizationforStandardization2017a}, the  psychoacoustic loudness is assessed in terms of the $N_5$ index in percentages. For instance, the percentage change in $N_5$ for \SI{65}{\decibelA} \texttt{AIR} noise after ANC is given by \{$1-(N^{\texttt{AIR}^*_\texttt{65dB(A)}}_\text{5,HATS}/N^{\texttt{AIR}_\texttt{65dB(A)}}_\text{5,HATS})\times100$\}\si{\percent} . In contrast to negligible attenuation in $L_\text{Aeq}$, a notable decrease in loudness was observed in all noise types noise types at \SI{65}{\decibelA} and \SI{70}{\decibelA} after ANC, in the absence of maskers ($N^{\texttt{AIR}^*_\texttt{65dB(A)}}_\text{5,HATS}$: \SI{10.91}{\percent}; 
$N^{\texttt{AIR}^*_\texttt{70dB(A)}}_\text{5,HATS}$: \SI{11.29}{\percent}; 
$N^{\texttt{MRT}^*_\texttt{65dB(A)}}_\text{5,HATS}$: \SI{14.46}{\percent}; 
$N^{\texttt{MRT}^*_\texttt{70dB(A)}}_\text{5,HATS}$: \SI{14.06}{\percent};
$N^{\texttt{TRA}^*_\texttt{65dB(A)}}_\text{5,HATS}$: \SI{11.72}{\percent}; 
$N^{\texttt{TRA}^*_\texttt{70dB(A)}}_\text{5,HATS}$: \SI{11.32}{\percent}).

Upon the addition of the bird masker the effect of ANC was diminished by more than half in \texttt{AIR}, and MRT noise types; and even increased the \texttt{TRA} loudness ($N^{\texttt{AIR}^*_\texttt{65dB(A)}+\text{B}}_\text{5,HATS}$: \SI{4.99}{\percent}; $N^{\texttt{MRT}^*_\texttt{65dB(A)}+\text{B}}_\text{5,HATS}$: \SI{4.91}{\percent}; $N^{\texttt{TRA}^*_\texttt{65dB(A)}+\text{B}}_\text{5,HATS}$: \SI{-4.04}{\percent}). 

Similarly, water masker further diminished the effect of ANC in \texttt{AIR} noise; and resulted in loudness increment in \texttt{MRT}, and \texttt{TRA} noise types ($N^{\texttt{AIR}^*_\texttt{65dB(A)}+\text{W}}_\text{5,HATS}$: \SI{3.74}{\percent}; $N^{\texttt{MRT}^*_\texttt{65dB(A)}+\text{W}}_\text{5,HATS}$: \SI{-0.32}{\percent}; $N^{\texttt{TRA}^*_\texttt{65dB(A)}+\text{W}}_\text{5,HATS}$: \SI{-4.04}{\percent}).

\subsubsection{Psychoacoustic sharpness}

Psychoacoustic sharpness is an objective representation of the amount of high-frequency content, without level dependence (i.e. sharpness doubles with an increase in sound level from \SI{30}{\decibel} to \SI{90}{\decibel}), and is negatively correlated to the pleasantness sensation \citet{Zwicker2013}. Percentage change in sharpness indices were computed similar to loudness in \Cref{sec:psyloud}.

Across all sharpness indices, ANC resulted in between ~7.10 to \SI{11}{\percent} 
increase in average sharpness across all noise types at both \SI{65}{\decibelA} and \SI{70}{\decibelA} ($S^{\texttt{AIR}^*_\texttt{65dB(A)}}_\text{HATS}$: \SI{-7.20}{\percent}; $S^{\texttt{AIR}^*_\texttt{70dB(A)}}_\text{HATS}$: \SI{-7.79}{\percent}; $S^{\texttt{MRT}^*_\texttt{65dB(A)}}_\text{HATS}$: \SI{-10.78}{\percent}; $S^{\texttt{MRT}^*_\texttt{70dB(A)}}_\text{HATS}$: \SI{-9.94}{\percent}; $S^{\texttt{TRA}^*_\texttt{65dB(A)}}_\text{HATS}$: \SI{-7.5}{\percent}; $S^{\texttt{TRA}^*_\texttt{70dB(A)}}_\text{HATS}$: \SI{-7.10}{\percent}). 

With predominantly high frequencies in the bird masker, its addition further increased the average sharpness of the residual noise after ANC, across all noise types ($S^{\texttt{AIR}^*_\texttt{65dB(A)}+\text{B}}_\text{HATS}$: \SI{-11.04}{\percent}; 
$S^{\texttt{MRT}^*_\texttt{65dB(A)}+\text{B}}_\text{HATS}$: \SI{-11.98}{\percent}; 
$S^{\texttt{TRA}^*_\texttt{65dB(A)}+\text{B}}_\text{HATS}$: \SI{-9.38}{\percent}). Moreover, due to greater high-frequency content in the water masker, the average sharpness was also notably increased when the water masker was added after ANC ($S^{\texttt{AIR}^*_\texttt{65dB(A)}+\text{W}}_\text{HATS}$: \SI{-14.72}{\percent}; 
$S^{\texttt{MRT}^*_\texttt{65dB(A)}+\text{W}}_\text{HATS}$: \SI{-13.77}{\percent}; 
$S^{\texttt{TRA}^*_\texttt{65dB(A)}+\text{W}}_\text{HATS}$: \SI{-13.75}{\percent}) 

\subsubsection{Psychoacoustic roughness and fluctuation strength}

Psychoacoustic roughness \textit{R} and fluctuation strength \textit{FS} indices quantitatively describe the perception due to temporal variations. Slow variations in the sound are characterized by the sensation of \textit{FS}, which gradually transitions to a sensation of \textit{R} that describes quickly varying sounds. Since both \textit{FS} and \textit{R} scores were marginal across all conditions ($<0.1$), they are not examined here for brevity. It is worth noting that experiments have shown that it is perceptually `no longer rough' at \SI{0.1}{\asper}, and a similar assumption can be made for \textit{FS} given close similarities in the underlying computational model with \textit{R}  \citep{Zwicker2013}.

\subsubsection{Psychoacoustic tonality}

Psychoacoustic tonality is the quantification of the amount of tonal components in the sound based on psychoacoustic loudness computation of tonal and non-tonal components. Similar to psychoacoustic loudness, percentage reduction in \si{\tuhms} is computed for $T_5$.

An increment in tonality was observed in the residual noise after ANC in \texttt{AIR} and \texttt{MRT} noise types at both \SI{65}{\decibelA} and \SI{70}{\decibelA}, whereas a slight reduction in tonality was observed in the residual \texttt{TRA} noise at both levels
($T^{\texttt{AIR}^*_\texttt{65dB(A)}}_\text{5,HATS}$: \SI{-10.81}{\percent}; 
$T^{\texttt{AIR}^*_\texttt{70dB(A)}}_\text{5,HATS}$: \SI{-4.35}{\percent};
$T^{\texttt{MRT}^*_\texttt{65dB(A)}}_\text{5,HATS}$: \SI{-10.00}{\percent}; 
$T^{\texttt{MRT}^*_\texttt{70dB(A)}}_\text{5,HATS}$: \SI{-9.375}{\percent};
$T^{\texttt{TRA}^*_\texttt{65dB(A)}}_\text{5,HATS}$: \SI{3.57}{\percent};
$T^{\texttt{TRA}^*_\texttt{70dB(A)}}_\text{5,HATS}$: \SI{3.03}{\percent}). 

The strong tonal components in the bird masker resulted in about 1.5 fold increase in tonality in the residual \texttt{AIR} noise after ANC when the bird masker was added; only ~\SI{20}{\percent} increase in \texttt{MRT} residual due to greater tonality components in \texttt{MRT} noise; and almost 2 fold increase in tonality in the residual \texttt{TRA} noise
($T^{\texttt{AIR}^*_\texttt{65dB(A)}+\text{B}}_\text{5,HATS}$: \SI{-143.24}{\percent}; 
$T^{\texttt{MRT}^*_\texttt{65dB(A)}+\text{B}}_\text{5,HATS}$: \SI{-28.75}{\percent}; 
$T^{\texttt{TRA}^*_\texttt{65dB(A)}+\text{B}}_\text{5,HATS}$: \SI{-196.43}{\percent}).

In contrast, the broadband nature and energetic masking properties of the water masker reduced overall tonality of the residual noise between \SI{8.11}{\percent} and \SI{21.43}{\percent} across the noise types.
($T^{\texttt{AIR}^*_\texttt{65dB(A)}+\text{W}}_\text{5,HATS}$: \SI{8.11}{\percent}; 
$T^{\texttt{MRT}^*_\texttt{65dB(A)}+\text{W}}_\text{5,HATS}$: \SI{12.5}{\percent}; 
$T^{\texttt{TRA}^*_\texttt{65dB(A)}+\text{W}}_\text{5,HATS}$: \SI{21.43}{\percent}).

\subsection{Annoyance}

\subsubsection{Degree of agreement on annoyance between verbal and numerical scale}

To examine the degree of agreement between the verbal and numerical annoyance scales with Bland-Altman statistics and plots, as shown in \Cref{fig:BA65dBA} in \labelcref{sec:BAstats}, the 5-point verbal category scale was first upscaled to between 0 and 100 to match the 101-point numerical scale, i.e. \{0, 25, 50, 75, 100\} \citep{Brink2016EffectsExperiment}. From the assessing the mean of the differences in the stimuli annoyance scores, a slight negative bias was observed only in \texttt{$\text{AIR}_{\text{65dB(A)}}+\text{B}$} for \texttt{AIR} noise; in \texttt{$\text{MRT}^{*}_{\text{65dB(A)}}$},  \texttt{$\text{MRT}_{\text{65dB(A)}}+\text{B}$},  \texttt{$\text{MRT}^{*}_{\text{65dB(A)}}+\text{B}$}, and \texttt{$\text{MRT}_{\text{70dB(A)}}$} for \texttt{MRT} noise; as well as in \texttt{$\text{TRA}_{\text{60dB(A)}}$} and \texttt{$\text{TRA}^{*}_{\text{65dB(A)}}+\text{W}$} for \texttt{TRA} noise, wherein the bias occurs if the equivalence line falls beyond the \SI{95}{\percent} mean confidence intervals.

Aside from several outliers beyond the narrow confidence intervals of the \SI{95}{\percent} upper and lower limits of agreement (LoA), most scores are within the LoA, and thus both scales largely agree. Hence, the 101-point numerical scale was chosen for statistical purposes \citep{InternationalOrganizationforStandardization2021}, and summarised in \Cref{tab:qnDataSummary} in \labelcref{sec:statsAppend}.

\subsubsection{Difference in annoyance}

The 3W-ART ANOVA analysis showed significant difference in annoyance between noise types ($p<0.001$), masker types ($p\ll0.001$), and ANC conditions ($p<0.01$), as shown in \Cref{tab:annoyARTstats}. Interaction effects were only observed in the two-way interaction between masker type and ANC conditions ($p<0.01$).

Posthoc ART contrast tests with Bonferroni correction for multiple pairs revealed that annoyance was significantly lowered with ANC with a large effect size ($p<0.01, d>0.7$). Between masker types, the bird and water were each significantly less annoying than without masker, and both to a large effect ($p_\text{\texttt{B-None}}<0.001, d_\text{\texttt{B-None}}<-0.7; p_\text{\texttt{B-None}}\ll0.001, d_\text{\texttt{B-None}}>0.7$). No significant differences were found between \texttt{B} and \texttt{W} maskers. 

In contrast tests of the masker-ANC interaction, bird masker with ANC ($\mathcal{N}^{*}_{\texttt{65\si{\decibelA}}}+\texttt{B}$) was significantly less annoying than (1) bird masker without ANC ($\mathcal{N}_{\texttt{65dB(A)}}+\texttt{B}$) with a large effect size ($p<0.01$, $d\ge0.8$); (2) no masker without ANC ($\mathcal{N}_{\texttt{l}}$) with a small effect ($p<0.001$, $0.2\le|d|<0.5$); and (3) no masker with ANC ($\mathcal{N}^{*}_{\texttt{l}}$) with a large effect ($p\ll0.001$, $d\ge0.8$). Water masker with ANC ($\mathcal{N}^{*}_{\texttt{65\si{\decibelA}}}+\texttt{W}$) was also significantly less annoying than (1) $\mathcal{N}_{\texttt{65dB(A)}}+\texttt{B}$ with a large effect ($p<0.01$, $d\ge0.8$); (2) $\mathcal{N}_{\texttt{l}}$ with a small effect ($p\ll0.001$, $0.2\le|d|<0.5$); and (3) $\mathcal{N}^{*}_{\texttt{l}}$ with a large effect ($p\ll0.001$, $d\ge0.8$). No significant differences were found in the remaining pairs. 

\subsection{Perceived affective quality}

\subsubsection{Circumplexity analysis}

Based on the high correspondence index (CI) of 0.861 (\SI{6.94}{\percent} violations; $p<0.001$) via the RTHOR method, the PAQ scores reflected a good adherence to the underlying circumplexity of the PAQ model. Visual inspection of the PCA loadings also exhibited sinusoidal tendencies, and thus further affirms the validity of the circumplexity assumption PAQ model in assessing the indoor environment in the context of environmental noise. This deviates from previous findings in an acute care setting, where the PAQ model was found to violate the circumplexity assumption \citep{Lam2022c}. For conciseness, the circumplexity analysis is described in detail in \labelcref{sec:circumplex}. 

\subsubsection{Effect on derived pleasantness and eventfulness}

With adherence to circumplexity, the derived \textit{pleasantness} and \textit{eventfulness} scores were computed from the 8 PAQ attributes according ISO 12913-3, denoted by \textit{ISOPL} and \textit{ISOEV}, respectively, to give
\begin{equation}
    \textit{ISOPL} = (p-a)+\cos{\SI{45}{\degree}}\cdot(ca-ch)+\cos{\SI{45}{\degree}}\cdot(v-m),
\end{equation}
and 
\begin{equation}
    \textit{ISOEV} = (e-u)+\cos{\SI{45}{\degree}}\cdot(ch-ca)+\cos{\SI{45}{\degree}}\cdot(v-m),
\end{equation}
respectively.

Despite the independence assumption of the \textit{ISOPL} and \textit{ISOEV} axes in the circumplex model, a multivariate analysis approach was adopted due to the lack of evidence in the applicability of the PAQ circumplex model in indoor environments. Since multivariate kurtosis is normal ($p>0.05$), and manageably skewed ($p<0.0001$; \citet{Knief2021ViolatingEvils}), the 3W-MANOVA test was employed to examine the effect of noise types, maskers, and ANC on the combined \textit{ISOPL} and \textit{ISOEV}. 
Results of the 3W-MANOVA analysis revealed significant differences in the combined \textit{ISOPL} and \textit{ISOEV} between the noise types ($p\ll0.001$, $0.06<\eta^2_\text{p}<0.14$) and maskers ($p\ll0.001$, $0.06<\eta^2_\text{p}<0.14$) both with moderate effects. Significant difference was also observed in ANC conditions, but with a small effect ($p\ll0.001$, $0.01<\eta^2_\text{p}<0.06$). Interaction effects were only significant between noise types and maskers with a small effect ($p<0.01$, $0.01<\eta^2_\text{p}<0.06$). 

Post-hoc univariate 3W-ANOVA for \textit{ISOPL} revealed significant differences between masker types with a moderate effect ($p\ll0.0001$, $0.06<\eta^2_\text{p}<0.14$) and ANC conditions with a small effect ($p<0.001$, $0.01<\eta^2_\text{p}<0.06$). No significant differences in \textit{ISOPL} between noise types were found. Pairwise comparison with Tukey's HSD between masker types showed that \textit{ISOPL} significantly increased on the addition of bird ($p\ll0.0001$, $0.05\le d<0.08$) and water ($p\ll0.0001$, $0.05\le d<0.08$) maskers, both with a moderate effect, regardless of noise type or ANC conditions. The \textit{ISOPL} was also significantly increased on the activation of ANC regardless of noise types and maskers, albeit to a small effect ($p<0.001$, $0.02\le d<0.05$). 

In terms of interaction between noise types and maskers without considering ANC condition, Tukey's HSD showed that 
the addition of water maskers regardless of ANC significantly increased the \textit{ISOPL} when added to \texttt{AIR} noise with a moderate effect ($p\ll0.0001$, $0.05\le d<0.08$), whereas the addition of bird maskers did not affect the \textit{ISOPL}. On the other hand, \textit{ISOPL} significantly increased with a large effect when either bird ($p\ll0.0001$, $d>0.08$) or water ($p\ll0.0001$, $d>0.08$) maskers were added to \texttt{TRA} noise regardless of ANC condition. In contrast, there was insufficient evidence to conclude that either maskers improved the \textit{ISOPL} of \texttt{MRT} noise. On the whole, \texttt{MRT} noise was also found to be significantly more pleasant than \texttt{TRA} with a moderate effect ($p<0.05$, $0.05\le d<0.08$), but no difference in \textit{ISOPL} between \texttt{AIR} and either \texttt{MRT} or \texttt{TRA} was observed, regardless of ANC condition. Within each noise type, no significant differences were found between bird and water maskers in terms of \textit{ISOPL}.

Post-hoc univariate 3W-ANOVA for \textit{ISOEV} revealed significant differences between noise types ($p\ll0.0001$, $\eta^2_\text{p}>0.14$) and maskers ($p\ll0.0001$, $0.06<\eta^2_\text{p}<0.14$), but no differences for ANC conditions. Post-hoc pairwise Tukey's HSD showed that \texttt{TRA} noise regardless of masker and ANC conditions, had significantly lower \textit{ISOEV} than \texttt{AIR} ($p\ll0.0001$, $d>0.08$) and \texttt{TRA} ($p\ll0.0001$, $d>0.08$), both with a large effect. Between maskers regardless of other conditions, the addition of bird ($p\ll0.0001$, $0.05\le d<0.08$) and water ($p<0.001$, $0.02\le d<0.05$) maskers significantly increased the \textit{ISOEV}, both with a moderate effect.

\subsubsection{Analysis of distribution of individual scores}

\begin{figure}[ht]
    \centering
    \subfigure[]{\includegraphics[width=0.45\linewidth]{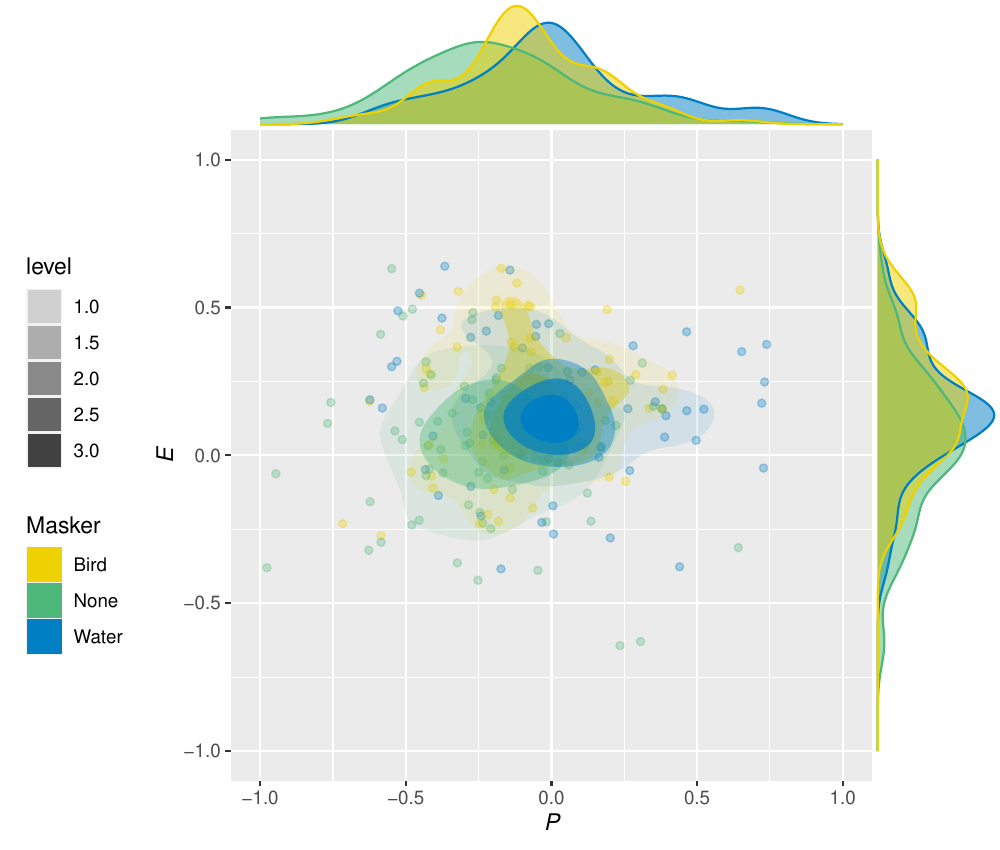}
    \label{fig:ISOPLEVdensityPlotsAIR}}
    \subfigure[]{\includegraphics[width=0.45\linewidth]{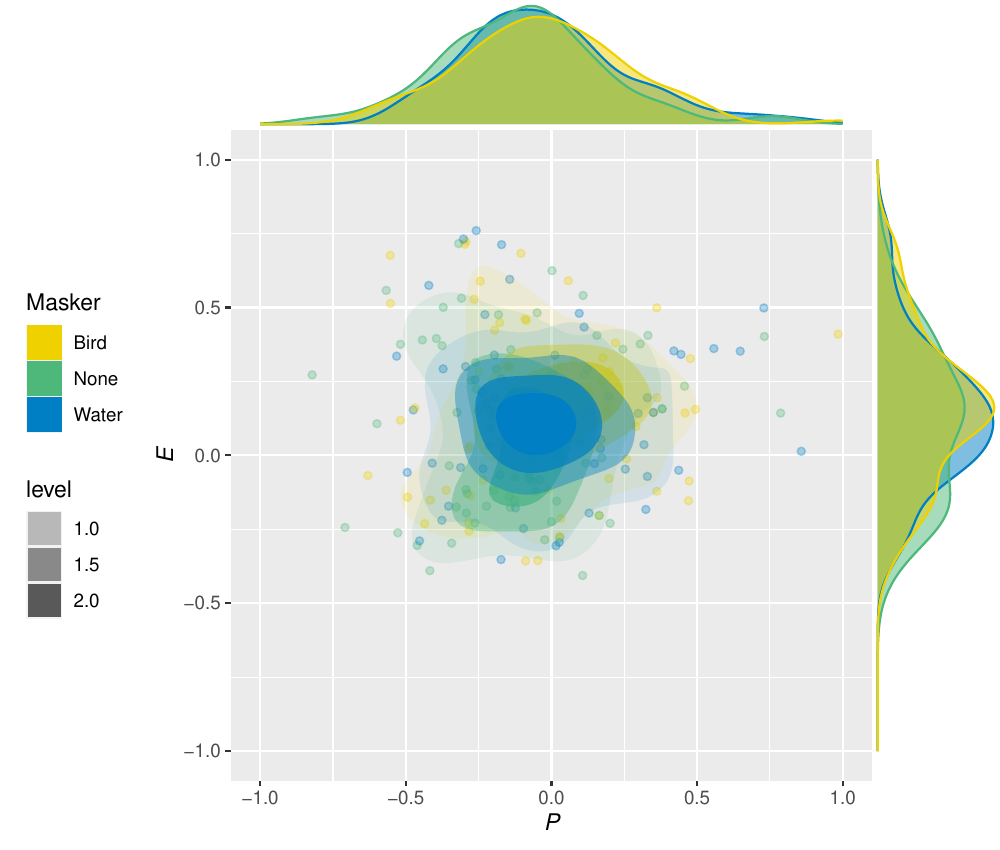}
    \label{fig:ISOPLEVdensityPlotsMRT}}
    \subfigure[]{\includegraphics[width=0.4\linewidth]{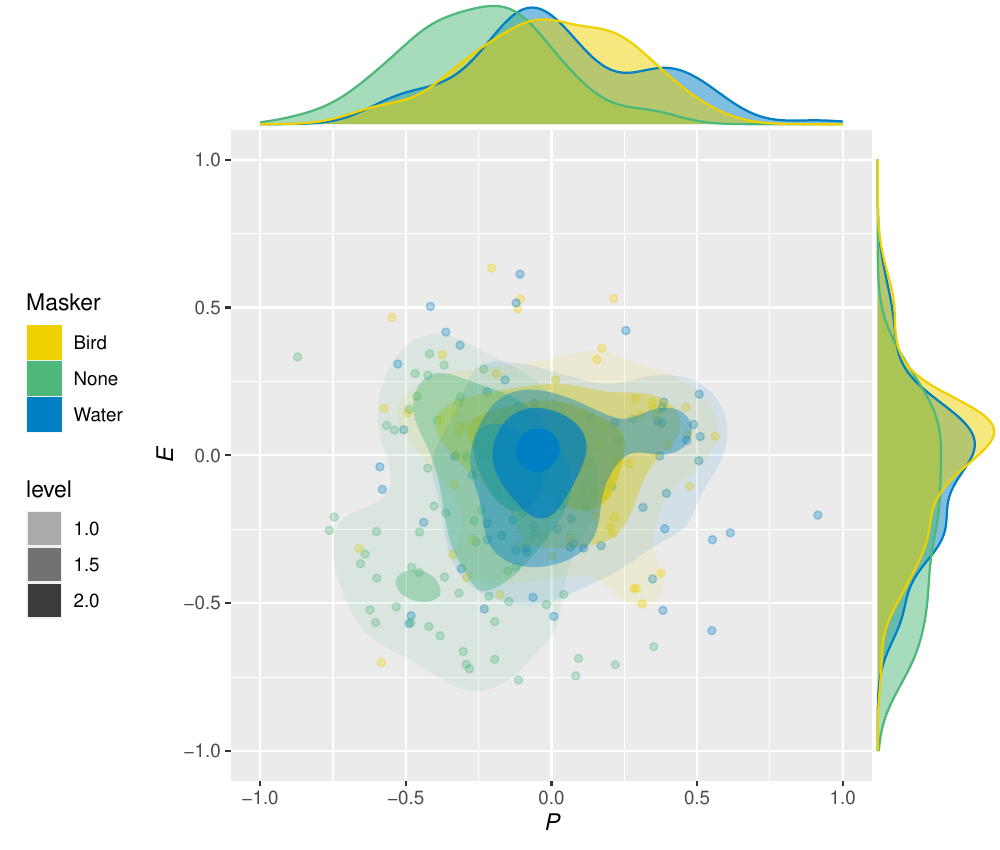}
    \label{fig:ISOPLEVdensityPlotsTRA}}
    \caption{Density plots of \textit{Eventfulness} (\textit{ISOEV}) as a function of \textit{Pleasantness} (\textit{ISOPL}) computed respectively with formulae A.2 and A.1 in ISO/TS 12913-3 across bird (\textcolor{den_1}{$\blacksquare$}), water (\textcolor{den_3}{$\blacksquare$}), and no maskers (\textcolor{den_2}{$\blacksquare$}). Plots are shown separately for (a) \texttt{AIR}, (b) \texttt{MRT}, and (c) \texttt{TRA} noise types at \SI{65}{\decibelA}.}
    \label{fig:ISOPLEVdensityPlots}
\end{figure}

To further analyse the interaction effect of noise types and maskers on \textit{ISOPL} and \textit{ISOEV}, the distribution of the \textit{ISOPL} and \textit{ISOEV} scores of individual participants were examined \citep{Mitchell2022HowData}. Since \textit{ISOPL} and \textit{ISOEV} is not influenced by ANC conditions, the distributions were aggregated over the ANC conditions, as shown in \Cref{fig:ISOPLEVdensityPlots}. Based on the compactness of the 2D density heat map and unimodality of the individual \textit{ISOPL} and \textit{ISOEV} density distributions, the participants tend to agree on both \textit{ISOPL} and \textit{ISOEV} for all conditions of \texttt{AIR}, albeit with some outliers, as shown in \Cref{fig:ISOPLEVdensityPlotsAIR}. The centrality of the 2D density heat maps indicates an overall neutrality of the \textit{ISOPL} and \textit{ISOEV} scores regardless of masker interventions for \texttt{MRT} noise, as shown in \Cref{fig:ISOPLEVdensityPlotsMRT}. It is worth noting that the neutrality of the \textit{ISOEV} scores in \texttt{MRT} without maskers could be attributed to the bimodal tendencies of the distribution around the centroid, as shown in \Cref{fig:ISOPLEVdensityPlotsTRA}. On the other hand, there is a larger spread in the 2D density plots across masker conditions for \texttt{TRA}, as a result of the bimodality of the  \textit{ISOPL} due to water maskers and the \textit{ISOEV} responses without maskers. The increment in \textit{ISOPL} upon the addition of water masker to \texttt{AIR} is evident in the shift in \textit{ISOPL} density distribution. Similarly, the significant increase in \textit{ISOPL} in \texttt{TRA} is exhibited in the shift of densities upon the addition of either bird or water maskers. 

\subsection{Perceived loudness}

Due to the non-normality of the loudness attribute across all noise types with the SWT ($p\ll0.001$), the non-parametric 2W-ART ANOVA procedure was employed to investigate the differences and interactions between the masker types and ANC conditions, as shown in \Cref{tab:PLNANOVAstats}. For each noise type, significant differences in \textit{PLN} were found between the masker types (\texttt{AIR}: $p<0.001$; \texttt{MRT}: $p\ll0.0001$; \texttt{TRA}: $p\ll0.0001$) and also between the ANC conditions (\texttt{AIR}: $p\ll0.0001$; \texttt{MRT}: $p<0.01$, \texttt{TRA}: $p\ll0.0001$). However, there was an absence of interaction effects between masker types and ANC conditions in all noise types. 

Posthoc ART contrast tests with Bonferroni correction for multiple pairs revealed that \textit{PLN} was significantly lower with than without ANC, across all noise types (\texttt{AIR}: $p\ll0.001$; \texttt{MRT}: $p<0.01$; \texttt{TRA}: $p\ll0.001$). A large effect size was observed in the \textit{PLN} differences across \texttt{AIR} and \texttt{TRA} ($d\ge0.8$), and a moderate effect was observed for \texttt{MRT} ($0.5\le d<0.8$).  

Between masker types, \textit{PLN} was only significantly increased when bird masker was added \texttt{MRT} without maskers with a moderate effect ($p<0.05$, $0.5\ge d<0.8$). However, \textit{PLN} was significantly increased with a large effect when water was added to \texttt{AIR} ($p<0.01$, $d>0.8$), \texttt{MRT} ($p\ll0.0001$, $d>0.8$), and \texttt{TRA} ($p\ll0.0001$, $d>0.8$) without maskers. Moreover, water maskers were also perceived to be significantly louder than bird maskers, with a large effect, across each noise type (\texttt{AIR}: $p\ll0.01$, $d>0.8$; \texttt{MRT}: $p<0.001$, $d>0.8$; \texttt{TRA}: $p\ll0.001$, $d>0.8$).


\subsection{Prediction of perceived annoyance and loudness}

Pearson's correlation across all the objective indicators and perceptual attributes (i.e. \textit{PAY}, \textit{PLN},  \textit{ISOPL}, \textit{ISOEV}) were tabulated in \Cref{tab:corrObjSubj}. Owing to the generality of \textit{PAY}, \textit{ISOPL} and \textit{ISOEV} assessments, they are not directly relatable to the noise-type dependent \textit{PLN} assessment, and thus \textit{PLN} is assumed to be uncorrelated with \textit{PAY}, \textit{ISOPL} and \textit{ISOEV}. 

The \textit{PAY} was correlated ($r>0.28$, $p<0.05$) with $L_\text{C,eq}$; $L_\text{A,eq}$; $R_\text{max}$, $R_{5}$, $R_{10}$, $R_{50}$; $N_\text{max}$, $N_{5}$, $N_{10}$, and $N_{50}$; all $S$ indicators; \textit{PA}; and \textit{ISOPL}. Besides \textit{PAY}, \textit{ISOPL} is also correlated with $L_\text{C,eq}$; $N_{10}$; all $S$ indicators; and all $R$ indicators except $R_{90}$ and $R_{95}$. On the other hand, \textit{ISOEV} is correlated to all $L_\text{C,eq}$, $L_\text{A,eq}$, $N$, and $T$ indicators; $S_\text{max}$, $S_{5}$, and $S_{10}$; $\textit{FS}_{10}$, $\textit{FS}_{50}$, $\textit{FS}_{90}$, and $\textit{FS}_{95}$; \textit{PA}; as well as level derivatives $L_\text{C,eq}-L_\text{A,eq}$, $L_\text{A10}-L_\text{A90}$, and $L_\text{C10}-L_\text{C90}$.

The \textit{PLN} of the \texttt{AIR} noise stimuli was correlated to all $L_\text{C,eq}$ indicators; $L_\text{A,eq}$, $L_\text{A,max}$, $L_\text{A5}$, and $L_\text{A10}$; $N_\text{max}$, $N_{5}$, $N_{10}$, and $N_{50}$; $S_\text{max}$; $R$ and $R_{50}$; $\textit{FS}_{90}$; and \textit{PA}. The \texttt{MRT} noise stimuli \textit{PLN} was correlated to all the $L_\text{C,eq}$ and $L_\text{A,eq}$ indicators except their maximum; all the $N$ indicators; $S_\text{max}$; \textit{PA}; and \textit{ISOEV}. Lastly, the \textit{PLN} of the \texttt{TRA} noise stimuli was correlated to all all the $L_\text{C,eq}$ and $L_\text{A,eq}$ indicators except their maximum; all $N$ indicators; all $R$ indicators; $S_\text{max}$; as well as \textit{PA}.

Due to inherent differences in assessment of \textit{PAY} and \textit{PLN}, step-wise linear regression (SWLR) models were constructed independently to predict \textit{PAY} and \textit{PLN}. Only correlated predictors were included in the SWLR process. In the prediction of \textit{PAY}, SWLR was conducted with and without subjective parameters (i.e. \textit{ISOPL} and \textit{ISOEV}). Through sequential replacement up to 6 predictors with a 5-fold cross validation strategy, the final model
\begin{equation}
\begin{split}
    \textit{PAY}_\text{mixed}=2.510~L_\text{C}
    -26.456~\mathrm{\textit{ISOPL}}-115.077,
\end{split}
\end{equation}
with mixed objective and subjective indicators ($r^2=0.934$, RMSE=2.285) seems to better predict \textit{PAY} over the final model 
\begin{equation}
\begin{split}
    \textit{PAY}_\text{obj}=-2.650~L_\text{Cmax}+4.148~L_\text{A10} -46.670~S_\text{ave}-36.538,
\end{split}
\end{equation}
with only objective indicators ($r^2=0.848$ and RMSE=2.997).

As \textit{PLN} was noise-type dependent, SWLR models were computed for stimuli in each of the noise types. To mitigate rank deficiencies the \textit{PLN} models for \texttt{AIR} and \texttt{MRT} noise were determined based on sequential replacement up to 3 predictors via a 2-fold cross validation (8 stimuli per noise type) to yield
\begin{equation}
    \textit{PLN}_\texttt{AIR}=6.606~L_\text{C}-336.288,
\end{equation}
and
\begin{equation}
    \textit{PLN}_\texttt{MRT}=3.196~L_\text{C95}+4.951~L_\text{A50}-414.119,
\end{equation}
with or without subjective indicators ($r_\text{AIR}^2=0.552$ and $\text{RMSE}_\text{AIR}=10.379$; $r_\text{MRT}^2=0.9735$ and $\text{RMSE}_\text{MRT}=4.99$). The \textit{PLN} model for \texttt{TRA} noise was similarly determined via sequential replacement up to 3 predictors but with a 3-fold cross validation (9 stimuli) to yield
\begin{equation}
    \textit{PLN}_\texttt{TRA}=10.076~N_\text{95}-55.785,
\end{equation}
with or without subjective indicators ($r_\text{TRA}^2=0.990$ and $\text{RMSE}_\text{TRA}=7.377$)

\subsection{Sentiment analysis of descriptive comparison}

To examine the qualitative effect of the stimuli as compared to the reference stimuli, a sentiment analysis was employed. A positive sentiment is graded with a positive score, whereas a negative sentiment is given a negative score. To account for negators, amplifiers or valence shifters that modify the sentiment, e.g. not annoying, a lexicon based analysis was conducted at sentence level. The descriptive comparisons were analyzed at sentence level based on comparisons with a dictionary of polarized words to yield a sentiment score \citep{Rinker2021}. When more than one sentence was used to describe a stimuli, the sentiment score was taken as the mean across all the sentence sentiments. 

Following non-normality of the sentiment scores for each noise type with SWT ($p\ll0.001$), the main effects in sentiment polarity due to masker types and ANC condition was examined with the 2W-ART ANOVA. Significant differences were found only in masker types for \texttt{MRT} noise ($p<0.01$). No significant differences in sentiment was observed in all other cases. Post-hoc ART contrast tests with Bonferroni correction for multiple pairs revealed that the addition of the bird masker resulted in a significantly positive sentiment over the addition of the water masker ($p<0.001$, $d\ge0.8$) or with no masker ($p<0.05$, $d\ge0.8$), regardless of ANC condition, as shown in \Cref{tab:sentiARTsummary} and \Cref{tab:sentiARTstats} in \labelcref{sec:statsAppend}.

\section{Discussion}
\label{sec:discussion}

\chadded[comment=R1.3]{The following discussion seeks to address the research questions established in \Cref{sec:intro-RQ}. \Cref{sec:disc-effectofANC} examines the impact of active reduction of low-frequencies in typical urban noise types by the ANW on comfort perception (RQ1). \Cref{sec:disc-maskereffectonANC} analyzes the objective performance and comfort perception of the ANW when combined with representative biophilic maskers in mitigating typical urban noise (RQ3). \Cref{sec:disc-objsubj} assesses the reliability of estimating perceived acoustic comfort in the evaluation of novel acoustic solutions using only objective parameters (RQ3). Finally, \Cref{sec:disc-limit} addresses the limitations of this study and and provide avenues for future research.}

\subsection{Effect of active noise control}
\label{sec:disc-effectofANC}

Objectively, the effect of ANC was evident in the spectral-based analysis of sound pressure measurements using both the HATS and EA microphones. A substantial reduction of approximately \SI{10}{\decibelA} was observed within the effective operating bandwidth of ANC (\SI{300}{\hertz} to \SI{1}{\kilo\hertz}). Although single-value decibel-based indicators showed minimal reductions on the full audio bandwidth (\SI{20}{\hertz} to \SI{20}{\kilo\hertz}), which is within the just noticeable difference of equivalent SPL (JND\textsubscript{SPL}) range of 1--\SI{3}{\decibel} \citep{White2018}, there was notable decrease in psychoacoustic loudness based on the HATS measurements across all noise types. This objective reduction in psychoacoustic loudness was reflected in the significant reduction in the subjective perception of perceived loudness within each noise type, which is much greater than the JND\textsubscript{\textit{N}} of 0.5 to \SI{0.8}{\soneGF} \citep{Fusaro2022AssessmentPerception}, regardless of masker intervention. 

Despite some general increases in other psychoacoustic parameters, such as sharpness and tonality, the subjective perception of perceived annoyance was consistently reduced with ANC, regardless of noise type and masker intervention. It is important to note that the increase in $S$ exceeded the JND\textsubscript{\textit{S}} of \SI{0.04}{\acum} \citep{Fusaro2022AssessmentPerception,Pedrielli2008}. The assessment of PAY according to ISO/TS 15666 aligns with the assessment of derived \textit{ISOPL} based on ISO/TS 12913-3, showing a significant increase in ISOPL with ANC, regardless of noise type and masker intervention. The interaction effects with ANC were only observed in PAY, where the addition of bird or water maskers after ANC resulted in a significant reduction in \textit{PAY} compared to when either maskers were added without ANC, with ANC alone, or no interventions. Therefore, the integration of ANC with IM using biophilic maskers has proven effective in reducing PAY in domestic settings. Qualitatively assessing the effect of ANC against the \SI{65}{\decibel} reference without intervention within each noise type, the presence of ANC did not alter the sentiment. 

The reduction in PAY and increased \textit{ISOPL}, despite the increment in psychoacoustic sharpness, deviate from previous findings on construction machinery \citep{Hong2022c} and typical urban sound sources \citep{Orga2021MultilevelRecordings}, where higher $S$ was associated with greater annoyance. Additionally, has been an indicator of unpleasantness, particularly for aircraft noise \citep{Torija2019}. These findings provide further evidence of the context-dependency when evaluating the acoustic environment and highlight the prominent role of low-frequency sounds in the overall perception of annoyance \citep{Kang2017}.

\subsection{Effect of informational masking on ANC} 
\label{sec:disc-maskereffectonANC}

The addition of maskers at \SI{3}{\decibelA} below urban noise $L_\text{Aeq}$ levels would technically result in a barely perceptible increase based on the borderline objective increment in $L_\text{Aeq}$ levels (<\SI{2}{\decibelA}) similar to the JND\textsubscript{SPL}, in both EA and HATS measurements, with or without ANC. In some instances both with and without ANC, however, the associated decibel-based indicators related to the temporality of the sound exhibited increments above the JND\textsubscript{SPL}. For instance, the intermittent, high-frequency bird masker increased associated A-weighted parameters (i.e. $L_\text{Amax}$, $L_\text{A5}$) in the HATS measurements only in predominantly lower frequency urban noise (i.e., \texttt{MRT}, \texttt{TRA}), whereas the broadband, low-frequency water masker resulted in increased associated A- and C-weighted parameters (i.e. $L_\text{A90}$, $L_\text{A95}$, $L_\text{C90}$, $L_\text{C95}$) only in non-stationary urban noise types (i.e. \texttt{AIR}, \texttt{MRT}). 

In terms of psychoacoustic loudness (i.e. $N_5$), the addition of the bird masker with ANC perceptibly halved the attenuation effect of ANC (> JND\textsubscript{\textit{N}}) in non-stationary urban noise types (i.e. \texttt{AIR}, \texttt{MRT}) and resulted in perceptible increment in \texttt{TRA}. Subjectively, however, the increase in PLN was only found to be significant when the bird masker was added to \texttt{MRT}. In terms of the water masker, its introduction perceptibly diminished the attenuation effect of ANC on $N_5$ to a third in \texttt{AIR}, resulted in perceptible increment in \texttt{TRA}, but did not affect the $N_5$ in \texttt{MRT}. Nevertheless, the subjective PLN was significantly increased regardless of noise type and ANC when water masker was introduced. The discrepancies between the psychoacoustic loudness, decibel-based metrics, and subjective PLN suggests that complex urban acoustic environments cannot be represented solely by energy-based acoustic indicators. 

The addition of either masker further increased the psychoacoustic sharpness after ANC beyond the JND\textsubscript{\textit{S}}, except when the bird masker was added to \texttt{MRT} and \texttt{TRA}. The significant reduction in PAY as described in \Cref{sec:disc-effectofANC}, despite further increment in $S$ upon the addition of either masker, is in further contrary to the previously reported positive correlation between $S$ and PAY. 

Owing to the tonal nature of the bird masker, $T_5$ was further increased significantly when bird maskers were added after ANC, whereas the addition of the broadband water masker after ANC substantially reduced the $T_5$. At least for \texttt{AIR}, the addition of the bird masker significantly decreased the \textit{PAY} and did not affect the \textit{ISOPL}, deviating from the literature wherein increased tonality was associated with increased annoyance \citep{Torija2019}.

\subsubsection{Subjective effects on aircraft noise}

Subjectively, addition of either masker to \texttt{AIR} after ANC significantly decreased the \textit{PAY}, increased \textit{ISOEV}, but did not impact sentiment scores. Water maskers also significantly increased the \textit{ISOPL} and \textit{PLN}, whereas bird maskers did not affect \textit{ISOPL} and \textit{PLN}.

Even though PAY reduction occurred to a large effect regardless of masker type when combined with ANC, its congruence with \textit{ISOPL} increment appears to provide stronger evidence for preference of water over bird sounds for masking \texttt{AIR} after ANC. However, it is important to note that introduction of water sounds resulted in confusion and increased disturbance in about \SI{8}{\percent} of the qualitative responses as compared to about \SI{3}{\percent} for bird, when added to \texttt{AIR}. This aligns with previous findings that appropriateness was dependent on the visibility of the sound source for water maskers in outdoor urban environments \citep{Lugten2018c,Hong2020b,Hong2020c,Li2020}, and thus water maskers should be cautiously employed over bird maskers. Moreover, the effect of masking was previously found to be dependent on the SPL of the target urban noise \citep{Hong2020h,Hong2021b}, and thus limits the applicability of the findings to \texttt{AIR} at \SI{65}{\decibelA}. The dearth of research on the perception of maskers in indoor environments exposed to \texttt{AIR} also warrants further investigation. All in all, With the evidence presented, it becomes apparent that adding bird maskers after ANC of \texttt{AIR} to further reduce the \textit{PAY} is more appropriate when an increase in \textit{PLN} is not desired. 

\subsubsection{Subjective effects on MRT noise}


When either maskers were added to \texttt{MRT} after ANC, there was a significant decrease in \textit{PAY} and increase in \textit{ISOEV}, but no change in \textit{ISOPL}. In contrast to \texttt{AIR}, both maskers increased the \textit{PLN} with \texttt{MRT}. Bird maskers also resulted in significantly positive shift in sentiments, whereas water maskers had no impact. 

Despite their ability to further reduce the \textit{PAY} after ANC of \texttt{MRT}, none of the maskers could reduce or at the least not impact the \textit{PLN}. The positive shift in sentiments as a result of the bird masker, however, merits further investigation, especially due to the lack of literature on \texttt{MRT} noise exposure indoors at wider range of SPL levels. 

\subsubsection{Subjective effects on traffic noise}


Either masker seems to the most effective when added to \texttt{TRA} after ANC, whereby \textit{PAY} was significantly decreased, both \textit{ISOPL} and \textit{ISOEV} were significantly increased, but both had no effect on the sentiment. Similar to all the other noise types, the water masker significantly increased the \textit{PLN}, whereas the bird masker did not affect the \textit{PLN}.   

Although the increase in \textit{ISOPL} corroborates with previous work on the augmentation of traffic noise with bird or water maskers \citep{Hong2020h,Hao2016}, the resultant increase in \textit{PLN} upon the addition of water maskers to \texttt{TRA} deviate from previous findings \citep{Coensel2011,Hong2020c}, albeit for outdoor urban acoustic environments. Likewise with \texttt{AIR}, only bird maskers appear to further reduce the \textit{PAY} without increasing the \textit{PLN} of \texttt{MRT} after ANC. 

\subsection{Role of objective and subjective metrics in evaluating perceived acoustic comfort} \label{sec:disc-objsubj}

Since perceived annoyance carries associated health risks \citep{WorldHealthOrganizationRegionalOfficeforEurope2018}, and perceived loudness influences PAY \citep{RadstenEkman2015a}, both PAY and PLN could be used to evaluate the overall indoor acoustic comfort. Thus, it is worthwhile to examine the extent to which PAY and PLN can be solely predicted by objective indicators. 

Overall, PAY appears to best predicted by a linear combination of a single objective ($L_\text{C}$) and a subjective (\textit{ISOPL}) parameter with a \SI{93.4}{\percent} fit, which is \SI{8.6}{\percent} better than the SWLR model with only objective parameters ($L_\text{Cmax}$, $L_\text{A10}$, $S_\text{ave}$). The inclusion of C-weighted parameters in both linear regression models imply that PAY is notably influenced by low-frequency perception, further exposing the limitations of A-weighted indicators. 

On the other hand, the best perceived loudness SWLR prediction models for each noise type did not include any subjective parameters. Notably, the goodness of fit for $\textit{PLN}_\text{\texttt{MRT}}$ and $\textit{PLN}_\text{\texttt{TRA}}$ prediction models were \SI{97.4}{\percent} and \SI{99.0}{\percent}, respectively, whereas the $\textit{PLN}_\text{\texttt{AIR}}$ was only \SI{55.2}{\percent}, which could be attributed to the difficulty in judging the \textit{PLN} of the non-stationary aircraft noise. Interestingly, $\textit{PLN}_\text{\texttt{MRT}}$ could be predicted by a combination of $L_\text{A50}$ and $L_\text{C95}$, but $\textit{PLN}_\text{\texttt{TRA}}$ could be fully predicted with $N_\text{95}$. 

Based on the PAY and PLN prediction models, it appears that almost all the psychoacoustic indicators were not useful for measuring the overall PAY, and the \textit{PLN} of \texttt{MRT} and \texttt{TRA}. It is worth noting that the \textit{ISOPL} derived from the \textit{PAQ} in ISO/TS 12913-3 still proves to be useful in determining \textit{PAY} in indoor soundscapes, in spite of recent concern regarding the validity of the \textit{PAQ} circumplex model for indoor soundscapes \citep{Torresin2020IndoorBuildings,Torresin2019a}. Recent developments in the prediction of \textit{ISOPL} with large subjective datasets could also pave the way for reliable prediction of \textit{PAQ} without requiring subjective input \citep{Ooi2023b,Ooi2023a,Watcharasupat2022,Ooi2022}.

\subsection{Limitations and future work} \label{sec:disc-limit}

\chadded[comment=R2.8]{This study acknowledges several limitations that warrant consideration, along with directions for future research. Firstly, it is important to note that the majority of participants were university students between the ages of 21 and 30 due to the remote location of the university campus. As a result, the generalizability of the findings is limited to the younger demographic, and follow-up studies should employ stratified sampling to ensure broader representation.} \chadded[comment=R2.3]{Furthermore, while temperature was effectively controlled to minimize its potential confounding effects, further investigation is needed to explore potential interaction effects that may arise under warmer temperature conditions \citep{Nitidara2022, Yin2022}. Recent evidence has underscored the significance of the acoustic environment in warmer conditions \citep{Guan2020}, thus emphasizing the potential of the ANC-IM approach to enhance overall indoor comfort.} \chadded[comment=R2.6]{Additionally, despite efforts to ensure accurate reproduction of the audio stimuli, real-world acoustic conditions could be substantially influenced by factors such as building orientation and elevation \citep{Sheikh2014a,Sheikh2014}.} 

In light of the perceptual differences observed with the ANW and its proposed integration with biophilic maskers, it is essential to acknowledge several limitations. Firstly, the PAY and PLN prediction models are valid for comparison to urban noise experienced at \SI{65}{\decibelA}, with the exception of \texttt{TRA}, which was also evaluated at \SI{60}{\decibelA}, i.e. $\texttt{TRA}_\text{65dB(A)}$. Secondly, the level at which the biophilic maskers were reproduced (i.e. $-\SI{3}{\decibelA}$) may not be optimal for indoor soundscapes due to the shift in context \citep{Torresin2020IndoorBuildings}, which together with the choice of masker is an important topic for future exploration. Thirdly, the spatial orientation of the maskers may also not be optimal although it is logical for biophilic sounds to impinge through the window \citep{Torresin2022AssociationsLockdown,Hong2020c}, which should be verified in in the future. 

\section{Conclusion}

Active control and interference are noise mitigation strategies that preserve natural ventilation and affords greater user control and flexibility over passive control methods. However, their perceptual effect on indoor acoustic comfort and potential for integration are not commonly investigated. Moreover, it is also worthwhile to determine the extent to which objective parameters could translate to perception such as perceived annoyance (PAY) and perceived loudness (PLN). This study evaluated an active-noise-control (ANC)-based ``anti-noise'' window (ANW) integrated with informational masking (IM) to enhance indoor acoustic comfort in a model bedroom. 

On the first research question to determine the effect of ANC alone, despite minimal objective reduction in decibel-based indicators, the ANW significantly reduced PAY and PLN across all noise types, while increasing the derived pleasantness (ISO 12913-3). Moreover, the reduction of low-frequency content in urban noise through the ANW, and resultant increase acoustic comfort but also psychoacoustic sharpness, underscores the importance of low-frequency noise perception which is lost in traditional ``A-weighting'' parameters and exposes the limitations of objective psychoacoustic indicators in accessing complex urban noises. 

Interaction discovered between the masker and the ANC effect of the ANW suggests that the proposed integration further increases the overall indoor acoustic comfort than with just the ANW or biophilic masking alone. Hence, there is great potential to further develop the ANC-IM approach and its associated human-centered evaluation for sustainable urban sound management in the built environment.

\section*{Declaration of competing interest}

The authors declare that they have no known competing financial interests or personal relationships that could have appeared to influence the work reported in this paper.

\section*{Acknowledgements}

This research is supported by the Singapore Ministry of National Development and the National Research Foundation, Prime Minister's Office under the Cities of Tomorrow Research Programme (Award No. COT-V4-2019-1 and COT-V4-2020-1). Any opinions, findings and conclusions or recommendations expressed in this material are those of the authors and do not reflect the view of National Research Foundation, Singapore, and Ministry of National Development, Singapore.

\section*{Data Availability}

The data that support the findings of this study are openly available in NTU research data repository DR-NTU (Data) at \url{https://doi.org/10.21979/N9/SEGEFM}. The replication code used in this study is available on GitHub at the following repository: \url{https://github.com/ntudsp/SPANR}. The code includes all the necessary scripts, functions, and instructions to reproduce the results reported in the study.

\printcredits

\setcounter{section}{0}
\renewcommand{\thesection}{Appendix \Alph{section}}

\setcounter{table}{0}
\renewcommand{\thetable}{\Alph{section}.\arabic{table}}

\setcounter{figure}{0}
\renewcommand{\thefigure}{\Alph{section}.\arabic{figure}}
\onecolumn

\section{Questionnaires}
\label{sec:questions}
\begin{sffamily}
\small%
\refstepcounter{table} \label{tab:question}
\noindent\textbf{\color{scolor}Table \thetable}\par%
\noindent{Pre-test assessment participant information questionnaire}%
\small%
\setlength\LTleft{0pt}
\setlength\LTright{0pt}

\begin{longtable}{@{\extracolsep{\fill}}
>{\raggedright\arraybackslash}p{0.075\textwidth}l
>{\raggedright\arraybackslash}p{0.35\textwidth}l
>{\raggedright\arraybackslash}p{0.4\textwidth}l
>{\raggedright\arraybackslash}p{0.075\textwidth}r@{}}

    \toprule
    Question Category
    & Instructions 
    & Specific Questions
    & Rating Scale\\
    \midrule
\endhead

    \multicolumn{4}{r}{[Continued on next page]} 
\endfoot

    \bottomrule
    
\endlastfoot

    \multirow[t]{21}{0.075\textwidth}{Individual Noise Sensitivity (INS)}
    & \multirow[t]{21}{0.35\textwidth}{To what extent you disagree/agree with the following sentences?} 
    & `I wouldn’t mind living on a noisy street if the apartment I had was nice.'
    & \multirow[t]{9}{0.075\textwidth}{Strongly Disagree--Strongly Agree (5-point categorical)}\\ \cline{3-3}

    & & `No one should mind much if someone turns up his or her stereo full blast once in awhile.' 
    & \\ \cline{3-3}

    & & `I get used to most noises without much difficulty.’ 
    & \\ \cline{3-3}

    & & `It would matter to me if an apartment I was interested in renting were located across from a fire station.’ 
    & \\ \cline{3-3}

    & & `It wouldn’t bother me to hear the sounds of everyday living from neighbors (footsteps, running water, etc.).’  
    & \\ \cline{3-3}

    & & `I’m good at concentrating no matter what is going on around me.' 
    & \\ \cline{3-3}

    & & `In a library, I don’t mind if people carry on a conversation if they do it quietly.’ 
    & \\ \cline{3-3}

    & & `I get mad at people who make noise that keeps me from falling asleep or getting work done.’ 
    & \\ \cline{3-3}

    & & `I wouldn’t mind living in an apartment with thin walls.’
    & \\ \cline{3-4}

    & & `I am more aware of noise than I used to be.’
    & \multirow[t]{12}{0.075\textwidth}{Strongly Agree--Strongly Disagree (5-point categorical)} \\ \cline{3-3}

    & & `At movies, whispering and crinkling candy wrappers disturb me.’
    & \\ \cline{3-3}

    & & `I am easily awakened by noise.’
    & \\ \cline{3-3}

    & & `If it’s noisy where I’m studying, I try to close the door or window or move someplace else.’
    & \\ \cline{3-3}

    & & `I get annoyed when my neighbors are noisy.’
    & \\ \cline{3-3}

    & & `Sometimes noises get on my nerves and get me irritated.’
    & \\ \cline{3-3}

    & & `Even music I normally like will bother me if I’m trying to concentrate.’
    & \\ \cline{3-3}

    & & `When I want to be alone, it disturbs me to hear outside noises.’
    & \\ \cline{3-3}

    & & `There are often times when I want complete silence.’
    & \\ \cline{3-3}

    & & `Motorcycles ought to be required to have bigger mufflers.’
    & \\ \cline{3-3}

    & & `I find it hard to relax in a place that’s noisy.’
    & \\ \cline{3-3}

    & & `I am sensitive to noise.’
    & \\ 

\midrule
    \multirow[t]{10}{0.08\textwidth}{Baseline Noise Annoyance (BNA)}
    & \multirow[t]{10}{0.35\textwidth}{Thinking about the last (12 months or so), when you are here at home, how much does noise from (noise source) bother, disturb or annoy you?} 
    & Road traffic 
    & \multirow[t]{5}{0.08\textwidth}{Not at all--Extremely (5-point categorical)} \\ \cline{3-3}

    & 
    & Aircraft  
    & \\ \cline{3-3}

    & 
    & MRT 
    & \\ \cline{3-3}

    & 
    & Construction Work (Work Site)
    & \\ \cline{3-3}

    & 
    & Construction Work (Renovations)
    & \\ \cline{3-3}

    & 
    & Any other noises (specify)
    & \\ \cline{3-4}
    
    &  
    & Road traffic 
    & \multirow[t]{5}{0.08\textwidth}{Not at all--Extremely (0-to-100 numerical scale)} \\ \cline{3-3}

    & 
    & Aircraft  
    & \\ \cline{3-3}

    & 
    & MRT 
    & \\ \cline{3-3}

    & 
    & Construction Work (Work Site)
    & \\ \cline{3-3}

    & 
    & Construction Work (Renovations)
    & \\ \cline{3-3}

    & 
    & Any other noises (specify)
    & \\

\midrule
    \multirow[t]{10}{0.08\textwidth}{10-item Perceived Stress Scale (PSS-10)}
    & \multirow[t]{10}{0.35\textwidth}{The questions in this scale ask you about your feelings and thoughts during the last month. In each case, you will be asked to indicate how often you felt or thought a certain way. Although some of the questions are similar, there are differences between them and you should treat each one as a separate question. The best approach is to answer each question fairly quickly. That is, don't try to count up the number of times you felt a particular way, but rather indicate the alternative that seems like a reasonable estimate. In the last month, how often have you... *} 
    & been upset because of something that happened unexpectedly? 
    & \multirow[t]{10}{0.075\textwidth}{Never--Very Often (5-point categorical)} \\ \cline{3-3}

    & & felt that you were unable to control the important things in your life
    & \\ \cline{3-3}

    & & felt nervous and "stressed"?
    & \\ \cline{3-3}

    & & felt confident about your ability to handle your personal problems?
    & \\ \cline{3-3}

    & & felt that things were going your way?
    & \\ \cline{3-3}

    & & found that you could not cope with all the things that you had to do?
    & \\ \cline{3-3}

    & & been able to control irritations in your life?
    & \\ \cline{3-3}

    & & felt that you were on top of things?
    & \\ \cline{3-3}

    & & been angered because of things that were outside of your control?
    & \\ \cline{3-3}

    & & felt difficulties were piling up so high that you could not overcome them?
    & \\ 

\midrule
    \multirow[t]{5}{0.08\textwidth}{WHO-Five Well-being Index (WHO-5)}
    & \multirow[t]{5}{0.35\textwidth}{For each of the statements below, which is the closest to how you have been feeling over the last two weeks?} 
    & I have felt cheerful and in good spirits. 
    & \multirow[t]{5}{0.075\textwidth}{At no time--All of the time (6-point categorical)} \\ \cline{3-3}

    & & I have felt calm and relaxed
    & \\ \cline{3-3}

    & & I have felt active and vigorous
    & \\ \cline{3-3}

    & & I woke up feeling fresh and rested.
    & \\ \cline{3-3}

    & & My daily life has been filled with things that interest me.
    & \\ 

\end{longtable}
\end{sffamily}
\hspace{2em}
\begin{sffamily}
\setcounter{table}{1}
\refstepcounter{table}\label{tab:stimuliquestion}

\small
\clearpage
\noindent\textbf{\color{scolor}Table \thetable}\par%

\noindent{Questionnaire to assess each stimuli. Each stimuli could be repeated as many times as required to answer the questions.}

\noindent\begin{tabularx}{\textwidth}{%
    @{}
    >{\setlength\hsize{0.1\hsize}}X 
    >{\setlength\hsize{0.55\hsize}}X
    >{\setlength\hsize{0.1\hsize}}X
    >{\setlength\hsize{0.25\hsize}}X
    @{}
}

\toprule
    Question Category
    & Instructions
    & Specific Questions
    & Rating Scale\\

\midrule
    \multirow[t]{2}{=}{Perceived Annoyance (PAY)}
    & \multicolumn{2}{>{\RaggedRight\arraybackslash}p{0.65\textwidth}}{Thinking about the noise you just heard, how much does the noise bother, disturb or annoy you?}
    & Not at all--Extremely (5-point categorical)\\ \cline{2-4}

    & \multicolumn{2}{>{\RaggedRight\arraybackslash}p{0.65\textwidth}}{This uses a 0-to-100 opinion scale for how much the noise you heard bothers, disturbs or annoys. If you are not at all annoyed choose 0; if you are extremely annoyed choose 10; if you are somewhere in between, choose a number between 0 and 100. Thinking about the noise you just heard what number from 0 to 100 best shows how much you are bothered, disturbed or annoyed by the noise?}
    & Not at all--Extremely (0-to-100 numerical scale)\\

\midrule
    \multirow[t]{8}{=}{Perceived Affective Quality (PAQ)}
    & \multirow[t]{8}{=}{For each of the 8 scales below, to what extent do you agree or disagree that the surrounding sound environment you heard is $\cdots$} 
    & Eventful
    & \multirow[t]{8}{=}{Strongly Disagree--Strongly Agree (0-to-100 numerical scale)}\\ \cline{3-3}

    & & Vibrant & \\ \cline{3-3}

    & & Pleasant & \\ \cline{3-3}

    & & Calm & \\ \cline{3-3}

    & & Uneventful & \\ \cline{3-3}

    & & Monotonous & \\ \cline{3-3}

    & & Annoying & \\ \cline{3-3}

    & & Chaotic & \\
\midrule
    Perceived Loudness (PLN)
    & \multicolumn{2}{>{\RaggedRight\arraybackslash}p{0.65\textwidth}}{This task compares the test track you just heard to a reference track. If the perceived loudness of the reference track is assigned a value of 100, rate the perceived loudness of the test track in comparison with the reference track. For example, if the test track is twice as loud, it should be rated as 200. And if it is half as loud, it should be rated as 50.}
    & Unbounded numerical rating \\

\midrule
    Open-ended Description
    & \multicolumn{2}{>{\RaggedRight\arraybackslash}p{0.65\textwidth}}{Describe the test track in comparison to the reference track in a few descriptive words.}
    & Text input \\

\bottomrule
\end{tabularx}
\end{sffamily}

\clearpage
\setcounter{table}{1}
\renewcommand{\thetable}{\Alph{section}.\arabic{table}}
\section{Bland-Altman statistics between 5-point categorical and 101-point numerical scale on perceived annoyance}
\label{sec:BAstats}

\begin{figure*}[hb]
    \centering
    \subfigure[]{\includegraphics[width=1\linewidth]{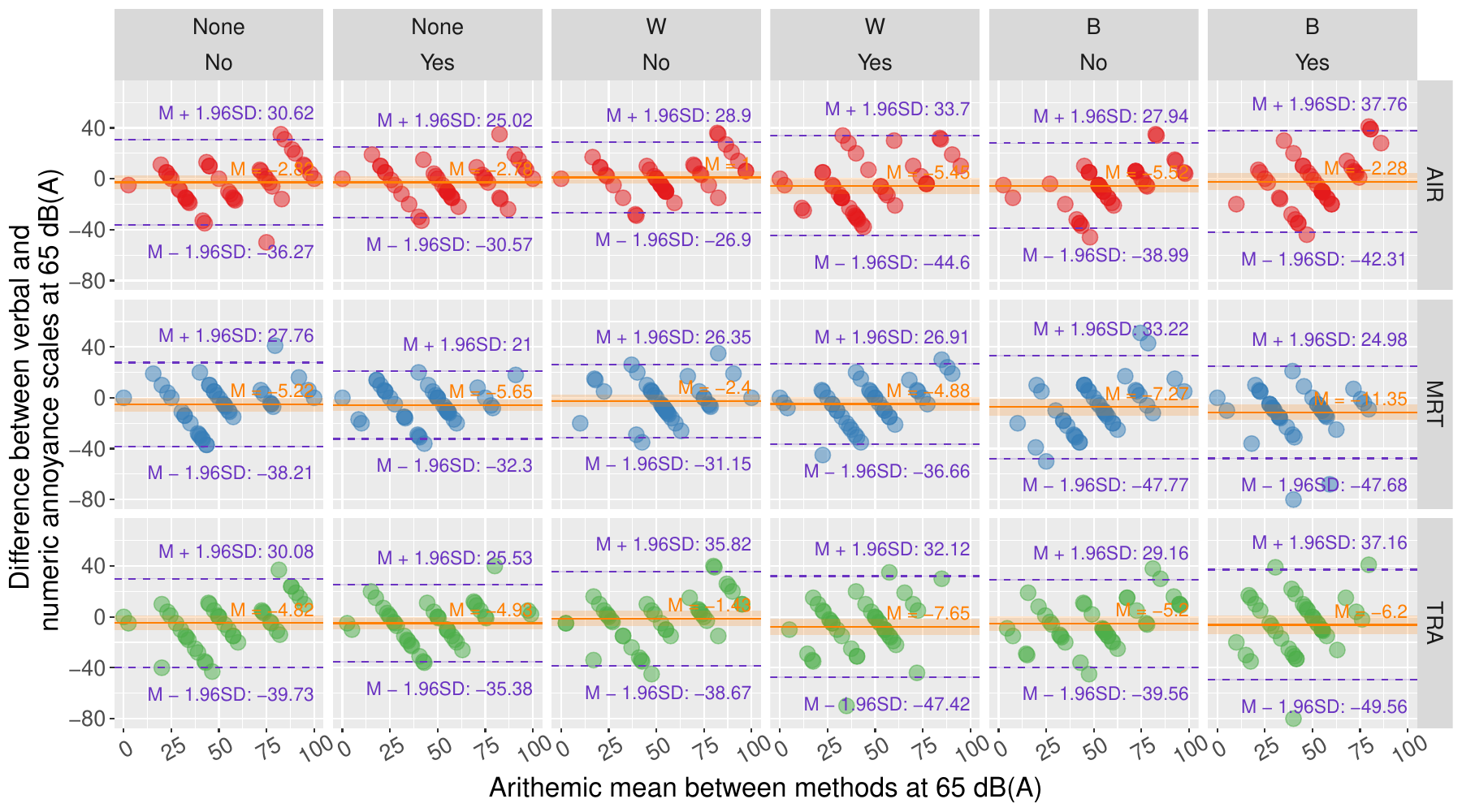}}
    \subfigure[]{\includegraphics[width=1\linewidth]{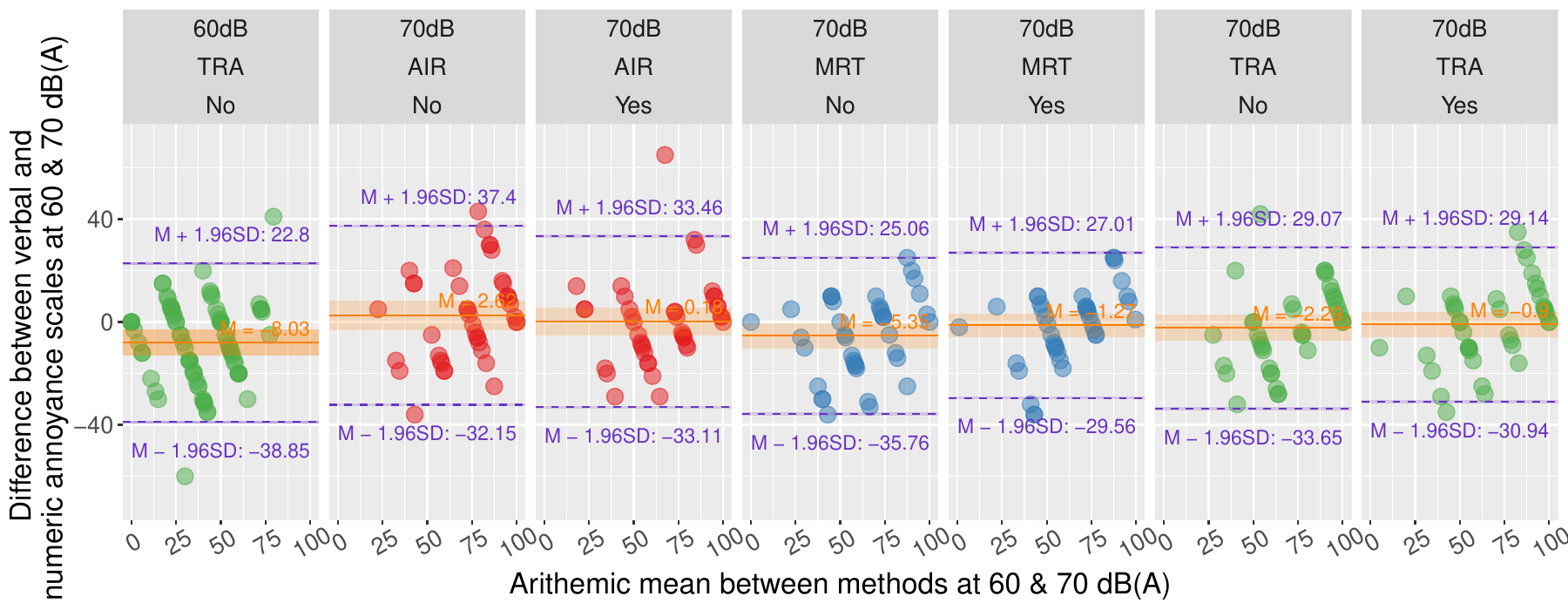}}
    \caption{Bland-Altman plot of agreement between annoyance scales (a) across masker types and ANC conditions (columns) and noise types (rows) at \SI{65}{\decibelA}, and (b) across noise types and ANC conditions at \SI{60}{\decibelA} and \SI{70}{\decibelA}. The average bias is estimated by the mean of the differences and bounded by the \SI{95}{\%} limits of agreement (LoA). Confidence intervals are based on \SI{95}{\%} confidence level for both LoAs (\textcolor{loaclr}{$\blacksquare$}) and the mean (\textcolor{mclr}{$\blacksquare$}).}
    \label{fig:BA65dBA}
\end{figure*}

\clearpage

\setcounter{table}{1}
\renewcommand{\thetable}{\Alph{section}.\arabic{table}}
\onecolumn

\section{Statistical results}
\label{sec:statsAppend}

\begin{sffamily}
\setcounter{table}{0}
\refstepcounter{table}\label{tab:qnDataSummary}

\noindent\textbf{\color{scolor}Table \thetable}\par%
\noindent{Summary of questionnaire scores on perceived annoyance (PAY), perceived affective quality (PAQ), and perceived loudness (PLN)}

\tiny

\centering
\begin{tabularx}{\linewidth}{ 
*{2}{>{\raggedright\arraybackslash}p{0.5cm}}
*{2}{>{\raggedright\arraybackslash}X}
*{20}{>{\raggedleft\arraybackslash}X}}

\toprule
\multicolumn{4}{c}{\textbf{ }} 
& \multicolumn{2}{c}{\textbf{PAY}} 
& \multicolumn{16}{c}{\textbf{PAQ}} 
& \multicolumn{2}{c}{\textbf{PLN}} \\
\cmidrule(l{1pt}r{1pt}){5-6} 
\cmidrule(l{1pt}r{1pt}){7-22} 
\cmidrule(l{1pt}r{1pt}){23-24}

\multicolumn{6}{c}{\textbf{ }} 
& \multicolumn{2}{c}{\textbf{\textit{e}}} 
& \multicolumn{2}{c}{\textbf{\textit{v}}} 
& \multicolumn{2}{c}{\textbf{\textit{p}}} 
& \multicolumn{2}{c}{\textbf{\textit{ca}}} 
& \multicolumn{2}{c}{\textbf{\textit{u}}} 
& \multicolumn{2}{c}{\textbf{\textit{m}}} 
& \multicolumn{2}{c}{\textbf{\textit{a}}} 
& \multicolumn{2}{c}{\textbf{\textit{ch}}} 
& \multicolumn{2}{c}{\textbf{ }} \\
\cmidrule(l{1pt}r{1pt}){7-8}
\cmidrule(l{1pt}r{1pt}){9-10}
\cmidrule(l{1pt}r{1pt}){11-12}
\cmidrule(l{1pt}r{1pt}){13-14}
\cmidrule(l{1pt}r{1pt}){15-16}
\cmidrule(l{1pt}r{1pt}){17-18}
\cmidrule(l{1pt}r{1pt}){19-20}
\cmidrule(l{1pt}r{1pt}){21-22}

\multirow[b]{-3}{=}{\raggedright\arraybackslash Noise Type} 
& \multirow[b]{-3}{=}{\raggedright\arraybackslash Masker} 
& \multirow[b]{-3}{=}{\raggedright\arraybackslash SPL, dB(A)} 
& \multirow[b]{-3}{=}{\raggedright\arraybackslash ANC}  
& $\mu$
& $\sigma$
& $\mu$ 
& $\sigma$ 
& $\mu$ 
& $\sigma$ 
& $\mu$ 
& $\sigma$ 
& $\mu$
& $\sigma$ 
& $\mu$ 
& $\sigma$ 
& $\mu$ 
& $\sigma$ 
& $\mu$ 
& $\sigma$ 
& $\mu$ 
& $\sigma$
& $\mu$ 
& $\sigma$\\

\midrule
 &  &  & No & 57.00 & 25.40 & 46.77 & 22.82 & 39.05 & 23.09 & 29.26 & 19.67 & 27.36 & 23.11 & 41.03 & 24.39 & 49.49 & 22.88 & 60.18 & 23.67 & 50.82 & 24.89 & 98.69 & 16.07\\
\cmidrule{4-24}
 &  & \multirow[t]{-2}{=}[1.8em]{\vfil \raggedright\arraybackslash 65} 
 & Yes & 56.95 & 26.44 & 46.92 & 21.91 & 36.31 & 23.99 & 33.51 & 20.72 & 28.36 & 21.14 & 47.31 & 21.94 & 52.97 & 22.05 & 55.46 & 26.64 & 47.03 & 25.38 & 93.31 & 28.33\\
\cmidrule{3-24}
 &  &  & No & 70.85 & 20.38 & 54.23 & 24.33 & 38.36 & 26.52 & 21.92 & 14.95 & 23.59 & 20.22 & 39.41 & 22.88 & 44.15 & 26.83 & 71.23 & 20.22 & 60.77 & 24.25 & 127.85 & 36.09\\
\cmidrule{4-24}
 & \multirow[t]{-4}{=}[3.2em]{\vfil\raggedright\arraybackslash None} 
 & \multirow[t]{-2}{=}[1.8em]{\vfil\raggedright\arraybackslash 70} 
 & Yes & 65.64 & 23.15 & 54.44 & 25.94 & 40.08 & 27.74 & 25.64 & 19.07 & 25.72 & 21.53 & 38.00 & 25.44 & 46.23 & 24.71 & 66.90 & 25.07 & 54.64 & 26.83 & 123.15 & 36.16\\
\cmidrule{2-24}
 &  &  & No & 61.05 & 18.73 & 56.92 & 19.26 & 52.33 & 23.33 & 37.49 & 20.05 & 33.56 & 20.58 & 38.79 & 22.00 & 37.51 & 20.47 & 55.69 & 20.25 & 55.85 & 18.45 & 100.38 & 28.41\\
\cmidrule{4-24}
 & \multirow[t]{-2}{=}[1.8em]{\vfil \raggedright\arraybackslash Bird} 
 & \multirow[t]{-2}{=}[1.8em]{\vfil \raggedright\arraybackslash 65}  
 & Yes & 52.74 & 16.84 & 54.08 & 17.24 & 51.23 & 22.96 & 37.13 & 17.23 & 33.00 & 20.01 & 38.87 & 21.88 & 35.79 & 20.80 & 53.21 & 18.50 & 51.67 & 19.13 & 91.00 & 26.24\\
\cmidrule{2-24}
 &  &  & No & 54.62 & 22.78 & 58.15 & 20.20 & 54.82 & 22.11 & 42.21 & 22.41 & 38.95 & 20.68 & 37.92 & 22.43 & 37.38 & 22.51 & 55.49 & 22.95 & 51.59 & 24.12 & 121.59 & 44.16\\
\cmidrule{4-24}
\multirow[t]{-8}{=}[6.4em]{\vfil \raggedright\arraybackslash Aircraft} 
& \multirow[t]{-2}{=}[1.8em]{\vfil \raggedright\arraybackslash Water} 
& \multirow[t]{-2}{=}[1.8em]{\vfil \raggedright\arraybackslash 65} 
& Yes 
& 48.08 & 22.19 & 56.08 & 17.73 & 53.23 & 22.32 & 48.33 & 23.11 & 46.69 & 24.81 & 34.77 & 20.91 & 34.36 & 20.29 & 46.54 & 23.68 & 43.56 & 21.21 & 104.69 & 40.64\\
\cmidrule{1-24}
 &  & 60 & No & 41.95 & 21.68 & 32.26 & 19.34 & 29.46 & 22.35 & 33.58 & 22.95 & 40.94 & 23.65 & 59.67 & 24.49 & 70.94 & 21.23 & 44.81 & 23.98 & 39.05 & 25.39 & 64.68 & 22.34\\
\cmidrule{3-24}
 &  &  & No & 54.62 & 23.06 & 36.08 & 24.38 & 30.23 & 20.87 & 25.97 & 17.79 & 34.08 & 23.85 & 61.05 & 27.39 & 69.79 & 21.24 & 55.79 & 24.84 & 46.51 & 25.19 & 99.92 & 15.37\\
\cmidrule{4-24}
 &  & \multirow[t]{-2}{=}[1.8em]{\vfil \raggedright\arraybackslash 65} 
 & Yes & 49.03 & 23.33 & 33.49 & 22.29 & 29.44 & 21.04 & 27.41 & 16.42 & 33.87 & 20.51 & 63.10 & 26.17 & 70.85 & 20.21 & 50.51 & 24.78 & 39.85 & 24.84 & 79.67 & 21.35\\
\cmidrule{3-24}
 &  &  & No & 71.64 & 19.64 & 38.49 & 27.92 & 28.46 & 21.09 & 19.31 & 16.05 & 22.00 & 18.43 & 55.05 & 29.10 & 72.03 & 23.94 & 71.77 & 21.07 & 55.54 & 26.05 & 152.79 & 34.35\\
\cmidrule{4-24}
 & \multirow[t]{-5}{=}[3.8em]{\vfil \raggedright\arraybackslash None} 
 & \multirow[t]{-2}{=}[1.8em]{\vfil \raggedright\arraybackslash 70} 
 & Yes & 67.85 & 22.93 & 37.26 & 27.93 & 29.74 & 23.17 & 21.08 & 18.01 & 27.54 & 23.03 & 59.79 & 28.35 & 69.74 & 24.45 & 66.10 & 22.54 & 49.49 & 28.55 & 138.41 & 35.12\\
\cmidrule{2-24}
 &  &  & No & 53.74 & 23.04 & 47.08 & 22.90 & 44.28 & 20.98 & 38.26 & 20.13 & 36.15 & 20.72 & 48.51 & 22.68 & 48.38 & 24.86 & 51.95 & 22.19 & 42.51 & 25.89 & 96.51 & 24.16\\
\cmidrule{4-24}
 & \multirow[t]{-2}{=}[1.8em]{\vfil \raggedright\arraybackslash Bird} 
 & \multirow[t]{-2}{=}[1.8em]{\vfil \raggedright\arraybackslash 65} 
 & Yes & 45.87 & 19.00 & 45.49 & 20.63 & 51.82 & 21.98 & 48.59 & 20.54 & 46.36 & 22.96 & 48.15 & 17.52 & 43.28 & 20.61 & 41.05 & 18.35 & 35.85 & 20.40 & 81.36 & 18.71\\
\cmidrule{2-24}
 &  &  & No & 51.82 & 24.41 & 44.46 & 20.25 & 45.67 & 22.57 & 42.31 & 23.86 & 41.36 & 23.30 & 53.49 & 23.50 & 55.59 & 22.28 & 53.08 & 24.08 & 43.82 & 23.64 & 131.74 & 38.64\\
\cmidrule{4-24}
\multirow[t]{-9}{=}[7em]{\vfil\raggedright\arraybackslash Traffic} 
& \multirow[t]{-2}{=}[1.8em]{\vfil\raggedright\arraybackslash Water} 
& \multirow[t]{-2}{=}[1.8em]{\vfil \raggedright\arraybackslash 65}
& Yes & 46.95 & 19.71 & 45.72 & 20.74 & 51.92 & 22.80 & 53.41 & 20.60 & 55.79 & 20.63 & 50.56 & 23.49 & 47.79 & 23.85 & 39.69 & 20.47 & 34.97 & 22.45 & 107.67 & 38.92\\
\cmidrule{1-24}
 &  &  & No & 53.21 & 22.35 & 56.79 & 21.64 & 47.51 & 22.19 & 36.95 & 19.97 & 32.62 & 21.36 & 42.56 & 23.61 & 42.87 & 22.64 & 54.56 & 22.39 & 51.10 & 22.48 & 101.49 & 20.91\\
\cmidrule{4-24}
 &  & \multirow[t]{-2}{=}[1.8em]{\vfil \raggedright\arraybackslash 65} 
 & Yes & 45.90 & 21.91 & 53.85 & 20.26 & 46.26 & 19.84 & 37.41 & 17.77 & 38.26 & 18.39 & 43.59 & 25.61 & 43.97 & 22.87 & 50.77 & 21.18 & 47.26 & 21.22 & 89.18 & 23.88\\
\cmidrule{3-24}
 &  &  & No & 63.82 & 21.92 & 60.69 & 19.51 & 46.05 & 22.68 & 29.41 & 16.58 & 25.21 & 16.09 & 37.41 & 22.30 & 42.31 & 23.43 & 63.82 & 22.25 & 60.72 & 20.64 & 133.90 & 39.97\\
\cmidrule{4-24}
 & \multirow[t]{-4}{=}[3.2em]{\vfil\raggedright\arraybackslash None} 
 & \multirow[t]{-2}{=}[1.8em]{\vfil\raggedright\arraybackslash 70} 
 & Yes & 62.18 & 19.04 & 56.28 & 20.94 & 46.33 & 22.90 & 28.79 & 15.60 & 24.31 & 16.13 & 38.28 & 22.74 & 35.38 & 23.44 & 63.38 & 20.45 & 57.08 & 20.64 & 126.38 & 39.57\\
\cmidrule{2-24}
 &  &  & No & 55.36 & 17.89 & 58.18 & 21.37 & 52.13 & 22.08 & 40.85 & 21.07 & 36.51 & 21.73 & 37.44 & 21.18 & 36.33 & 16.89 & 53.92 & 21.61 & 50.41 & 19.64 & 103.74 & 27.85\\
\cmidrule{4-24}
 & \multirow[t]{-2}{=}[1.8em]{\vfil \raggedright\arraybackslash Bird} 
 & \multirow[t]{-2}{=}[1.8em]{\vfil \raggedright\arraybackslash 65} 
 & Yes & 46.56 & 21.91 & 58.03 & 19.06 & 53.82 & 20.86 & 44.21 & 21.31 & 40.97 & 20.32 & 37.79 & 20.99 & 38.72 & 19.75 & 48.23 & 23.58 & 45.10 & 22.16 & 98.36 & 25.03\\
\cmidrule{2-24}
 &  &  & No & 55.92 & 20.15 & 57.92 & 19.41 & 52.85 & 22.65 & 39.33 & 20.46 & 39.64 & 19.37 & 38.46 & 20.93 & 35.79 & 22.11 & 56.36 & 20.79 & 49.08 & 22.53 & 121.87 & 35.52\\
\cmidrule{4-24}
\multirow[t]{-8}{=}[6.2em]{\vfil \raggedright\arraybackslash Train} 
& \multirow[t]{-2}{=}[1.8em]{\vfil \raggedright\arraybackslash Water} 
& \multirow[t]{-2}{=}[1.8em]{\vfil \raggedright\arraybackslash 65} & Yes & 47.44 & 20.57 & 53.44 & 19.89 & 49.51 & 21.83 & 45.82 & 20.95 & 43.15 & 21.87 & 39.10 & 20.89 & 37.54 & 21.36 & 45.46 & 23.32 & 47.00 & 22.58 & 116.08 & 38.72\\
\bottomrule
\end{tabularx}
\end{sffamily}

\hspace{2em}

\clearpage
\begin{sffamily}

\setcounter{table}{1}
\refstepcounter{table}\label{tab:annoyARTstats}

\noindent\textbf{\color{scolor}Table \thetable}\par%
\noindent{Summary of ART ANOVA and relevant post-hoc contrast tests for annoyance analysis. Interaction and pairwise comparisons are denoted by the colon and en-dash terms, respectively. Combination of variables is denoted by a comma.}

\scriptsize 

\centering
\begin{tabularx}{\linewidth}{@{\extracolsep{\fill}}
*{1}{>{\raggedright\arraybackslash}p{4.5cm}}
*{1}{>{\raggedright\arraybackslash}p{7.5cm}}
*{1}{>{\raggedright\arraybackslash}p{2cm}}
*{1}{>{\raggedleft\arraybackslash}X}
@{}}
\toprule
\textbf{Term} 
& \textbf{Test} 
& \textbf{$p$-value }
& \textbf{d}\\

\midrule
\textbf{Noise Type} & 3W-ART ANOVA & \textbf{0.0003}*** & \\
AIR--TRA & ART Contrast & \textbf{0.0006}*** & 3.4071\\
AIR--MRT & ART Contrast & \textbf{0.0044}** & 2.9624\\
TRA--MRT & ART Contrast & 1.0000 & -0.4447\\

\midrule
\textbf{Masker} & 3W-ART ANOVA & 0.0000 & \\
Bird--None & ART Contrast & \textbf{0.0005}*** & -3.2629\\
Bird--Water & ART Contrast & 1.0000 & 0.7757\\
None--Water & ART Contrast & \textbf{0.0000}**** & 4.0386\\

\midrule
\textbf{ANC} & 3W-ART ANOVA & \textbf{0.0094}** & \\
No--Yes & ART Contrast & \textbf{0.0094}** & 1.9551\\

\midrule
\textbf{Noise Type:Masker} & 3W-ART ANOVA & 0.2329 & \\
\textbf{Noise Type:ANC} & 3W-ART ANOVA & 0.2886 & \\

\midrule
\textbf{Masker:ANC} & 3W-ART ANOVA & \textbf{0.0011}** & \\
Bird,No--Bird,Yes & ART Contrast & \textbf{0.0052}** & 3.4071\\
Bird,No--None,No & ART Contrast & 1.0000 & 2.9624\\
Bird,No--None,Yes & ART Contrast & 1.0000 & -0.4447\\
Bird,No--Water,No & ART Contrast & 1.0000 & 3.4071\\
Bird,No--Water,Yes & ART Contrast & \textbf{0.0029}** & 2.9624\\
Bird,Yes--None,No & ART Contrast & \textbf{0.0001}*** & -0.4447\\
Bird,Yes--None,Yes & ART Contrast & \textbf{0.0000}**** & 3.4071\\
Bird,Yes--Water,No & ART Contrast & 0.1328 & 2.9624\\
Bird,Yes--Water,Yes & ART Contrast & 1.0000 & -0.4447\\
None,No--None,Yes & ART Contrast & 1.0000 & 3.4071\\
None,No--Water,No & ART Contrast & 1.0000 & 2.9624\\
None,No--Water,Yes & ART Contrast & \textbf{0.0000}**** & -0.4447\\
None,Yes--Water,No & ART Contrast & 0.6988 & 3.4071\\
None,Yes--Water,Yes & ART Contrast & \textbf{0.0000}**** & 2.9624\\
Water,No--Water,Yes & ART Contrast & 0.0839 & -0.4447\\

\midrule
\textbf{Noise Type:Masker:ANC} & 3W-ART ANOVA & 0.4323 & \\
\bottomrule
\multicolumn{4}{l}{\rule{0pt}{1em}*:p<0.05, **:p<0.01, ***:p<0.001, ****:p<0.0001}\\
\end{tabularx}
\end{sffamily}

\clearpage
\begin{sffamily}

\setcounter{table}{2}
\refstepcounter{table}\label{tab:PAQANOVAstats}

\noindent\textbf{\color{scolor}Table \thetable}\par%
\noindent{Summary of 3WMANOVA, post-hoc univariate 3WANOVA, and relevant post-hoc Tukey HSD tests for ISOPL and ISOEV. Interaction and pairwise comparisons are denoted by the colon and en-dash terms, respectively. Combination of variables is denoted by a comma. Effect size for 3WMANOVA and 3WANOVA are based on $\eta^2_\text{p}$, while the effect size for Tukey's HSD is based on Cohen's $d$.}

\tiny

\centering
\begin{longtable}{@{\extracolsep{\fill}}
*{1}{>{\raggedright\arraybackslash}p{1.5cm}}
*{1}{>{\raggedright\arraybackslash}p{4.5cm}}
*{1}{>{\raggedright\arraybackslash}p{5cm}}
*{1}{>{\raggedright\arraybackslash}p{2cm}}
*{1}{>{\raggedleft\arraybackslash}p{1.5cm}}
@{}}

    \toprule
    Variable 
    & Term 
    & Test 
    & p-value 
    & Effect\\
    \midrule
\endhead


 & \scriptsize Noise Type & \scriptsize 3WMANOVA & \scriptsize\textbf{0.0000}**** & \scriptsize 0.0739\\
\cmidrule{2-5}
 & \scriptsize Masker & \scriptsize 3WMANOVA & \scriptsize\textbf{0.0000}**** & \scriptsize 0.0615\\
\cmidrule{2-5}
 & \scriptsize ANC & \scriptsize 3WMANOVA & \scriptsize\textbf{0.0001}*** & \scriptsize 0.0272\\
\cmidrule{2-5}
 & \scriptsize Noise Type:Masker & \scriptsize 3WMANOVA & \scriptsize\textbf{0.0086}** & \scriptsize 0.0148\\
\cmidrule{2-5}
 & \scriptsize Noise Type:ANC & \scriptsize 3WMANOVA & \scriptsize  0.4501 & \scriptsize 0.0027\\
\cmidrule{2-5}
 & \scriptsize Masker:ANC & \scriptsize 3WMANOVA & \scriptsize  0.4680 & \scriptsize 0.0026\\
\cmidrule{2-5}
\scriptsize\multirow[t]{-7}{=}[5.3em]{\vfil \raggedright\arraybackslash ISOPL:ISOEV}
& \scriptsize Noise Type:Masker:ANC & \scriptsize 3WMANOVA & \scriptsize  0.9792 & \scriptsize 0.0015\\
\midrule

 & \scriptsize Noise Type & \scriptsize 3WANOVA & \scriptsize 0.1203 & \scriptsize 0.0062\\
\cmidrule{2-5}
 & \scriptsize Masker & \scriptsize 3WANOVA & \scriptsize\textbf{0.0000}**** & \scriptsize 0.0818\\
 & Bird--None & Tukey HSD & \textbf{0.0000}**** & 0.5336\\
 & Bird--Water & Tukey HSD & 0.2191 & -0.1540\\
 & None--Water & Tukey HSD & \textbf{0.0000}**** & -0.6876\\
\cmidrule{2-5}
 & \scriptsize ANC & \scriptsize 3WANOVA & \scriptsize\textbf{0.0001}*** & \scriptsize 0.0225\\
 & No--Yes & Tukey HSD & \textbf{0.0001}*** & -0.2996\\
\cmidrule{2-5}
 & \scriptsize Noise Type:Masker & \scriptsize 3WANOVA & \scriptsize\textbf{0.0088}** & \scriptsize 0.0196\\
 & Aircraft,Bird--Aircraft,None & Tukey HSD & 0.1434 & 0.4352\\
 & Aircraft,Bird--Aircraft,Water & Tukey HSD & 0.3847 & -0.3577\\
 & Aircraft,Bird--Traffic,Bird & Tukey HSD & 0.6193 & -0.3030\\
 & Aircraft,Bird--Traffic,None & Tukey HSD & \textbf{0.0075}** & 0.5903\\
 & Aircraft,Bird--Traffic,Water & Tukey HSD & 0.2682 & -0.3895\\
 & Aircraft,Bird--Train,Bird & Tukey HSD & 0.8395 & -0.2455\\
 & Aircraft,Bird--Train,None & Tukey HSD & 1.0000 & 0.0266\\
 & Aircraft,Bird--Train,Water & Tukey HSD & 0.7793 & -0.2634\\
 & Aircraft,None--Aircraft,Water & Tukey HSD & \textbf{0.0000}**** & -0.7929\\
 & Aircraft,None--Traffic,Bird & Tukey HSD & \textbf{0.0002}*** & 0.7382\\
 & Aircraft,None--Traffic,None & Tukey HSD & 0.9885 & 0.1551\\
 & Aircraft,None--Traffic,Water & Tukey HSD & \textbf{0.0000}**** & -0.8247\\
 & Aircraft,None--Train,Bird & Tukey HSD & \textbf{0.0008}*** & 0.6808\\
 & Aircraft,None--Train,None & Tukey HSD & 0.2095 & -0.4086\\
 & Aircraft,None--Train,Water & Tukey HSD & \textbf{0.0005}*** & -0.6987\\
 & Aircraft,Water--Traffic,Bird & Tukey HSD & 1.0000 & -0.0547\\
 & Aircraft,Water--Traffic,None & Tukey HSD & \textbf{0.0000}**** & -0.9480\\
 & Aircraft,Water--Traffic,Water & Tukey HSD & 1.0000 & -0.0317\\
 & Aircraft,Water--Train,Bird & Tukey HSD & 0.9988 & -0.1122\\
 & Aircraft,Water--Train,None & Tukey HSD & 0.2853 & -0.3844\\
 & Aircraft,Water--Train,Water & Tukey HSD & 0.9997 & 0.0943\\
 & Traffic,Bird--Traffic,None & Tukey HSD & \textbf{0.0000}**** & 0.8933\\
 & Traffic,Bird--Traffic,Water & Tukey HSD & 0.9998 & -0.0864\\
 & Traffic,Bird--Train,Bird & Tukey HSD & 1.0000 & 0.0575\\
 & Traffic,Bird--Train,None & Tukey HSD & 0.5027 & 0.3297\\
 & Traffic,Bird--Train,Water & Tukey HSD & 1.0000 & 0.0396\\
 & Traffic,None--Traffic,Water & Tukey HSD & \textbf{0.0000}**** & -0.9797\\
 & Traffic,None--Train,Bird & Tukey HSD & \textbf{0.0000}**** & 0.8358\\
 & Traffic,None--Train,None & Tukey HSD & \textbf{0.0136}* & -0.5636\\
 & Traffic,None--Train,Water & Tukey HSD & \textbf{0.0000}**** & -0.8537\\
 & Traffic,Water--Train,Bird & Tukey HSD & 0.9930 & -0.1439\\
 & Traffic,Water--Train,None & Tukey HSD & 0.1890 & -0.4161\\
 & Traffic,Water--Train,Water & Tukey HSD & 0.9972 & 0.1260\\
 & Train,Bird--Train,None & Tukey HSD & 0.7466 & 0.2722\\
 & Train,Bird--Train,Water & Tukey HSD & 1.0000 & -0.0179\\
 & Train,None--Train,Water & Tukey HSD & 0.6745 & -0.2901\\
\cmidrule{2-5}
 & \scriptsize Noise Type:ANC & \scriptsize 3WANOVA & \scriptsize  0.1780 & \scriptsize 0.0050\\
\cmidrule{2-5}
 & \scriptsize Masker:ANC & \scriptsize 3WANOVA & \scriptsize  0.2001 & \scriptsize 0.0047\\
\cmidrule{2-5}
\scriptsize\multirow[t]{-47}{=}[6em]{\vfil \raggedright\arraybackslash ISOPL} 
& \scriptsize Noise Type:Masker:ANC & \scriptsize 3WANOVA & \scriptsize  0.7487 & \scriptsize 0.0028\\
\midrule
 & \scriptsize Noise Type & \scriptsize 3WANOVA & \scriptsize\textbf{0.0000}**** & \scriptsize 0.1416\\
 & Aircraft--Traffic & Tukey HSD & \textbf{0.0000}**** & 0.8302\\
 & Aircraft--Train & Tukey HSD & 0.9059 & -0.0392\\
 & Traffic--Train & Tukey HSD & \textbf{0.0000}**** & -0.8693\\
\cmidrule{2-5}
 & \scriptsize Masker & \scriptsize 3WANOVA & \scriptsize\textbf{0.0000}**** & \scriptsize 0.0461\\
 & Bird--None & Tukey HSD & \textbf{0.0000}**** & 0.5116\\
 & Bird--Water & Tukey HSD & 0.3303 & 0.1314\\
 & None--Water & Tukey HSD & \textbf{0.0001}*** & -0.3802\\
\cmidrule{2-5}
 & \scriptsize ANC & \scriptsize 3WANOVA & \scriptsize 0.0630 & \scriptsize 0.0050\\
\cmidrule{2-5}
 & \scriptsize Noise Type:Masker & \scriptsize 3WANOVA & \scriptsize  0.1353 & \scriptsize 0.0102\\
\cmidrule{2-5}
 & \scriptsize Noise Type:ANC & \scriptsize 3WANOVA & \scriptsize  0.8937 & \scriptsize 0.0003\\
\cmidrule{2-5}
 & \scriptsize Masker:ANC & \scriptsize 3WANOVA & \scriptsize  0.8445 & \scriptsize 0.0005\\
\cmidrule{2-5}
\scriptsize\multirow[t]{-13}{=}[5.5em]{\vfil \raggedright\arraybackslash ISOEV} 
& \scriptsize Noise Type:Masker:ANC & \scriptsize 3WANOVA & \scriptsize  0.9981 & \scriptsize 0.0002\\
\bottomrule
\multicolumn{5}{l}{\scriptsize \rule{0pt}{1em}*:p<0.05, **:p<0.01, ***:p<0.001, ****:p<0.0001}\\
\end{longtable}
\end{sffamily}

\twocolumn
\begin{sffamily}
\setcounter{table}{3}
\refstepcounter{table}\label{tab:PLNANOVAstats}

\scriptsize

\noindent\textbf{\color{scolor}Table \thetable}\par%
\noindent{Summary of 2W-ART ANOVA, and relevant post-hoc ART contrast tests for PLN.}  

\centering
\begin{tabularx}{\linewidth}{@{\extracolsep{\fill}}
>{\raggedright\arraybackslash}p{0.8cm}
>{\raggedright\arraybackslash}p{1.5cm}
>{\raggedright\arraybackslash}p{2.2cm}
>{\raggedright\arraybackslash}X
>{\raggedleft\arraybackslash}X
@{}}

\toprule
Variable & Term & Test & $p$-value & $d$\\
\midrule 
& \textbf{Masker} & 2W-ART ANOVA & \textbf{0.0009}*** & \\
& Bird--None & ART Contrast & 1.0000 & -0.1377\\
& Bird--Water & ART Contrast & \textbf{0.0016}** & -1.0288\\
& None--Water & ART Contrast & \textbf{0.0080}** & -0.8911\\
\cmidrule{2-5}
& \textbf{ANC} & 2W-ART ANOVA & \textbf{0.0000}**** & \\
& No--Yes & ART Contrast & \textbf{0.0000}**** & 1.0035\\
\cmidrule{2-5}
\multirow[t]{-7}{=}[2.2em]{\vfil\raggedright\arraybackslash AIR}  
& Masker:ANC & 2W-ART ANOVA & 0.5642 & \\

\midrule
& \textbf{Masker} & 2W-ART ANOVA & \textbf{0.0000}**** & \\
& Bird--None & ART Contrast & \textbf{0.0286}* & 0.7693\\
& Bird--Water & ART Contrast & \textbf{0.0004}*** & -1.1468\\
& None--Water & ART Contrast & \textbf{0.0000}**** & -1.916\\
\cmidrule{2-5}
& \textbf{ANC} & 2W-ART ANOVA & \textbf{0.0066}** & \\
& No--Yes & ART Contrast & \textbf{0.0066}** & 0.6972\\
\cmidrule{2-5}
\multirow[t]{-7}{=}[2.2em]{\vfil\raggedright\arraybackslash MRT} & Masker:ANC & 2W-ART ANOVA & 0.5939 & \\

\midrule
 & \textbf{Masker} & 2W-ART ANOVA & \textbf{0.0000}**** & \\
 & Bird--None & ART Contrast & 1.0000 & 0.08\\
 & Bird--Water & ART Contrast & \textbf{0.0000}**** & -2.3868\\
 & None--Water & ART Contrast & \textbf{0.0000}**** & -2.4668\\
\cmidrule{2-5}
 & \textbf{ANC} & 2W-ART ANOVA & \textbf{0.0000}**** & \\
 & No--Yes & ART Contrast & \textbf{0.0000}**** & 2.2363\\
\cmidrule{2-5}
\multirow[t]{-7}{=}[2.2em]{\vfil\raggedright\arraybackslash TRA} & Masker:ANC & 2W-ART ANOVA & 0.1177 & \\
\bottomrule

\multicolumn{5}{l}{\rule{0pt}{1em}*:p<0.05, **:p<0.01, ***:p<0.001, ****:p<0.0001}\\
\end{tabularx}
\end{sffamily}

\hspace{2em}
\vfill\eject

\begin{sffamily}
\setcounter{table}{4}
\refstepcounter{table}\label{tab:sentiARTsummary}

\scriptsize

\noindent\textbf{\color{scolor}Table \thetable}\par%
\noindent{Summary statistics of sentiment scores.}  

\centering
\begin{tabularx}{\linewidth}{@{\extracolsep{\fill}}
>{\raggedright\arraybackslash}p{1.8cm}
*{2}{>{\raggedright\arraybackslash}X}
*{2}{>{\raggedleft\arraybackslash}X}
@{}}
\toprule
Variable 
& Masker 
& ANC 
& Mean, $\mu$ 
& SD, $\sigma$\\

\midrule
 &  & No 
 & -0.13 
 & 0.36\\
\cmidrule{3-5}
 & \multirow[t]{-2}{=}[1.7em]{\vfil \raggedright\arraybackslash Bird} 
 & Yes & 0.11 & 0.42\\
\cmidrule{2-5}
 &  & No & -0.07 & 0.33\\
\cmidrule{3-5}
 & \multirow[t]{-2}{=}[1.7em]{\vfil \raggedright\arraybackslash None} 
 & Yes & 0.03 & 0.31\\
\cmidrule{2-5}
 &  & No & 0.00 & 0.43\\
\cmidrule{3-5}
\multirow[t]{-6}{=}[3.7em]{\vfil \raggedright\arraybackslash AIR} 
& \multirow[t]{-2}{=}[1.7em]{\vfil \raggedright\arraybackslash Water} 
& Yes & -0.05 & 0.46\\
\cmidrule{1-5}
 &  & No & -0.07 & 0.31\\
\cmidrule{3-5}
 & \multirow[t]{-2}{=}[1.7em]{\vfil \raggedright\arraybackslash Bird} 
 & Yes & 0.11 & 0.27\\
\cmidrule{2-5}
 &  & No & -0.02 & 0.29\\
\cmidrule{3-5}
 & \multirow[t]{-2}{=}[1.7em]{\vfil \raggedright\arraybackslash None} 
 & Yes & -0.04 & 0.41\\
\cmidrule{2-5}
 &  & No & -0.19 & 0.47\\
\cmidrule{3-5}
\multirow[t]{-6}{=}[3.7em]{\vfil \raggedright\arraybackslash MRT} 
& \multirow[t]{-2}{=}[1.7em]{\vfil \raggedright\arraybackslash Water} 
& Yes & -0.10 & 0.46\\
\cmidrule{1-5}
 &  & No & 0.05 & 0.49\\
\cmidrule{3-5}
 & \multirow[t]{-2}{=}[1.7em]{\vfil \raggedright\arraybackslash Bird} 
 & Yes & -0.03 & 0.31\\
\cmidrule{2-5}
 &  & No & -0.02 & 0.35\\
\cmidrule{3-5}
 & \multirow[t]{-2}{=}[1.7em]{\vfil \raggedright\arraybackslash None} 
 & Yes & -0.05 & 0.31\\
\cmidrule{2-5}
 &  & No & -0.21 & 0.37\\
\cmidrule{3-5}
\multirow[t]{-6}{=}[3.7em]{\vfil \raggedright\arraybackslash TRA} 
& \multirow[t]{-2}{=}[1.7em]{\vfil \raggedright\arraybackslash Water} 
& Yes & -0.02 & 0.43\\
\bottomrule
\end{tabularx}
\end{sffamily}
\vfill\null
\newpage
\begin{sffamily}
\setcounter{table}{5}
\refstepcounter{table}\label{tab:sentiARTstats}

\scriptsize

\noindent\textbf{\color{scolor}Table \thetable}\par%
\noindent{Summary of ART ANOVA and relevant post-hoc contrast tests for sentiment analysis. Interaction and pairwise comparisons are denoted by the colon and en-dash terms, respectively.}  

\centering
\begin{tabularx}{\linewidth}{@{\extracolsep{\fill}}
>{\raggedright\arraybackslash}p{0.8cm}
>{\raggedright\arraybackslash}p{1.5cm}
>{\raggedright\arraybackslash}p{2.2cm}
>{\raggedright\arraybackslash}X
>{\raggedleft\arraybackslash}X
@{}}

\toprule
Variable 
& Test 
& Term 
& $p$-value 
& Effect size, $d$\\

\midrule
 & Masker & 2W-ART ANOVA & 0.5025 & \\
\cmidrule{2-5}
 & ANC & 2W-ART ANOVA & 0.0648 & \\
\cmidrule{2-5}
\multirow[t]{-3}{=}[2.3em]{\vfil \raggedright\arraybackslash AIR} 
& Masker:ANC & 2W-ART ANOVA & 0.2964 & \\

\midrule
 & \textbf{Masker} & 2W-ART ANOVA & \textbf{0.0008}*** & \\
 & Bird--None & ART Contrast & \textbf{0.0216}* & 80.58\\
 & Bird--Water & ART Contrast & \textbf{0.0006}*** & 129.11\\
 & None--Water & ART Contrast & 0.3126 & 48.53\\
\cmidrule{2-5}
 & ANC & 2W-ART ANOVA & 0.0934 & \\
\cmidrule{2-5} 
\multirow[t]{-6}{=}[2.3em]{\vfil \raggedright\arraybackslash MRT} 
& Masker:ANC & 2W-ART ANOVA & 0.2297 & \\

\midrule
 & Masker & 2W-ART ANOVA & 0.1082 & \\
\cmidrule{2-5}
 & ANC & 2W-ART ANOVA & 0.3741 & \\
\cmidrule{2-5}
\multirow[t]{-3}{=}[2.3em]{\vfil \raggedright\arraybackslash TRA} 
 & Masker:ANC & 2W-ART ANOVA & 0.1334 & \\
\bottomrule
\multicolumn{5}{l}{\rule{0pt}{1em} *:$p<0.05$, **:$p<0.01$, ***:$p<0.001$, ****:$p<0.0001$}\\
\end{tabularx}
\end{sffamily}

\onecolumn
\begin{sffamily}

\setcounter{table}{6}
\refstepcounter{table}\label{tab:corrObjSubj}
\setlength{\tabcolsep}{1pt}

\noindent\textbf{\color{scolor}Table \thetable}\par%
\noindent{Pearson's correlation across all objective and subjective indices. Values smaller than 0.28 have been removed for conciseness.}

\tiny
\centering
\begin{longtable}
{@{\extracolsep{\fill}}
>{\raggedright\arraybackslash}l
*{18}{>{\raggedright\arraybackslash}r}
@{}}
\toprule

& \textit{PAY} & \textit{PLN} & $L_{\text{Ceq}}$ & $L_{\text{Cmax}}$ & $L_{\text{C5}}$ & $L_{\text{C10}}$ 
& $L_{\text{C50}}$ & $L_{\text{C90}}$ & $L_{\text{C95}}$ & $L_{\text{Aeq}}$  & $L_{\text{Amax}}$ & $L_{\text{A5}}$ 
& $L_{\text{A10}}$ & $L_{\text{A50}}$ & $L_{\text{A90}}$ & $L_{\text{A95}}$ & $N_{\text{max}}$ & $N_{\text{5}}$ \\
\midrule
\textit{PAY} & 1.00 &  & 0.32 &  & 0.28 & 0.29 &  &  &  & 0.30 &  &  &  &  &  &  &  & 0.30\\
\textit{PLN} &  & 1.00 & 0.51 &  & 0.37 & 0.39 & 0.45 & 0.32 & 0.29 & 0.49 &  & 0.35 & 0.37 & 0.44 &  &  & 0.31 & 0.41\\
$L_{\text{Ceq}}$ & 0.32 & 0.51 & 1.00 & 0.76 & 0.88 & 0.90 & 0.71 & 0.37 & 0.33 & 0.99 & 0.73 & 0.85 & 0.88 & 0.68 &  & 0.13 & 0.79 & 0.92\\
$L_{\text{Cmax}}$ &  &  & 0.76 & 1.00 & 0.96 & 0.95 &  &  &  & 0.80 & 0.98 & 0.96 & 0.96 & 0.13 & -0.36 & -0.40 & 0.97 & 0.90\\
$L_{\text{C5}}$ & 0.28 & 0.37 & 0.88 & 0.96 & 1.00 & 1.00 & 0.33 &  &  & 0.90 & 0.95 & 0.99 & 0.99 & 0.30 &  & -0.30 & 0.97 & 0.97\\
$L_{\text{C10}}$ & 0.29 & 0.39 & 0.90 & 0.95 & 1.00 & 1.00 & 0.36 &  &  & 0.92 & 0.93 & 0.98 & 0.99 & 0.33 &  &  & 0.96 & 0.98\\
$L_{\text{C50}}$ &  & 0.45 & 0.71 &  & 0.33 & 0.36 & 1.00 & 0.76 & 0.69 & 0.66 &  & 0.31 & 0.34 & 0.99 & 0.70 & 0.62 &  & 0.40\\
$L_{\text{C90}}$ &  & 0.32 & 0.37 &  &  &  & 0.76 & 1.00 & 0.99 & 0.32 &  &  &  & 0.75 & 0.96 & 0.95 &  & \\
$L_{\text{C95}}$ &  & 0.29 & 0.33 &  &  &  & 0.69 & 0.99 & 1.00 &  &  &  &  & 0.68 & 0.95 & 0.96 &  & \\
$L_{\text{Aeq}}$ & 0.30 & 0.49 & 0.99 & 0.80 & 0.90 & 0.92 & 0.66 & 0.32 &  & 1.00 & 0.79 & 0.90 & 0.92 & 0.65 &  &  & 0.82 & 0.91\\
$L_{\text{Amax}}$ &  &  & 0.73 & 0.98 & 0.95 & 0.93 &  &  &  & 0.79 & 1.00 & 0.97 & 0.96 &  & -0.35 & -0.39 & 0.95 & 0.86\\
$L_{\text{A5}}$ &  & 0.35 & 0.85 & 0.96 & 0.99 & 0.98 & 0.31 &  &  & 0.90 & 0.97 & 1.00 & 1.00 & 0.30 &  & -0.32 & 0.95 & 0.93\\
$L_{\text{A10}}$ &  & 0.37 & 0.88 & 0.96 & 0.99 & 0.99 & 0.34 &  &  & 0.92 & 0.96 & 1.00 & 1.00 & 0.32 &  & -0.28 & 0.95 & 0.95\\
$L_{\text{A50}}$ &  & 0.44 & 0.68 & 0.13 & 0.30 & 0.33 & 0.99 & 0.75 & 0.68 & 0.65 &  & 0.30 & 0.32 & 1.00 & 0.71 & 0.63 &  & 0.36\\
$L_{\text{A90}}$ &  &  &  & -0.36 &  &  & 0.70 & 0.96 & 0.95 &  & -0.35 &  &  & 0.71 & 1.00 & 0.99 & -0.33 & \\
$L_{\text{A95}}$ &  &  & 0.13 & -0.40 & -0.30 &  & 0.62 & 0.95 & 0.96 &  & -0.39 & -0.32 & -0.28 & 0.63 & 0.99 & 1.00 & -0.36 & \\
$N_{\text{max}}$ &  & 0.31 & 0.79 & 0.97 & 0.97 & 0.96 &  &  &  & 0.82 & 0.95 & 0.95 & 0.95 &  & -0.33 & -0.36 & 1.00 & 0.95\\
$N_{\text{5}}$ & 0.30 & 0.41 & 0.92 & 0.90 & 0.97 & 0.98 & 0.40 &  &  & 0.91 & 0.86 & 0.93 & 0.95 & 0.36 &  &  & 0.95 & 1.00\\
$N_{\text{10}}$ & 0.31 & 0.44 & 0.94 & 0.86 & 0.95 & 0.96 & 0.46 &  &  & 0.92 & 0.82 & 0.91 & 0.93 & 0.41 &  &  & 0.92 & 1.00\\
$N_{\text{50}}$ &  & 0.45 & 0.63 &  &  &  & 0.96 & 0.88 & 0.84 & 0.57 &  &  &  & 0.93 & 0.82 & 0.77 &  & 0.33\\
$N_{\text{90}}$ &  & 0.29 & 0.28 & -0.30 &  &  & 0.70 & 0.99 & 0.99 &  & -0.30 &  &  & 0.69 & 0.96 & 0.97 &  & \\
$N_{\text{95}}$ &  &  &  & -0.32 &  &  & 0.64 & 0.98 & 0.99 &  & -0.32 &  &  & 0.63 & 0.95 & 0.97 &  & \\
$S_{\text{max}}$ &  & -0.31 & -0.48 &  &  &  & -0.54 & -0.67 & -0.69 & -0.36 &  &  &  & -0.46 & -0.50 & -0.53 &  & -0.32\\
$S$ &  &  & -0.37 &  &  &  &  &  &  &  &  &  &  &  &  &  &  & -0.38\\
$S_{\text{5}}$ &  &  &  &  &  &  & -0.28 & -0.40 & -0.43 &  &  &  &  &  &  & -0.29 &  & \\
$S_{\text{10}}$ &  &  &  &  &  &  & -0.29 & -0.39 & -0.41 &  &  &  &  &  &  &  &  & \\
$S_{\text{50}}$ &  &  & -0.36 &  &  &  &  &  &  &  &  &  &  &  &  &  &  & -0.37\\
$S_{\text{90}}$ &  &  & -0.45 & -0.44 & -0.47 & -0.47 &  &  &  & -0.34 & -0.33 & -0.36 & -0.37 &  &  &  & -0.49 & -0.56\\
$S_{\text{95}}$ &  &  & -0.47 & -0.48 & -0.50 & -0.51 &  &  &  & -0.36 & -0.36 & -0.40 & -0.41 &  &  &  & -0.52 & -0.59\\
$R_{\text{max}}$ &  & 0.33 & 0.63 & 0.45 & 0.56 & 0.56 & 0.43 & 0.29 &  & 0.57 & 0.41 & 0.49 & 0.49 & 0.37 &  &  & 0.54 & 0.64\\
$R$ &  & 0.33 & 0.50 &  &  &  & 0.65 & 0.68 & 0.66 & 0.38 &  &  & 0.13 & 0.56 & 0.58 & 0.57 &  & 0.39\\
$R_{\text{5}}$ &  & 0.35 & 0.61 & 0.31 & 0.43 & 0.45 & 0.54 & 0.46 & 0.44 & 0.49 &  & 0.32 & 0.35 & 0.44 & 0.31 & 0.29 & 0.43 & 0.58\\
$R_{\text{10}}$ &  & 0.34 & 0.59 &  & 0.39 & 0.41 & 0.54 & 0.50 & 0.49 & 0.47 &  & 0.28 & 0.31 & 0.44 & 0.35 & 0.34 & 0.38 & 0.56\\
$R_{\text{50}}$ &  & 0.33 & 0.48 &  &  &  & 0.67 & 0.71 & 0.70 & 0.36 &  &  &  & 0.59 & 0.62 & 0.61 & 0.13 & 0.35\\
$R_{\text{90}}$ &  & 0.28 & 0.33 &  &  &  & 0.64 & 0.79 & 0.79 &  &  &  &  & 0.58 & 0.75 & 0.75 &  & \\
$R_{\text{95}}$ &  &  & 0.31 &  &  &  & 0.62 & 0.80 & 0.80 &  &  &  &  & 0.56 & 0.76 & 0.76 &  & 0.13\\
$T_{\text{max}}$ &  &  &  & 0.46 & 0.37 & 0.32 &  & -0.33 & -0.41 &  & 0.53 & 0.44 & 0.38 &  & -0.33 & -0.41 & 0.35 & \\
$T_{\text{5}}$ &  &  &  & 0.29 & 0.28 &  &  &  & -0.35 &  & 0.37 & 0.35 & 0.31 & 0.30 &  & -0.32 &  & \\
$T_{\text{10}}$ &  &  &  &  &  &  & 0.35 &  & -0.32 &  &  & 0.29 &  & 0.38 &  &  &  & \\
$T_{\text{50}}$ &  &  &  &  &  &  & 0.43 &  &  &  &  &  & 0.13 & 0.46 &  &  &  & \\
$T_{\text{90}}$ &  &  & 0.13 &  &  &  & 0.44 &  &  &  &  &  &  & 0.46 &  &  &  & \\
$T_{\text{95}}$ &  &  & 0.13 &  &  &  & 0.43 &  &  &  &  &  &  & 0.46 &  &  &  & \\
$FS_{\text{max}}$ &  &  &  & 0.50 & 0.46 & 0.43 &  & -0.34 & -0.34 &  & 0.46 & 0.43 & 0.40 &  & -0.50 & -0.50 & 0.54 & 0.45\\
$FS_{\text{5}}$ &  &  &  & 0.43 & 0.30 &  &  & -0.34 & -0.38 &  & 0.46 & 0.33 &  &  & -0.40 & -0.44 & 0.37 & \\
$FS_{\text{10}}$ &  &  &  & 0.53 & 0.38 & 0.32 & -0.32 & -0.55 & -0.59 &  & 0.56 & 0.41 & 0.35 & -0.31 & -0.58 & -0.62 & 0.46 & \\
$FS_{\text{50}}$ &  &  &  & 0.51 & 0.36 & 0.30 &  & -0.53 & -0.60 &  & 0.55 & 0.41 & 0.34 &  & -0.55 & -0.61 & 0.43 & \\
$FS_{\text{90}}$ &  &  &  & 0.46 & 0.41 & 0.37 &  & -0.57 & -0.62 &  & 0.48 & 0.43 & 0.39 &  & -0.56 & -0.61 & 0.41 & 0.30\\
$FS_{\text{95}}$ &  &  &  & 0.45 & 0.41 & 0.38 &  & -0.50 & -0.55 &  & 0.47 & 0.44 & 0.40 &  & -0.48 & -0.54 & 0.40 & 0.31\\
\textit{PA} & 0.29 & 0.41 & 0.92 & 0.90 & 0.97 & 0.98 & 0.40 &  &  & 0.92 & 0.87 & 0.94 & 0.95 & 0.37 &  &  & 0.96 & 1.00\\
\textit{ISOPL} & -0.60 &  &  &  &  &  &  &  &  &  &  &  &  &  &  &  &  & \\
\textit{ISOEV} & 0.36 &  &  & 0.33 & 0.34 & 0.33 &  &  &  & 0.28 & 0.35 & 0.36 & 0.35 &  &  &  & 0.32 & 0.29\\
$L_{\text{Ceq}}-L_{\text{Aeq}}$ &  &  & -0.49 & -0.64 & -0.62 & -0.60 &  &  &  & -0.63 & -0.73 & -0.72 & -0.70 &  &  &  & -0.57 & -0.48\\
$L_{\text{A10}}-L_{\text{A90}}$ & 0.13 &  & 0.31 & 0.76 & 0.70 & 0.67 & -0.35 & -0.75 & -0.76 & 0.37 & 0.75 & 0.71 & 0.69 & -0.36 & -0.87 & -0.88 & 0.73 & 0.58\\
$L_{\text{C10}}-L_{\text{C90}}$ &  &  & 0.36 & 0.79 & 0.74 & 0.71 & -0.28 & -0.73 & -0.75 & 0.41 & 0.78 & 0.74 & 0.72 & -0.30 & -0.82 & -0.85 & 0.77 & 0.63\\
\midrule
\midrule
& $N_{\text{10}}$ & $N_{\text{50}}$ & $N_{\text{90}}$ & $N_{\text{95}}$ & $S_{\text{max}}$ & $S$ 
& $S_{\text{5}}$ & $S_{\text{10}}$ & $S_{\text{50}}$ & $S_{\text{90}}$ & $S_{\text{95}}$ & $R_{\text{max}}$ 
& $R$ & $R_{\text{5}}$ & $R_{\text{10}}$ & $R_{\text{50}}$ & $R_{\text{90}}$ & $R_{\text{95}}$ \\

\midrule
\textit{\textit{PA}Y} & 0.31 &  &  &  &  &  &  &  &  &  &  &  &  &  &  &  &  & \\
\textit{PLN} & 0.44 & 0.45 & 0.29 &  & -0.31 &  &  &  &  &  &  & 0.33 & 0.33 & 0.35 & 0.34 & 0.33 & 0.28 & \\
$L_{\text{Ceq}}$ & 0.94 & 0.63 & 0.28 &  & -0.48 & -0.37 &  &  & -0.36 & -0.45 & -0.47 & 0.63 & 0.50 & 0.61 & 0.59 & 0.48 & 0.33 & 0.31\\
$L_{\text{Cmax}}$ & 0.86 &  & -0.30 & -0.32 &  &  &  &  &  & -0.44 & -0.48 & 0.45 &  & 0.31 &  &  &  & \\
$L_{\text{C5}}$ & 0.95 &  &  &  &  &  &  &  &  & -0.47 & -0.50 & 0.56 &  & 0.43 & 0.39 &  &  & \\
$L_{\text{C10}}$ & 0.96 &  &  &  &  &  &  &  &  & -0.47 & -0.51 & 0.56 &  & 0.45 & 0.41 &  &  & \\
$L_{\text{C50}}$ & 0.46 & 0.96 & 0.70 & 0.64 & -0.54 &  & -0.28 & -0.29 &  &  &  & 0.43 & 0.65 & 0.54 & 0.54 & 0.67 & 0.64 & 0.62\\
$L_{\text{C90}}$ &  & 0.88 & 0.99 & 0.98 & -0.67 &  & -0.40 & -0.39 &  &  &  & 0.29 & 0.68 & 0.46 & 0.50 & 0.71 & 0.79 & 0.80\\
$L_{\text{C95}}$ &  & 0.84 & 0.99 & 0.99 & -0.69 &  & -0.43 & -0.41 &  &  &  &  & 0.66 & 0.44 & 0.49 & 0.70 & 0.79 & 0.80\\
$L_{\text{Aeq}}$ & 0.92 & 0.57 &  &  & -0.36 &  &  &  &  & -0.34 & -0.36 & 0.57 & 0.38 & 0.49 & 0.47 & 0.36 &  & \\
$L_{\text{Amax}}$ & 0.82 &  & -0.30 & -0.32 &  &  &  &  &  & -0.33 & -0.36 & 0.41 &  &  &  &  &  & \\
$L_{\text{A5}}$ & 0.91 &  &  &  &  &  &  &  &  & -0.36 & -0.40 & 0.49 &  & 0.32 & 0.28 &  &  & \\
$L_{\text{A10}}$ & 0.93 &  &  &  &  &  &  &  &  & -0.37 & -0.41 & 0.49 & 0.13 & 0.35 & 0.31 &  &  & \\
$L_{\text{A50}}$ & 0.41 & 0.93 & 0.69 & 0.63 & -0.46 &  &  &  &  &  &  & 0.37 & 0.56 & 0.44 & 0.44 & 0.59 & 0.58 & 0.56\\
$L_{\text{A90}}$ &  & 0.82 & 0.96 & 0.95 & -0.50 &  &  &  &  &  &  &  & 0.58 & 0.31 & 0.35 & 0.62 & 0.75 & 0.76\\
$L_{\text{A95}}$ &  & 0.77 & 0.97 & 0.97 & -0.53 &  & -0.29 &  &  &  &  &  & 0.57 & 0.29 & 0.34 & 0.61 & 0.75 & 0.76\\
$N_{\text{max}}$ & 0.92 &  &  &  &  &  &  &  &  & -0.49 & -0.52 & 0.54 &  & 0.43 & 0.38 & 0.13 &  & \\
$N_{\text{5}}$ & 1.00 & 0.33 &  &  & -0.32 & -0.38 &  &  & -0.37 & -0.56 & -0.59 & 0.64 & 0.39 & 0.58 & 0.56 & 0.35 &  & 0.13\\
$N_{\text{10}}$ & 1.00 & 0.40 &  &  & -0.39 & -0.40 &  &  & -0.39 & -0.56 & -0.59 & 0.64 & 0.44 & 0.61 & 0.59 & 0.40 &  & \\
$N_{\text{50}}$ & 0.40 & 1.00 & 0.85 & 0.81 & -0.66 & -0.28 & -0.37 & -0.36 & -0.28 &  &  & 0.40 & 0.76 & 0.60 & 0.63 & 0.79 & 0.80 & 0.79\\
$N_{\text{90}}$ &  & 0.85 & 1.00 & 1.00 & -0.65 &  & -0.40 & -0.37 &  &  &  &  & 0.68 & 0.44 & 0.49 & 0.71 & 0.82 & 0.83\\
$N_{\text{95}}$ &  & 0.81 & 1.00 & 1.00 & -0.65 &  & -0.40 & -0.38 &  &  &  &  & 0.66 & 0.42 & 0.47 & 0.69 & 0.80 & 0.82\\
$S_{\text{max}}$ & -0.39 & -0.66 & -0.65 & -0.65 & 1.00 & 0.75 & 0.88 & 0.84 & 0.73 & 0.57 & 0.55 & -0.44 & -0.75 & -0.68 & -0.72 & -0.75 & -0.71 & -0.71\\
$S$ & -0.40 & -0.28 &  &  & 0.75 & 1.00 & 0.93 & 0.95 & 1.00 & 0.94 & 0.92 & -0.56 & -0.68 & -0.77 & -0.77 & -0.65 & -0.52 & -0.51\\
$S_{\text{5}}$ &  & -0.37 & -0.40 & -0.40 & 0.88 & 0.93 & 1.00 & 0.99 & 0.92 & 0.76 & 0.74 & -0.43 & -0.68 & -0.68 & -0.71 & -0.67 & -0.61 & -0.60\\
$S_{\text{10}}$ &  & -0.36 & -0.37 & -0.38 & 0.84 & 0.95 & 0.99 & 1.00 & 0.95 & 0.79 & 0.77 & -0.45 & -0.68 & -0.69 & -0.71 & -0.67 & -0.59 & -0.59\\
$S_{\text{50}}$ & -0.39 & -0.28 &  &  & 0.73 & 1.00 & 0.92 & 0.95 & 1.00 & 0.93 & 0.92 & -0.54 & -0.67 & -0.76 & -0.76 & -0.64 & -0.51 & -0.49\\
$S_{\text{90}}$ & -0.56 &  &  &  & 0.57 & 0.94 & 0.76 & 0.79 & 0.93 & 1.00 & 1.00 & -0.66 & -0.59 & -0.77 & -0.76 & -0.55 & -0.37 & -0.35\\
$S_{\text{95}}$ & -0.59 &  &  &  & 0.55 & 0.92 & 0.74 & 0.77 & 0.92 & 1.00 & 1.00 & -0.66 & -0.58 & -0.76 & -0.76 & -0.54 & -0.35 & -0.33\\
$R_{\text{max}}$ & 0.64 & 0.40 &  &  & -0.44 & -0.56 & -0.43 & -0.45 & -0.54 & -0.66 & -0.66 & 1.00 & 0.63 & 0.77 & 0.74 & 0.58 & 0.44 & 0.43\\
$R$ & 0.44 & 0.76 & 0.68 & 0.66 & -0.75 & -0.68 & -0.68 & -0.68 & -0.67 & -0.59 & -0.58 & 0.63 & 1.00 & 0.94 & 0.96 & 1.00 & 0.95 & 0.94\\
$R_{\text{5}}$ & 0.61 & 0.60 & 0.44 & 0.42 & -0.68 & -0.77 & -0.68 & -0.69 & -0.76 & -0.77 & -0.76 & 0.77 & 0.94 & 1.00 & 0.99 & 0.91 & 0.79 & 0.77\\
$R_{\text{10}}$ & 0.59 & 0.63 & 0.49 & 0.47 & -0.72 & -0.77 & -0.71 & -0.71 & -0.76 & -0.76 & -0.76 & 0.74 & 0.96 & 0.99 & 1.00 & 0.94 & 0.83 & 0.82\\
$R_{\text{50}}$ & 0.40 & 0.79 & 0.71 & 0.69 & -0.75 & -0.65 & -0.67 & -0.67 & -0.64 & -0.55 & -0.54 & 0.58 & 1.00 & 0.91 & 0.94 & 1.00 & 0.97 & 0.96\\
$R_{\text{90}}$ &  & 0.80 & 0.82 & 0.80 & -0.71 & -0.52 & -0.61 & -0.59 & -0.51 & -0.37 & -0.35 & 0.44 & 0.95 & 0.79 & 0.83 & 0.97 & 1.00 & 1.00\\
$R_{\text{95}}$ &  & 0.79 & 0.83 & 0.82 & -0.71 & -0.51 & -0.60 & -0.59 & -0.49 & -0.35 & -0.33 & 0.43 & 0.94 & 0.77 & 0.82 & 0.96 & 1.00 & 1.00\\
$T_{\text{max}}$ &  &  & -0.41 & -0.45 & 0.56 &  & 0.40 & 0.29 &  &  &  &  &  &  &  &  & -0.34 & -0.37\\
$T_{\text{5}}$ &  &  & -0.34 & -0.40 & 0.49 &  & 0.41 & 0.32 &  &  &  &  &  &  &  &  & -0.29 & -0.32\\
$T_{\text{10}}$ &  &  & -0.29 & -0.36 & 0.42 &  & 0.40 & 0.33 &  &  &  &  &  &  &  &  &  & -0.28\\
$T_{\text{50}}$ &  &  &  &  &  &  & 0.30 & 0.28 &  &  &  &  &  &  &  &  &  & \\
$T_{\text{90}}$ &  &  &  &  &  &  &  &  &  &  &  &  &  &  &  &  &  & \\
$T_{\text{95}}$ &  &  &  &  &  &  &  &  &  &  &  &  &  &  &  &  &  & \\
$FS_{\text{max}}$ & 0.40 & -0.29 & -0.36 & -0.34 &  & -0.28 &  &  &  & -0.48 & -0.49 & 0.58 &  & 0.28 &  &  &  & \\
$FS_{\text{5}}$ & 0.13 & -0.34 & -0.39 & -0.38 & 0.43 &  &  &  &  &  &  & 0.38 &  &  &  &  & -0.29 & -0.30\\
$FS_{\text{10}}$ &  & -0.48 & -0.61 & -0.61 & 0.56 &  & 0.30 &  &  &  &  &  & -0.30 &  &  & -0.34 & -0.46 & -0.48\\
$FS_{\text{50}}$ &  & -0.43 & -0.59 & -0.61 & 0.60 &  & 0.37 & 0.28 &  &  &  &  &  &  &  & -0.31 & -0.43 & -0.45\\
$FS_{\text{90}}$ &  & -0.28 & -0.58 & -0.60 & 0.56 &  & 0.46 & 0.41 &  &  &  &  &  &  &  &  & -0.30 & -0.32\\
$FS_{\text{95}}$ &  &  & -0.51 & -0.54 & 0.52 &  & 0.45 & 0.41 &  &  &  &  &  &  &  &  &  & \\
\textit{PA} & 0.99 & 0.33 &  &  & -0.29 & -0.33 &  &  & -0.33 & -0.52 & -0.55 & 0.63 & 0.37 & 0.57 & 0.54 & 0.33 &  & \\
\textit{ISOPL} &  &  &  &  & 0.32 & 0.34 & 0.34 & 0.33 & 0.33 & 0.31 & 0.31 &  &  &  &  &  &  & \\
\textit{ISOEV} & 0.28 &  &  &  &  &  &  &  &  &  &  &  &  &  &  &  &  & \\
$L_{\text{Ceq}}-L_{\text{Aeq}}$ & -0.44 &  &  &  & -0.36 & -0.50 & -0.62 & -0.59 & -0.49 & -0.33 & -0.30 &  & 0.39 & 0.28 & 0.32 & 0.40 & 0.46 & 0.46\\
$L_{\text{A10}}-L_{\text{A90}}$ & 0.53 & -0.50 & -0.80 & -0.80 & 0.33 &  &  &  &  & -0.33 & -0.37 &  & -0.36 &  &  & -0.41 & -0.61 & -0.62\\
$L_{\text{C10}}-L_{\text{C90}}$ & 0.57 & -0.44 & -0.78 & -0.79 & 0.34 &  &  &  &  & -0.32 & -0.36 &  & -0.31 &  &  & -0.36 & -0.56 & -0.58\\
\midrule
\midrule
& $T_{\text{max}}$ & $T_{\text{5}}$ & $T_{\text{10}}$ & $T_{\text{50}}$ & $T_{\text{90}}$ & $T_{\text{95}}$ 
& $FS_{\text{max}}$ & $FS_{\text{5}}$ & $FS_{\text{10}}$ & $FS_{\text{50}}$ & $FS_{\text{90}}$ & $FS_{\text{95}}$ 
& \textit{PA} & \textit{\textit{ISOPL}} & \textit{\textit{ISOEV}} & $L_{\text{Ceq}}$-$L_{\text{Aeq}}$  
& $L_{\text{A10}}$-$L_{\text{A90}}$ & $L_{\text{C10}}$-$L_{\text{C90}}$\\

\midrule
\textit{PAY} &  &  &  &  &  &  &  &  &  &  &  &  & 0.29 & -0.60 & 0.36 &  & 0.13 & \\
\textit{PLN} &  &  &  &  &  &  &  &  &  &  &  &  & 0.41 &  &  &  &  & \\
$L_{\text{Ceq}}$ &  &  &  &  & 0.13 & 0.13 &  &  &  &  &  &  & 0.92 &  &  & -0.49 & 0.31 & 0.36\\
$L_{\text{Cmax}}$ & 0.46 & 0.29 &  &  &  &  & 0.50 & 0.43 & 0.53 & 0.51 & 0.46 & 0.45 & 0.90 &  & 0.33 & -0.64 & 0.76 & 0.79\\
$L_{\text{C5}}$ & 0.37 & 0.28 &  &  &  &  & 0.46 & 0.30 & 0.38 & 0.36 & 0.41 & 0.41 & 0.97 &  & 0.34 & -0.62 & 0.70 & 0.74\\
$L_{\text{C10}}$ & 0.32 &  &  &  &  &  & 0.43 &  & 0.32 & 0.30 & 0.37 & 0.38 & 0.98 &  & 0.33 & -0.60 & 0.67 & 0.71\\
$L_{\text{C50}}$ &  &  & 0.35 & 0.43 & 0.44 & 0.43 &  &  & -0.32 &  &  &  & 0.40 &  &  &  & -0.35 & -0.28\\
$L_{\text{C90}}$ & -0.33 &  &  &  &  &  & -0.34 & -0.34 & -0.55 & -0.53 & -0.57 & -0.50 &  &  &  &  & -0.75 & -0.73\\
$L_{\text{C95}}$ & -0.41 & -0.35 & -0.32 &  &  &  & -0.34 & -0.38 & -0.59 & -0.60 & -0.62 & -0.55 &  &  &  &  & -0.76 & -0.75\\
$L_{\text{Aeq}}$ &  &  &  &  &  &  &  &  &  &  &  &  & 0.92 &  & 0.28 & -0.63 & 0.37 & 0.41\\
$L_{\text{Amax}}$ & 0.53 & 0.37 &  &  &  &  & 0.46 & 0.46 & 0.56 & 0.55 & 0.48 & 0.47 & 0.87 &  & 0.35 & -0.73 & 0.75 & 0.78\\
$L_{\text{A5}}$ & 0.44 & 0.35 & 0.29 &  &  &  & 0.43 & 0.33 & 0.41 & 0.41 & 0.43 & 0.44 & 0.94 &  & 0.36 & -0.72 & 0.71 & 0.74\\
$L_{\text{A10}}$ & 0.38 & 0.31 &  & 0.13 &  &  & 0.40 &  & 0.35 & 0.34 & 0.39 & 0.40 & 0.95 &  & 0.35 & -0.70 & 0.69 & 0.72\\
$L_{\text{A50}}$ &  & 0.30 & 0.38 & 0.46 & 0.46 & 0.46 &  &  & -0.31 &  &  &  & 0.37 &  &  &  & -0.36 & -0.30\\
$L_{\text{A90}}$ & -0.33 &  &  &  &  &  & -0.50 & -0.40 & -0.58 & -0.55 & -0.56 & -0.48 &  &  &  &  & -0.87 & -0.82\\
$L_{\text{A95}}$ & -0.41 & -0.32 &  &  &  &  & -0.50 & -0.44 & -0.62 & -0.61 & -0.61 & -0.54 &  &  &  &  & -0.88 & -0.85\\
$N_{\text{max}}$ & 0.35 &  &  &  &  &  & 0.54 & 0.37 & 0.46 & 0.43 & 0.41 & 0.40 & 0.96 &  & 0.32 & -0.57 & 0.73 & 0.77\\
$N_{\text{5}}$ &  &  &  &  &  &  & 0.45 &  &  &  & 0.30 & 0.31 & 1.00 &  & 0.29 & -0.48 & 0.58 & 0.63\\
$N_{\text{10}}$ &  &  &  &  &  &  & 0.40 & 0.13 &  &  &  &  & 0.99 &  & 0.28 & -0.44 & 0.53 & 0.57\\
$N_{\text{50}}$ &  &  &  &  &  &  & -0.29 & -0.34 & -0.48 & -0.43 & -0.28 &  & 0.33 &  &  &  & -0.50 & -0.44\\
$N_{\text{90}}$ & -0.41 & -0.34 & -0.29 &  &  &  & -0.36 & -0.39 & -0.61 & -0.59 & -0.58 & -0.51 &  &  &  &  & -0.80 & -0.78\\
$N_{\text{95}}$ & -0.45 & -0.40 & -0.36 &  &  &  & -0.34 & -0.38 & -0.61 & -0.61 & -0.60 & -0.54 &  &  &  &  & -0.80 & -0.79\\
$S_{\text{max}}$ & 0.56 & 0.49 & 0.42 &  &  &  &  & 0.43 & 0.56 & 0.60 & 0.56 & 0.52 & -0.29 & 0.32 &  & -0.36 & 0.33 & 0.34\\
$S$ &  &  &  &  &  &  & -0.28 &  &  &  &  &  & -0.33 & 0.34 &  & -0.50 &  & \\
$S_{\text{5}}$ & 0.40 & 0.41 & 0.40 & 0.30 &  &  &  &  & 0.30 & 0.37 & 0.46 & 0.45 &  & 0.34 &  & -0.62 &  & \\
$S_{\text{10}}$ & 0.29 & 0.32 & 0.33 & 0.28 &  &  &  &  &  & 0.28 & 0.41 & 0.41 &  & 0.33 &  & -0.59 &  & \\
$S_{\text{50}}$ &  &  &  &  &  &  &  &  &  &  &  &  & -0.33 & 0.33 &  & -0.49 &  & \\
$S_{\text{90}}$ &  &  &  &  &  &  & -0.48 &  &  &  &  &  & -0.52 & 0.31 &  & -0.33 & -0.33 & -0.32\\
$S_{\text{95}}$ &  &  &  &  &  &  & -0.49 &  &  &  &  &  & -0.55 & 0.31 &  & -0.30 & -0.37 & -0.36\\
$R_{\text{max}}$ &  &  &  &  &  &  & 0.58 & 0.38 &  &  &  &  & 0.63 &  &  &  &  & \\
$R$ &  &  &  &  &  &  &  &  & -0.30 &  &  &  & 0.37 &  &  & 0.39 & -0.36 & -0.31\\
$R_{\text{5}}$ &  &  &  &  &  &  & 0.28 &  &  &  &  &  & 0.57 &  &  & 0.28 &  & \\
$R_{\text{10}}$ &  &  &  &  &  &  &  &  &  &  &  &  & 0.54 &  &  & 0.32 &  & \\
$R_{\text{50}}$ &  &  &  &  &  &  &  &  & -0.34 & -0.31 &  &  & 0.33 &  &  & 0.40 & -0.41 & -0.36\\
$R_{\text{90}}$ & -0.34 & -0.29 &  &  &  &  &  & -0.29 & -0.46 & -0.43 & -0.30 &  &  &  &  & 0.46 & -0.61 & -0.56\\
$R_{\text{95}}$ & -0.37 & -0.32 & -0.28 &  &  &  &  & -0.30 & -0.48 & -0.45 & -0.32 &  &  &  &  & 0.46 & -0.62 & -0.58\\
$T_{\text{max}}$ & 1.00 & 0.88 & 0.75 & 0.47 & 0.44 & 0.43 & 0.37 & 0.77 & 0.84 & 0.85 & 0.61 & 0.56 &  &  & 0.31 & -0.45 & 0.44 & 0.45\\
$T_{\text{5}}$ & 0.88 & 1.00 & 0.97 & 0.80 & 0.77 & 0.75 &  & 0.43 & 0.53 & 0.58 & 0.54 & 0.54 &  &  & 0.33 & -0.40 & 0.32 & 0.36\\
$T_{\text{10}}$ & 0.75 & 0.97 & 1.00 & 0.93 & 0.89 & 0.87 &  &  & 0.35 & 0.42 & 0.47 & 0.51 &  &  & 0.33 & -0.35 &  & 0.30\\
$T_{\text{50}}$ & 0.47 & 0.80 & 0.93 & 1.00 & 0.99 & 0.97 & -0.35 &  &  &  &  & 0.35 &  &  &  &  &  & 0.13\\
$T_{\text{90}}$ & 0.44 & 0.77 & 0.89 & 0.99 & 1.00 & 1.00 & -0.38 &  &  &  &  & 0.29 &  &  &  &  &  & \\
$T_{\text{95}}$ & 0.43 & 0.75 & 0.87 & 0.97 & 1.00 & 1.00 & -0.38 &  &  &  &  &  &  &  &  &  &  & \\
$FS_{\text{max}}$ & 0.37 &  &  & -0.35 & -0.38 & -0.38 & 1.00 & 0.78 & 0.67 & 0.62 & 0.47 & 0.35 & 0.44 &  &  &  & 0.58 & 0.53\\
$FS_{\text{5}}$ & 0.77 & 0.43 &  &  &  &  & 0.78 & 1.00 & 0.93 & 0.91 & 0.55 & 0.44 &  &  &  &  & 0.44 & 0.40\\
$FS_{\text{10}}$ & 0.84 & 0.53 & 0.35 &  &  &  & 0.67 & 0.93 & 1.00 & 0.96 & 0.61 & 0.51 &  &  &  & -0.32 & 0.61 & 0.60\\
$FS_{\text{50}}$ & 0.85 & 0.58 & 0.42 &  &  &  & 0.62 & 0.91 & 0.96 & 1.00 & 0.75 & 0.67 &  &  &  & -0.35 & 0.59 & 0.58\\
$FS_{\text{90}}$ & 0.61 & 0.54 & 0.47 &  &  &  & 0.47 & 0.55 & 0.61 & 0.75 & 1.00 & 0.98 & 0.31 &  & 0.34 & -0.30 & 0.62 & 0.65\\
$FS_{\text{95}}$ & 0.56 & 0.54 & 0.51 & 0.35 & 0.29 &  & 0.35 & 0.44 & 0.51 & 0.67 & 0.98 & 1.00 & 0.33 & 0.13 & 0.35 & -0.30 & 0.56 & 0.61\\
\textit{PA} &  &  &  &  &  &  & 0.44 &  &  &  & 0.31 & 0.33 & 1.00 &  & 0.30 & -0.51 & 0.58 & 0.63\\
\textit{ISOPL} &  &  &  &  &  &  &  &  &  &  &  & 0.13 &  & 1.00 &  &  &  & \\
\textit{ISOEV} & 0.31 & 0.33 & 0.33 &  &  &  &  &  &  &  & 0.34 & 0.35 & 0.30 &  & 1.00 & -0.30 & 0.30 & 0.34\\
$L_{\text{Ceq}}-L_{\text{Aeq}}$ & -0.45 & -0.40 & -0.35 &  &  &  &  &  & -0.32 & -0.35 & -0.30 & -0.30 & -0.51 &  & -0.30 & 1.00 & -0.47 & -0.48\\
$L_{\text{A10}}-L_{\text{A90}}$ & 0.44 & 0.32 &  &  &  &  & 0.58 & 0.44 & 0.61 & 0.59 & 0.62 & 0.56 & 0.58 &  & 0.30 & -0.47 & 1.00 & 0.99\\
$L_{\text{C10}}-L_{\text{C90}}$ & 0.45 & 0.36 & 0.30 & 0.13 &  &  & 0.53 & 0.40 & 0.60 & 0.58 & 0.65 & 0.61 & 0.63 &  & 0.34 & -0.48 & 0.99 & 1.00\\

\bottomrule
\end{longtable}
\end{sffamily}

\setcounter{table}{1}
\renewcommand{\thetable}{\Alph{section}.\arabic{table}}
\twocolumn

\section{Circumplexity analysis}
\label{sec:circumplex}

Circumplexity tests are based on inherent characteristics of a circumplex model \citep{Tracey2000,Locke2019}. In a circumplex model, intercorrelation should conform to the inequality requirement of $P_1>P_2>P_3>P_4$, where correlations of adjacent variables (\textit{$P_1$}) must be greater than orthogonal variable correlations (\textit{$P_2$}), and \textit{$P_2$} itself must be greater than correlations of variables \SI{135}{\degree} apart (\textit{$P_3$}), and \textit{$P_3$} in turn must be greater than the correlations of opposing variables on each axis (\textit{$P_4$}). An 8-attribute model consists of a total of 28 correlation pairs. All PAQ correlation pairs were computed via the Pearson's method and examined for compliance to the correlation inequality \citep{Lam2022c}, as shown in \Cref{tab:corPCA}. 

The circumplex inequality criteria is further examined through the RTHOR correspondence index (CI), wherein –1 indicates complete violation, 0 indicates chance, and 1.0 indicates a perfect fit \citep{Tracey2000}. A score of 0.5 indicates that \SI{75}{\percent} of the predictions satisfied the inequality criteria. A total of 288 predictions were tested for violation \citep{Zeigler-Hill2010,Locke2019}, where 20 of the 288 predictions violated the inequality criteria, yielding a CI of 0.861 ($p<0.001$). This reflects a low rate of inequality violation (\SI{6.94}{\percent} violation) and signifies a good adherence to the underlying circumplexity of the PAQ model.

Principal components analysis of the 8-attribute responses yield two components corresponding to the \textit{p-a} and \textit{p-a} axes, respectively, which together explains \SI{67}{\percent} of the total variance (\textit{p-a}:\SI{33.6}{\percent}; \textit{e-u}: \SI{33.4}{\percent}). On the \textit{p-a} axis, the loadings exhibited expected behaviour, i.e. positive loadings for \textit{v}, \textit{p}, and \textit{ca}; and negative loadings for \textit{m}, \textit{a}, and \textit{ch}. Similarly for the \textit{e-u} axis, loadings for \textit{ch}, \textit{e} and \textit{v} were positive, while loadings for \textit{ca}, \textit{u}, and \textit{m} were negative. The sinusoidality of the loadings lends further evidence to the circumplexity of the PAQ model \citep{Locke2019}.

\hspace{2em}
\begin{sffamily}
\setcounter{table}{0}
\refstepcounter{table}\label{tab:corPCA}

\scriptsize
\setlength{\tabcolsep}{1pt}
\noindent\textbf{\color{scolor}Table \thetable}\par%

\noindent{Pearson's correlation and principal component analysis loadings of PAQ attributes.} 

\centering
\begin{tabularx}{\linewidth}{@{\extracolsep{\fill}}X*{10}{>{\raggedleft\arraybackslash}X}@{}}
\toprule
\multicolumn{1}{c}{\textbf{ }} 
& \multicolumn{8}{c}{\textbf{Pearson's Correlation}} 
& \multicolumn{2}{c}{\textbf{Loadings}} \\
\cmidrule(l{3pt}r{3pt}){2-9} \cmidrule(l{3pt}r{3pt}){10-11}
  
& \textit{e} 
& \textit{v} 
& \textit{p} 
& \textit{ca} 
& \textit{u} 
& \textit{m} 
& \textit{a} 
& \textit{ch} 
& p-a 
& e-u\\

\midrule
\textit{e} & 1.00 & $P_1$ & $P_2$ & $P_3$ & $P_4$ & $P_3$ & $P_2$ & $P_1$ & 0.54 & 0.71\\
\textit{v} & 0.59 & 1 & $P_1$ & $P_2$ & $P_3$ & $P_4$ & $P_3$ & $P_2$ & 0.71 & 0.38\\
\textit{p} & 0.15 & 0.48 & 1 & $P_1$ & $P_2$ & $P_3$ & $P_4$ & $P_3$ & 0.81 & -0.36\\
\textit{ca} & 0.04 & 0.31 & 0.76 & 1 & $P_1$ & $P_2$ & $P_3$ & $P_4$ & 0.68 & -0.48\\
\textit{u} & -0.69 & -0.45 & -0.12 & 0.01 & 1 & $P_1$ & $P_2$ & $P_3$ & -0.54 & -0.61\\
\textit{m} & -0.40 & -0.32 & -0.23 & -0.08 & 0.49 & 1 & $P_1$ & $P_2$ & -0.56 & -0.31\\
\textit{a} & 0.18 & -0.03 & -0.54 & -0.52 & 0 & 0.17 & 1 & $P_1$ & -0.53 & 0.67\\
\textit{ch} & 0.34 & 0.16 & -0.33 & -0.34 & -0.2 & 0.01 & 0.67 & 1 & -0.25 & 0.76\\
\bottomrule
\end{tabularx}
\end{sffamily}

\twocolumn

\bibliographystyle{cas-model2-names}
\bibliography{references}





\end{document}